\documentclass[longbibliography,
twocolumn,
showpacs,
floatfix,
aps,
prb,
amsmath,
amssymb,
superscriptaddress,
10pt]{revtex4-2}

\usepackage[utf8]{inputenc}

\usepackage{amsmath} 
\usepackage{graphicx}
\usepackage{dcolumn}
\usepackage{bm}
\usepackage{braket} 
\usepackage{epsfig} 
\usepackage[pdfencoding=auto, psdextra,colorlinks=true,citecolor=blue,linkcolor=magenta,hypertexnames=false]{hyperref} 
\usepackage{color} 
\usepackage[normalem]{ulem} 
\usepackage{cancel} 
\usepackage{stmaryrd}
\usepackage{comment}
\usepackage[capitalize]{cleveref}
\usepackage{dsfont}
\usepackage{lineno} 

\usepackage{amsfonts}
\usepackage[bbgreekl]{mathbbol}
\DeclareSymbolFontAlphabet{\mathbbm}{bbold}
\DeclareSymbolFontAlphabet{\mathbb}{AMSb}

\newcommand{\cjs}[1]{\textcolor{cyan}{\ifmmode\mathrm{\cancel{\ensuremath{#1}}}\else\sout{#1}\fi}}

\newcommand{\btilde}[1]{\bar{\tilde{#1}}}
\newcommand\abs[1]{\left|#1\right|}
\newcommand{\anth}[0]{\alpha_{\mathrm{nth}}}
\newcommand{\etaB}[1]

\renewcommand{\Im}{\operatorname{\mathrm{Im}}}
\renewcommand{\Re}{\operatorname{\mathrm{Re}}}

\newcommand{\pref}[2]{Fig.~\hyperref[#1]{\ref*{#1}(#2)}}

\begin{document}
\title{Non-thermal pairing glue of electrons in the steady state}
\author{Michele Pini}
\thanks{These two authors contributed equally.}
\affiliation{Theoretical Physics III, Center for Electronic Correlations and Magnetism,
Institute of Physics, University of Augsburg, 86135 Augsburg, Germany}
\affiliation{Max Planck Institute for the Physics of Complex Systems, 01187 Dresden, Germany}
\author{Christian H. Johansen}
\thanks{These two authors contributed equally.}
\affiliation{Pitaevskii BEC Center, CNR-INO and Department of Physics, University of Trento, Italy}
\affiliation{Max Planck Institute for the Physics of Complex Systems, 01187 Dresden, Germany}
\author{Francesco Piazza}
\affiliation{Theoretical Physics III, Center for Electronic Correlations and Magnetism,
Institute of Physics, University of Augsburg, 86135 Augsburg, Germany}
\affiliation{Max Planck Institute for the Physics of Complex Systems, 01187 Dresden, Germany}
\date{\today}

\begin{abstract}
The study of mechanisms for enhancing superconductivity has been a central topic in condensed matter physics due to the combination of fundamental and technological interests. One promising route is to exploit non-equilibrium effects in the steady state. Efforts in this direction have so far focused on enhancing the pairing mechanism known from thermal equilibrium through modified distributions for the electrons or the bosons mediating the electron-electron interaction.
In this work, we identify an additional pairing mechanism that is active only outside thermal equilibrium. By generalizing Eliashberg theory to non-equilibrium steady states using the Keldysh formalism, we derive a set of Eliashberg equations that capture the effect of this genuinely non-thermal pairing glue even in the weak-coupling regime. We discuss two examples where this mechanism has a major impact.
First, in a temperature-bias setup, we find that superconductivity is enhanced when the boson mediator is colder than the electrons. Second, we find that an incoherent drive of the boson mediator at energies much greater than the temperature pushes the system far from thermal equilibrium but leaves the critical coupling essentially unchanged, owing to a competition between electron heating and the enhancement of pairing by the non-thermal glue.

\end{abstract}
\maketitle

\section{Introduction}
\label{sec:introduction}
In recent years, a series of experiments on transient light-induced superconductivity \cite{Fausti2011,Nicoletti2014,Hu2014,Mitrano2016,Cantaluppi2018,Budden2021,Isoyama2021,Rowe2023} has reignited the interest in discovering non-thermal mechanisms to enhance or induce superconductivity, either in a transient or a non-equilibrium steady state (NESS). 
Focusing on systems in the steady state, the idea of using photons confined within a cavity and coupled to quantum materials \cite{Garcia-Vidal2021,Schlawin2022,Bloch2022} appears as a promising direction, as the confinement-enhanced light-matter coupling allows to avoid intense driving, still appreciably affecting the material. In particular, experiments have indicated that this might be used
to modify superconducting pairing \cite{Thomas2019,keren2025cavity}. The simplest model for a light-induced non-thermal steady-state of matter that enhances superconductivity was first considered in the context of the Eliashberg \emph{effect} \cite{Eliashberg1970,*Eliashbergrus,IvlevEliashberg1973,Wyatt1966,Dayem1967,Klapwijk1977}, recently explored also in cavity settings \cite{CurtisGalitski2019,Islam2025}. The idea is to directly drive the electrons into a non-thermal distribution that is more prone to electron-pair formation. An alternative route is to instead focus on the boson mediating electron-electron interactions that lead to pairing and superconductivity, either by directly employing cavity photons as mediators  \cite{Schlawin2019,Gao2020,Chakraborty2021,Andolina2024,Chakraborty2025}, or by coupling them to a degree of freedom in the material (phonon, surface plasmon, etc.) \cite{SentefRubio2018,LuRubio2024,Eckhardt2024,Valerii2025}. The latter option, albeit with intense laser light (instead of cavity light) and thus confined to a transient, is also the one employed by the experiments mentioned at the beginning.

When considering all the above scenarios, we find that the attention has so far focused on enhancing the ``glue'' that is responsible for pairing also at thermal equilibrium. Indeed, even in the cases where non-equilibrium effects are directly considered, the main approach consists of replacing the thermal distributions of the electrons \cite{Eliashberg1970,IvlevEliashberg1973,Wyatt1966,Dayem1967,Klapwijk1977,CurtisGalitski2019,Islam2025} or of the boson mediators \cite{Eckhardt2024,Chakraborty2025} in the gap equation with their non-equilibrium counterparts. This amounts to modifying the pairing mechanism that is already active at thermal equilibrium. In this work, we identify instead an additional pairing mechanism that is active only when boson mediators and electrons are not in mutual thermal equilibrium.
This mechanism manifests itself as a new independent equation that couples to the regular gap equation. Out of equilibrium, the extended system of equations has to be solved simultaneously, giving rise to a genuinely non-thermal pairing mechanism that \emph{cannot} be described by a gap equation that differs from the thermal equilibrium one just by the functional form of energy distributions or by the value of the coupling constants. 

A good starting point to describe a system of boson mediators and superconducting electrons is Eliashberg \emph{theory} of phonon-mediated superconductivity \cite{Eliashberg1960,Eliashberg1961} (for recent reviews, see~\cite{MarsiglioReview2020,ChubukovReview2020}). 
This theory improves on BCS theory by including retardation effects (i.e.~a frequency dependence) in the electron-electron interaction due to the dynamics of the boson mediator. 
At thermal equilibrium, these retardation effects are typically important in intermediate and strong-coupling regimes \cite{Marsiglio1991,Combescot1995}, but introduce only minor corrections to BCS theory in the weak-coupling regime \cite{WangChubukov2013,Marsiglio2018,MirabiMarsiglio2020}.
In contrast, we find that, outside of thermal equilibrium, retardation effects are essential also in the weak-coupling regime, as they capture important non-thermal contributions in the gap equation that are missing in BCS theory. 

As a first step, we generalize Eliashberg theory to the case of a time-translation-invariant NESS employing the Keldysh diagrammatic formalism \cite{Sieberer2016,KamenevBook}.
Differently from previous dynamical implementations, which considered momentum-averaged models of transient \cite{Kemper2015,Sentef2016,Grunwald2024} or quasistationary Floquet-driven \cite{Murakami2017,Babadi2017} superconductors,
we also go beyond momentum-averaging, which turns out to be necessary to generate the new non-thermal contribution to the pairing.
As a specific implementation, we consider the case of a forward-scattering electron-electron interaction, which can describe phonon-mediated interactions in FeSe thin films on dielectric substrates \cite{Rademaker2016,Wang2016,SentefRubio2018} and photon-mediated interactions in driven cavity materials \cite{Gao2020,Chakraborty2021}. In particular, we consider the perfect forward-scattering limit, where the momentum transferred by the boson mediator is the smallest momentum scale in the system. This limit makes our equations analytically tractable in the weak-coupling regime, while still qualitatively reproducing the behavior of small but finite momentum transfer \cite{Verelogiannis1996,Danylenko1999,KunSondhi2000,Rademaker2016}.
While we find that this model is particularly suited for highlighting non-thermal pairing effects, we remark that the non-equilibrium pairing mechanism that we identified is not limited to the perfect forward-scattering case. 
For completeness, we also note that there are subtleties regarding the extent to which beyond-mean-field effects may modify the nature of the phase transition in this model. Further details are discussed in \cref{sec:model}.

Applying our formalism to this perfect forward-scattering model, we compute both analytically and numerically the steady-state properties at the superconducting transition for two different examples: (i) a temperature-bias setup with bosons and electrons coupled to two different thermal baths (similarly to  Ref.~\cite{FloresCalderon2025}) and (ii) a setup where the bosons are incoherently driven at energies much greater than the temperature (similarly to Ref.~\cite{Eckhardt2024}).

In both setups, the non-thermal pairing mechanism significantly affects the physics of the superconducting phase transition. Specifically, the critical coupling $g_c$ in the weak-coupling regime can be expressed in terms of three self-consistently determined parameters: the effective electron temperature $T_e$, the electron spectral width at the Fermi surface (FS) $\eta$, and the non-thermal pairing parameter $\anth$. The first two describe modifications of the electron distribution and spectrum, respectively. 
The last parameter describes instead a genuinely non-thermal effect, as it arises from the breaking of the thermal fluctuation-dissipation relations in the anomalous self-energy. It is non-zero only outside of thermal equilibrium and its sign determines whether it contributes to enhance or suppress electron pairing with respect to the thermal case.

The paper is organized as follows. In \cref{sec:key_results} we summarize the key results of the paper and verify their validity for the two examples. 
\cref{sec:model} introduces the model. \cref{sec:NESS-Eliasberg_eqs} discusses the structure of the NESS-Eliashberg equations.
In \cref{sec:nonThDeriv}, expressions for the spectral gap and the critical coupling in the NESS are derived.
\cref{sec:num_sol} discusses in more detail the two considered examples, and finally \cref{sec:conclusion} contains our conclusions. Additional details on the derivation and formal structure of the theory, as well as details on the methods to solve the NESS-Eliashberg equations, are given in \cref{app:pathIntegral,app:th_eq,app:num_methods,app:causal_structure_MF,app:GK_FK,app:boson_self-energy,app:QuadInt,app:cryostat_model,app:T-bias_wc,app:incoherent_wc,app:quasi-classical_approx}, which are summarized at the end of the main text. 

Throughout this paper, we will work in units where the Boltzmann constant $k_\mathrm{B}$ and the reduced Planck constant $\hbar$ are equal to unity.

\section{Key results}
\label{sec:key_results}

In this Section, we present our key result: an analytical expression for the critical coupling of the superconducting phase transition in the NESS, which explicitly contains an additional contribution from the genuinely non-thermal pairing mechanism, as illustrated in \pref{fig:setup}{a}. We then briefly discuss the main predictions and the underlying physical picture for two different examples, whereby we also benchmark our analytics with the numerical solution of the NESS Eliashberg equations.

\subsection{Superconducting critical point in a NESS at weak coupling}
\label{subsec:gc_KR}
\begin{figure}[tbp]
\centering
\includegraphics[width=0.95\columnwidth]{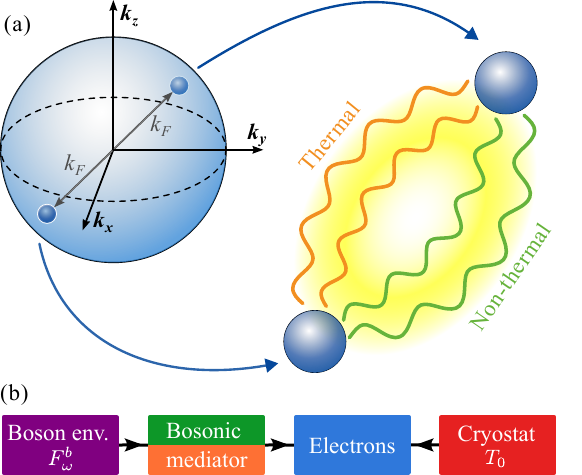}
\caption{(a) Illustration of the coexistence of both thermal and non-thermal pairing mechanisms between two electrons at opposite sides of the FS in the NESS. Both  mechanisms originate from interactions mediated by the bosonic mode. (b) Sketch of the generic setup used to consider both thermal and NESS superconductivity. It consists of a system composed of electrons which interact through a bosonic mediator. 
The system is coupled to two different environments: one coupled to the bosons, defined by the distribution $F^b_\omega$, and a second coupled directly to the electrons. The latter is a thermal bath at a fixed temperature $T_0$ and therefore referred to as the ``cryostat''.}
\label{fig:setup}
\end{figure}

Consider an interacting system of bosons and electrons in a NESS that is generated by linearly coupling them to two different environments, as sketched in \pref{fig:setup}{b}.
The bosons have a spectrum with a characteristic frequency $\omega_0$, spectral width $\kappa$ and mediate an attractive forward-scattering interaction for the electrons with coupling $g$ (see \cref{sec:model} for more details on the model).
We consider a regime where the coupling between bosons and electrons is small compared to $\omega_0$ and $\kappa$, such that the feedback of the electrons on the bosons is negligible (see \cref{app:boson_self-energy}).

The key result of our paper is that the critical coupling $g_c$ of the superconducting phase transition in the NESS is modified by a genuinely non-thermal contribution.
This becomes evident in the weak-coupling regime with well-defined electronic quasiparticles, where the critical coupling can be approximated as (see \cref{subsec:critical_coupling})
\begin{equation}
    \label{eq:gc}
    g_c \approx \left[\frac{1}{g_{c,\mathrm{th}}(T_e)}-\frac{\anth }{4\eta \bar{\omega}_0 }\right]^{-1},
\end{equation}
where
\begin{equation}
    \label{eq:gc_th}
    g_{c,\mathrm{th}}(T_e)=  4 T_e \bar{\omega}_0
\end{equation}
is the weak-coupling expression for the critical coupling at thermal equilibrium and $\bar{\omega}_0=\omega_0+\kappa^2/\omega_0$ is the renormalized frequency of the boson due to the finite spectral width. In these expressions, $\eta$ denotes the spectral width (or inverse damping rate) of the electrons at the FS induced by the interaction with the bosons and the eventual coupling to the cryostat, while $T_e$ is the effective temperature of the electrons in the NESS which is determined by the slope of the electron distribution close to the FS. Finally, $\anth$ is a non-thermal pairing parameter that appears only away from thermal equilibrium, resulting in $g_c \neq g_{c,\mathrm{th}}$ whenever $\anth \neq 0$. 
Having $\anth\neq 0$, therefore, signifies that the non-thermal mechanism illustrated in \pref{fig:setup}{a} is activated.
These three quantities generally depend on properties of the NESS and they need to be self-consistently determined by solving the NESS-Eliashberg equations (see \cref{sec:NESS-Eliasberg_eqs,sec:nonThDeriv}). We show this by considering two different protocols for generating the NESS.
For these two examples, we have derived analytical weak-coupling expressions describing the properties of the NESS (see \cref{sec:num_sol}). 

\begin{figure*}[tbp]
\centering
\includegraphics[width=0.8\textwidth]{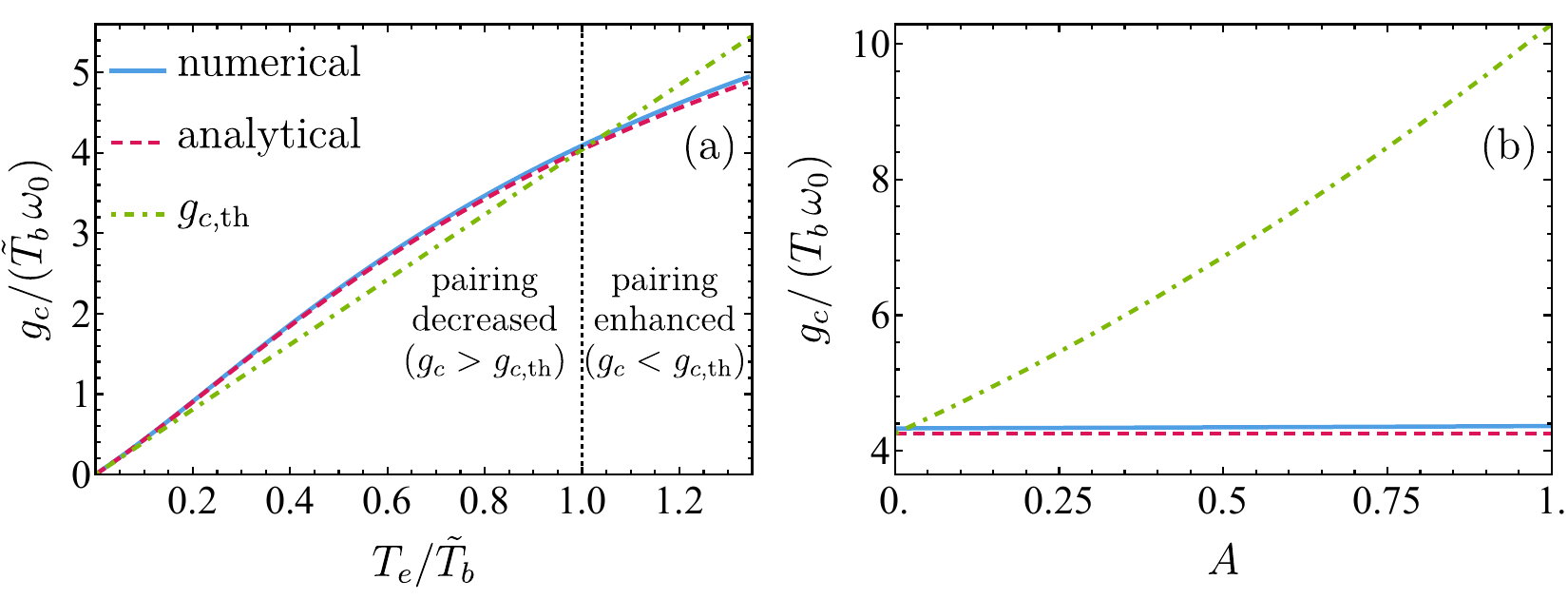}
\caption{
Critical coupling $g_c$ for the two different example settings for NESS superconductivity. The ``numerical'' (blue solid) lines are the numerically converged self-consistent solutions of \cref{eq:selfEnergy_EQs} in the NESS. The ``analytical'' (red dashed) lines correspond to the self-consistent analytical solutions in the weak-coupling regime. The ``$g_{c,\mathrm{th}}$'' lines (green dot-dashed) correspond to the thermal equilibrium weak-coupling approximation given by \cref{eq:gc_th} evaluated at $T_e$.
(a) Critical coupling $g_c$ in the temperature-bias setting as a function of the effective electron temperature $T_e$ at fixed boson temperature $T_b=\tilde{T}_b=10^{-3}\omega_0$.
The two curves intersect when $T_e=\tilde{T}_b$ in the NESS (vertical black dotted line). The analytical weak-coupling solution is calculated via \cref{eq:gc_closedform_T-bias_wc}. 
The bath-induced spectral widths for bosons and electrons are $\kappa/\omega_0=0.1$ and $\eta_0/\omega_0=10^{-6}$, respectively. (b) Critical coupling $g_c$ in the example with incoherent driving of the bosons as a function of the driving strength $A$. 
The boson spectral width is $\kappa/\omega_0=0.25$ and the thermal (low energy) part of the boson distribution is described by the temperature $T_b/\omega_0=10^{-3}$. The incoherent drive [parametrized in \cref{eq:Fd_inc}] is centered at the boson frequency $\omega_0$ and has a Gaussian envelope with a width $v/\omega_0=0.05$.}
\label{fig:KR}
\end{figure*}

Before turning to the specific examples, let us stress a property of our newly introduced pairing mechanism that demonstrates its genuinely non-thermal nature. It is known from Eliashberg theory at thermal equilibrium, that in the weak-coupling limit the usual BCS result has to be obtained, apart from minor corrections \cite{Combescot1990,Marsiglio2018,MirabiMarsiglio2020}. In our \cref{eq:gc}, $g_{c,\text{th}}$ indeed corresponds to the BCS result in a perfect forward-scattering case \cite{KunSondhi2000,Rademaker2016}. However, in our NESS, we find that the additional pairing mechanism containing $\anth$ remains finite in the weak-coupling limit. This immediately tells us that it is not smoothly connected to any thermal phenomena, because the latter are fully contained in the BCS expression.

\subsection{NESS examples}
\label{subsec:KR_examples}
To test the validity of \cref{eq:gc} and get a better understanding of the non-thermal effects, we have carefully analyzed two qualitatively different scenarios for generating NESSs.

{\bf Temperature bias} --
In the first example we consider a situation where the system is subjected to a temperature bias, i.e.~the boson environment and the cryostat in \pref{fig:setup}{b} are two thermal baths with temperatures $T_b$ and $T_0$, respectively. 
 We assume that the coupling of the bosons to their bath is strong enough for them to be treated as thermal at temperature $T_b$. The  effective temperature $T_e$ of the electrons generically ends up between the two external bath temperatures. 
Within this setting, we can either generate a NESS by keeping $T_b$ fixed to a constant value and vary $T_0$, or remain in thermal equilibrium under simultaneous variation of the baths’ temperatures $T_b=T_0$.
Such a setting can be realized, for example, in the case of cavity implementations, where the electrons are in contact with a cryostat, while the cavity photons are in contact with an external electromagnetic environment, as theoretically considered in Refs.~\cite{CurtisGalitski2019,FloresCalderon2025,Islam2025} and experimentally realized in Ref.~\cite{JarcFausti2023}. 

Within this setting, weak-coupling analytical self-consistent solutions for $\eta$, $T_e$ and $\anth$ can be found (see \cref{subsec:results_T-bias}). In particular, in the NESS close to thermal equilibrium ($T_0\approx T_e \approx T_b$) we find that 
\begin{equation}
    \anth \approx \mathcal{C} \, \frac{T_b-T_e}{T_e},
    \label{eq:alpha_expansion}
\end{equation}
where $\mathcal{C}$ is a positive constant that depends on $T_b$ and the spectrum of both bosons and electrons. Following the discussion of \cref{eq:gc}, we see that the pairing is non-thermally enhanced ($\anth<0$) for bosons cooler than electrons ($T_b<T_e$) and decreased ($\anth>0$) for hotter bosons ($T_b>T_e$).

To highlight this result, \pref{fig:KR}{a} shows the numerical result for the critical coupling $g_c$ in the NESS against the effective electron temperature $T_e$ obtained by keeping the temperature of the bosonic bath fixed to $T_b=\tilde{T}_b=10^{-3}\omega_0$ and varying the cryostat temperature $T_0$,
compared to the thermal equilibrium result where the temperatures $T_b=T_e=T_0$ are all varied together.
Note that, by plotting $g_c$ against $T_e$ instead of against $T_0$, only genuinely non-thermal effects lead to deviations from the thermal equilibrium result, as ``thermal-like'' equilibration effects are already absorbed in the value of $T_e\neq T_0$. 
In \pref{fig:KR}{a}, we observe genuinely non-thermal deviations
throughout the whole temperature range, with the pairing in the NESS being non-thermally enhanced for $T_b<T_e$ and decreased for $T_b>T_e$, in line with the result sketched in \cref{eq:alpha_expansion}.

This example clearly shows that one can separate two-types of non-equilibrium effects: 1) ``thermal-like'' effects, whereby objects that were defined already at thermal equilibrium still exist in the NESS but with a modified value, and 2) ``genuinely non-thermal'' effects, whereby new objects that did not exist at thermal equilibrium appear. In the present case, the NESS modification of the electron temperature and damping belong to category 1), while $\anth$ belongs to 2), and so does in particular its effect on the critical coupling, corresponding to the difference between the green dot-dashed and the blue curve in \pref{fig:KR}{a}.

Combining the weak-coupling analytical self-consistent solutions for $\eta$, $T_e$ and $\anth$ with \cref{eq:gc}, we also find a closed analytical expression for the critical coupling in the NESS [see \cref{eq:gc_closedform_T-bias_wc}]. This analytical solution is also plotted in \pref{fig:KR}{a} and agrees very well with the numerical result, confirming the validity of \cref{eq:gc} to capture the main contributions to the critical coupling. 

More detailed results and discussions, including numerical and analytical results for $\eta$, $T_e$ and $\anth$, are given in \cref{subsec:results_T-bias}.

{\bf Incoherent driving of bosons} --
As a complementary example, we consider decoupling the cryostat from the electrons and instead generate a NESS by driving bosons at a temperature $T_b$ incoherently around their resonance frequency $\omega_0$ with a driving strength $A$ and a width $v$ [see the parametrization in \cref{eq:Fd_inc}].
Such protocol should be possible within the setting of cavity-mediated superconductivity, where the cavity \cite{Gao2020} can be driven with a high level of control. Incoherent driving of the bosons in superconductors was considered in \cite{Eckhardt2024}, but their model was fundamentally different and they did not include non-thermal effects at a self-consistent level. 

In this case, the physics is different from the temperature-bias case, as the drive ($\omega\approx\omega_0$) and thermal ($\omega\approx0$) effects are energetically well separated.  
By increasing the drive strength the electrons heat up [see \cref{eq:incTe}] resulting in an increase of the thermal part of the critical coupling $g_{c,\mathrm{th}}$ and making it harder to reach a superconducting state.
This is seen in \pref{fig:KR}{b} where both $g_c$ and $g_{c,\mathrm{th}}$ are plotted as a function of the drive strength.
Taking into account the genuinely non-thermal effects, one observes that $g_c$ does not follow the same behavior as $g_{c,\text{th}}$ but is instead essentially unperturbed by the drive
\begin{equation}
    g_c\approx g_{c,\text{th}}(T_b)=g_{c}(A=0).
\end{equation}
The stability of $g_c$ against incoherent driving can be understood from the closed-form expressions for $T_e$, $\eta$, and $\anth$ (see \cref{sec:inc}), combined with \cref{eq:gc}. 
Using the same language as for the discussion of the previous example, we can describe the physics of the present example as follows. While the ``thermal-like'' effects increase $T_e$ and $\eta$, and thus also $g_c$, the ``genuinely non-thermal'' mechanism $\anth$ enhances the bosons’ ability to pair electrons in equal measure, resulting in a net cancellation of the effects of the incoherent drive.  
Additionally, we are able to show that this result is independent of the exact protocol for the incoherent drive, as long as the energy separation and weak-coupling criteria are satisfied. 
The incoherent driving of the bosons at energies much larger than $T_e$ pushes the system into a highly non-thermal state while only weakly affecting the critical coupling.

In summary, we find that our approximation in \cref{eq:gc} works well in the weak-coupling regime.
In the two qualitatively different protocols we have analyzed, we find that the genuinely non-thermal pairing mechanism that we propose here can have a substantial impact.
Furthermore, our approach allows for a deep understanding of the weak-coupling regime, as we can derive analytical expression for the relevant quantities. 
Our result highlight the genuinely non-thermal pairing as an additional tuning knob for controlling the superconducting threshold.     

\section{Model}
\label{sec:model}
Having presented our key results, we now turn to a more detailed discussion of the underlying model.
To investigate superconductivity in a non-equilibrium steady state (NESS), we consider a system coupled to external environments, as illustrated in \pref{fig:setup}{b}. The total Hamiltonian reads
\begin{equation}
    \hat{H}_{\mathrm{tot}}=\hat{H}_{\mathrm{sys}}+\hat{H}_{\mathrm{env}},
    \label{eq:H_tot}
\end{equation}
where $\hat{H}_{\mathrm{sys}}$ describes an interacting electron–boson system and $\hat{H}_{\mathrm{env}}$ accounts for the environments and their coupling to the system.

The system Hamiltonian can be decomposed as
\begin{equation}
    \hat{H}_{\mathrm{sys}}=\hat{H}_{e,0}+\hat{H}_{b,0}+\hat{H}_{e-b},
    \label{eq:H_sys}
\end{equation}
where
\begin{equation}
    \hat{H}_{e,0}=\sum_{\bm{k}}\sum_{\sigma=\uparrow,\downarrow}
    \xi_{\bm{k}}
    \hat{c}^{\dag}_{\sigma, \bm{k}} \hat{c}_{\sigma, \bm{k}}
    \label{H_0e}
\end{equation}
is the Hamiltonian for free electrons with mass $m$ and chemical potential $\mu$, such that their dispersion is
$\xi_{\bm{k}}=\bm{k}^2/2m-\mu$. The Hamiltonian for the free bosons with dispersion $\omega_\mathbf{q}$ is
\begin{equation}
    \hat{H}_{b,0}=\sum_{\bm{q}} \omega_{\bm{q}} {\hat{a}^\dagger}_{\bm{q}} \hat{a}_{\bm{q}},
    \label{eq:H_0b_fs}
\end{equation}
and 
\begin{equation}
    \hat{H}_{e-b}=2 \sum_{\bm{q},\bm{k},\sigma} g_{\bm{q}} \,
    \hat{c}^\dagger_{\sigma, \bm{k}+\bm{q}}
    \hat{c}_{\sigma, \bm{k}}
    (\hat{a}^\dagger_{\bm{q}} + \hat{a}_{-\bm{q}})
    \label{eq:H_bf_fs}
\end{equation}
is the Hamiltonian for the electron-boson interaction, where the coupling constant assumes the forward-scattering form $g_{\bm{q}}=g_0 \, e^{-|\bm{q}|/q_0}$.
Such a form for the coupling has been employed to describe phonon-mediated interactions in monolayer FeSe on SrTiO$_3$ substrates~\cite{Rademaker2016,Wang2016}, a material that has also been considered in a cavity setup~\cite{SentefRubio2018}. 
Realistic values of $q_0$ in FeSe thin films are of the order of $0.1 k_F$ (where $k_F$ is the Fermi momentum), i.e.~they are rather small compared to electronic momentum scales \cite{SentefRubio2018}.
To simplify our calculation, we focus on the perfect forward-scattering limit $q_0\to0$, meaning that we assume that $1/q_0$ is by far the largest length scale in the system. This limit has the advantage of providing an analytically tractable model in the weak-coupling regime, while retaining the same qualitative features of the finite $q_0$ case \cite{Verelogiannis1996,Danylenko1999,KunSondhi2000,Rademaker2016}.
This perfect forward-scattering model has also been considered in the context of cavity-driven materials to describe photon-mediated superconductivity, since also in that case the momentum $q_0$ transferred by optical photons is much smaller than the electronic momentum scales~\cite{Gao2020,Chakraborty2021}.
In this limit, \cref{eq:H_0b_fs,eq:H_bf_fs} simplify to
\begin{equation}
    \hat{H}_{b,0}=\omega_0 \hat{a}^\dagger \hat{a},
    \label{eq:H_0b}
\end{equation}
where we dropped the momentum dependence and considered only the boson mode with frequency $\omega_0=
\omega_{\bm{q}=0}$, and 
\begin{equation}
    \hat{H}_{e-b}=2 g_0 \sum_{\bm{k},\sigma}
    \hat{c}^\dagger_{\sigma, \bm{k}}
    \hat{c}_{\sigma, \bm{k}}
    (\hat{a}^\dagger + \hat{a}),
    \label{eq:H_bf}
\end{equation}
where $g_0=g_{\bm{q}=0}$ is the boson-electron coupling strength at zero momentum transfer.

We note that, within this forward-scattering model, beyond-mean-field effects (not included in Eliashberg theory) might suppress superconductivity to low temperatures due to thermal fluctuations of collective modes~\cite{KunSondhi2000}. Even if this is the case, we remark that the phase transition predicted within our mean-field Eliashberg theory would anyway retain the physical meaning of a ``pseudogap-opening'' transition \cite{KunSondhi2000}, and be observed as such. This kind of transition has been directly observed in recent experiments on FeSe thin films, independently from the  ``zero-resistance'' superconducting phase transition \cite{Faeth2021}. Since our mean-field theory predicts only the former, for simplicity we will keep referring to it as ``superconducting phase transition'' throughout the paper.

To generate a NESS, we assume linear coupling between the system and the external environments (see \cref{app:QuadInt,app:cryostat_model}). The linearity allows the environmental degrees of freedom to be integrated out exactly, yielding renormalized spectra and distribution functions for both electrons and bosons [see \cref{eq:D_omega_RAK,eq:cothDist,eq:Sigma^cryo,eq:cryo_sub} in \cref{sec:NESS-Eliasberg_eqs}]. When the environments are not mutually thermalized, the system settles into a NESS.

For the bosons, we focus on the regime in which the electron–boson coupling $g_0$ is weak compared to the boson–environment coupling. In this case, the back-action of the electrons on the bosons is negligible (see \cref{app:boson_self-energy}). For the electrons, we consider the opposite regime: their coupling to the cryostat is weak relative to $g_0$. In this limit, the properties of the electrons in the interacting NESS can differ substantially from those determined solely by the cryostat. 

\section{NESS-Eliashberg equations}
\label{sec:NESS-Eliasberg_eqs}
 In this section, we present the structure of the NESS-Eliashberg equations. A detailed derivation, based on a real-time path-integral formalism, is provided in \cref{app:pathIntegral}. 
We focus on the two-point correlation functions, which contain information about both the spectral properties and the particles distribution. 
Since the NESS is non-thermal, neither the spectrum nor the distribution of the electrons are known a priori, and both must be determined self-consistently. 
To this end, the electron propagators acquire a $2\times 2$ matrix structure \cite{KamenevBook}
\begin{equation}
    \mathcal{G}_{x,x'}=
    \begin{pmatrix}
        \mathcal{G}^{K} & \mathcal{G}^{R} \\
        \mathcal{G}^{A} & 0 \\
    \end{pmatrix}_{x,x'},
    \label{eq:Full_G}
 \end{equation}
where $x=(\bm{r},t)$ denotes the space-time coordinate. We assume a spin-balanced system, such that the propagators for both spin-components are identical.
The retarded propagator describes the linear response of the system to single-particle excitations. For the electrons, it is defined via the expectation value of the anti-commutator
\begin{equation}
    \mathcal{G}^R_{x,x'}=-i\theta_{t-t'}\left\langle \left\{\hat{c}^{}_{\sigma,x},\hat{c}^\dagger_{\sigma,x'}\right\}\right\rangle,
\end{equation}
where $\theta_{t-t'}$ is the Heaviside step function.
The advanced propagator is related to the retarded propagator through the relation $\mathcal{G}^A_{x,x'}=\left(\mathcal{G}^R_{x',x}\right)^*$, with the asterisk denoting complex conjugation.
Assuming a state that is invariant under translations in both space and time, the space-time dependence of $\mathcal{G}_{x,x'}$ can be diagonalized by transforming to momentum and frequency space.
Additionally, the perfect forward-scattering in \cref{eq:H_bf}, decouples different momentum modes, allowing us to focus solely on the Fermi momentum $|\bm{k}|=k_F$, which is the first mode to develop an instability in the pairing channel \cite{Altland&Simons}.
Consequently, we will discuss the physics at $k_F$ for the remainder of paper.  
A more general momentum-resolved description is provided in \cref{app:pathIntegral}. 

The retarded propagator provides direct access to the steady-state electron spectral function $\mathcal{A}_\omega$, which encodes the density of single-particle excitations, through the propagator's imaginary part 
\begin{equation}\label{eq:A_el}
\mathcal{A}_{\omega}=i\left(\mathcal{G}^R_{\omega}-\mathcal{G}^A_{\omega}\right)=-2\Im\mathcal{G}^R_{\omega}.
\end{equation}

While the retarded propagator only contains information about the spectrum, the Keldysh propagator contains information about how this spectrum is occupied. It is defined via the expectation value of the commutator
\begin{equation}
    \mathcal{G}^K_{x,x'}=-i\left\langle \left[\hat{c}^{}_{\sigma,x},\hat{c}^\dagger_{\sigma,x'}\right]\right\rangle.
\end{equation}
To make the connection to the occupation explicit, the steady-state Keldysh propagator in frequency space at $k_F$ can be reparameterized  in terms of the spectral function
\begin{equation}\label{eq:electronFe}
\mathcal{G}^K_{\omega}=-iF^e_{\omega}\mathcal{A}_{\omega},
\end{equation}
where the distribution function $F^e_\omega$ fully characterizes the occupation of the spectrum. In thermal equilibrium, it is fixed by the thermal FDRs, taking the familiar form $F^e_{\omega}=\tanh\left(\omega/2T\right)$.

In the superconducting phase the bosonic mediator leads to two types of interaction channels. One is the exchange channel, which gives rise to the normal self-energy ($\mathbbm{\Sigma}$), the other is the pairing channel, which generates the anomalous self-energy ($\mathbbm{\Delta}$).
Out of equilibrium both self-energies acquire a $2\times 2$ matrix structure
\begin{equation}\label{eq:Causal_self-energy}
\mathbbm{\Sigma}_{\omega}=\begin{pmatrix}
        0 & \Sigma^{A} \\
        \Sigma^{R} & \Sigma^{K} \\
    \end{pmatrix}_{\omega},\quad \mathbbm{\Delta}_{\omega}=\begin{pmatrix}
        0 & \Delta^{A} \\
        \Delta^{R} & \Delta^{K} \\
    \end{pmatrix}_{\omega},
\end{equation}
where the retarded and advanced components are connected by complex conjugation \footnote{The conjugation relation between anomalous retarded $(\Delta^R_\omega)$ and advanced $(\Delta^A_\omega)$ self-energies is true only at $\abs{\bm{k}}=k_F$ and for the gauge choice $\Delta^R_{\omega=0} \in \mathbb{R}$. This is discussed in detail in \cref{app:causal_structure_MF}.}.
Using the retarded component of the normal self-energy, the Fermi momentum can be defined self-consistently as
\begin{equation}\label{eq:kF}
    k_F=\sqrt{2m (\mu-\Re\Sigma^{R}_{\bm{k}=\bm{k}_F,\omega=0} )}\,.
\end{equation}
For convenience it is useful to define the normal-state propagator (i.e.~the normal propagator without the anomalous self-energy) as
\begin{equation}\label{eq:normalElectronProp}
    G_\omega=\left[(\omega-\xi_{\bm{k}_F})\sigma_x - \mathbbm{\Sigma}_\omega\right]^{-1}=\left(\omega \sigma_x - \mathbbm{\Sigma}_\omega+\Re\mathbbm{\Sigma}_0 \right)^{-1},
\end{equation}
where $\sigma_x$ is the first Pauli-matrix. In the second equality we used $\xi_{\bm{k}_F}=k_F^2/2m-\mu$ and \cref{eq:kF} to simplify the expression. 
As the system enters the superconducting phase, the anomalous self-energy becomes finite and the normal electron propagator becomes
\begin{equation}\label{eq:mathcalG}
    \mathcal{G}_\omega=(G^{-1}_{\omega}+\mathbbm{\Delta}_{\omega} G^{ T}_{-\omega} \bar{\mathbbm{\Delta}}^{T}_{\omega})^{-1},
\end{equation}
where (see \cref{app:causal_structure_MF})
\begin{equation}\label{eq:barDelta_def}
    \bar{\mathbbm{\Delta}}_{\omega}
    =\begin{pmatrix}
        0 & \bar{\Delta}^{A} \\
        \bar{\Delta}^{R} & \bar{\Delta}^{K} \\
    \end{pmatrix}_{\omega}
    =\begin{pmatrix}
        0 & (\Delta^{A})^* \\
        ({\Delta}^{R})^* & -({\Delta}^{K})^* \\
    \end{pmatrix}_{\omega},
\end{equation}
resulting in the retarded component of the normal propagator taking the form 
\begin{equation}\label{eq:fullGR}
    \mathcal{G}^R_\omega=\frac{1}{\left(G^R_\omega\right)^{-1}+G^A_{-\omega}({\Delta}^A_\omega)^* \Delta^R_\omega}.
\end{equation}
Considering electrons at $k_F$ allows us to fix a gauge in which $(\Delta^A_\omega)^*=\Delta^R_\omega$ (cf.~\cref{app:causal_structure_MF}).
As follows from \cref{eq:A_el}, the electron spectrum is fully determined by the retarded propagator. Combined with \cref{eq:fullGR}, this implies that the spectral gap does not explicitly depend on $\Delta^K_\omega$.
In thermal equilibrium and the weak-coupling regime, $\Delta^R_\omega$  reduces to the conventional BCS gap. More generally, weak coupling implies that $\Delta^R_\omega$ varies slowly in the vicinity of the FS.
Under this assumption, the energy gap in the electron spectrum is set by $\Delta^R_{0}$, which is purely real in the chosen gauge~\footnote{Note that in principle the spectral gap should be self-consistently evaluated at the gap edge via the equation $\Delta_0=\Re \Delta^R_{\omega=\Delta_0}$ \cite{MirabiMarsiglio2020}, but $\Delta_0 \ll \omega_0, \kappa$ in weak-coupling regime, so that $\Delta_0=\Delta^R_{\omega\approx0}$ is a good approximation.}.

The spectral gap arises in the superconducting phase due to the condensation of Cooper pairs, giving rise to finite anomalous propagators
\begin{equation}\label{eq:anom_opDef}
\begin{aligned}
    \mathcal{F}^R_{x,x'}&=-i\theta_{t-t'}\left\langle\left\{\hat{c}_{\uparrow,x},\hat{c}_{\downarrow,x'}\right\}\right\rangle,\\
     \mathcal{F}^A_{x,x'}&=i\theta_{t'-t}\left\langle\left\{\hat{c}_{\uparrow,x},\hat{c}_{\downarrow,x'}\right\}\right\rangle,\\
     \mathcal{F}^K_{x,x'}&=-i\left\langle\left[\hat{c}_{\uparrow,x},\hat{c}_{\downarrow,x'}\right]\right\rangle,
\end{aligned}
\end{equation}
which can be directly computed from the normal and anomalous self-energies 
\begin{equation}\label{eq:anomalousProp}
    \mathcal{F}_\omega=G_\omega \mathbbm{\Delta}_\omega \mathcal{G}_{-\omega}^T=\mathcal{G}_\omega \mathbbm{\Delta}_\omega G_{-\omega}^T,
\end{equation}
with a $2\times2$ matrix structure analogous to \cref{eq:Full_G}.
\begin{figure}[tbp]
\centering
\includegraphics[width=0.85\columnwidth]{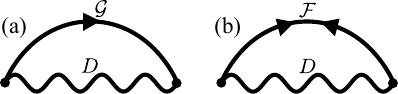}
\caption{Structure of (a) normal and (b) anomalous self-energy diagrams. The wavy line represents the boson propagator, while the straight line with a single arrow (two opposing arrows) denote the normal (anomalous) electron propagator.}
\label{fig:SE}
\end{figure}
Within the path-integral formulation outlined in \cref{app:pathIntegral}, the boson-mediated interaction gives rise to normal and anomalous self-energies of a one-loop form. Their diagrammatic representation is shown in \cref{fig:SE}, while the corresponding NESS-Eliashberg equations read
\begin{equation}\label{eq:selfEnergy_EQs}
\begin{aligned}
\Sigma^R_\omega&=ig\int_\epsilon \left(D^R_{\omega-\epsilon}\mathcal{G}^K_\epsilon+D^K_{\omega-\epsilon}\mathcal{G}^R_\epsilon\right),\\
\Sigma^K_\omega&=ig\int_\epsilon \left(D^R_{\omega-\epsilon}\mathcal{G}^R_\epsilon+D^A_{\omega-\epsilon}\mathcal{G}^A_\epsilon+D^K_{\omega-\epsilon}\mathcal{G}^K_\epsilon\right),\\
\Delta^R_\omega&=ig\int_\epsilon \left(D^R_{\omega-\epsilon}\mathcal{F}^K_\epsilon+D^K_{\omega-\epsilon}\mathcal{F}^R_\epsilon\right),\\
\Delta^K_\omega&=ig\int_\epsilon \left(D^R_{\omega-\epsilon}\mathcal{F}^R_\epsilon+D^A_{\omega-\epsilon}\mathcal{F}^A_\epsilon+D^K_{\omega-\epsilon}\mathcal{F}^K_\epsilon\right),
\end{aligned}
\end{equation}
where we have introduced the notation $\int_\epsilon=\int_{-\infty}^\infty\frac{d\epsilon}{2\pi}$. The effective electron-electron interaction strength is $g=8g_0^2\sqrt{1+\kappa^2/\omega_0^2}$ and $D^{R/A/K}_\omega$ is the propagator for the real quadrature of the boson.
In thermal equilibrium, the Keldysh components $\Sigma^K_\omega$ and $\Delta^K_\omega$ satisfy FDRs analogous to those of the electron propagator, and Eqs.~\eqref{eq:selfEnergy_EQs} become fully equivalent to the real-frequency analytic continuation of the Eliashberg equations in Matsubara frequencies, as shown in \cref{app:th_eq}. 

In Eqs.~\eqref{eq:selfEnergy_EQs}, no equations for the boson self-energies have been included. 
This is justified by the bosons being much more strongly coupled to their environment than they are to the electrons, such that the back-action of the electrons can be neglected (see \cref{app:boson_self-energy} for justification).  
Within this approximation, only the environment-induced self-energies have to be included in the boson propagators.
As shown in \cref{app:QuadInt}, including these results in the boson propagators taking the form
\begin{equation}
    \begin{aligned}
        D^{R(A)}_\omega &= \frac{1}{2} \frac{\omega_0}{(\omega \pm i \kappa)^2-\omega_0^2},\\
        D^K_\omega &= 2i F^{b}_\omega \,\Im D^{R}_\omega,
    \end{aligned}    
    \label{eq:D_omega_RAK}
\end{equation}
where $\kappa$ is an environment-induced spectral width and the renormalization of the boson energy has been reabsorbed into $\omega_0$.
The bosonic distribution function, $F^b_\omega=n^{E}_\omega-n^E_{-\omega}$, is fully determined by the odd part of the bosonic environment's occupation function ($n^E_\omega$). For instance, if the boson environment is a thermal bath at a temperature $T_b$, the boson distribution function becomes
\begin{equation}\label{eq:cothDist}
    F^{b,\mathrm{th}}_{T_b,\omega}=\coth \frac{\omega}{2 T_b}.
\end{equation}

Analogously, we include the possibility of weakly coupling the electrons to an external environment, which we model as a cryostat.
The cryostat drives the electron distribution toward a thermal one at temperature $T_0$ 
\footnote{We use a linear coupling to a fermionic bath to describe this process. Such a simplified description avoids the additional complexity of a bosonic bath, and since we are anyway focusing on a weakly-coupled cryostat the precise modeling details have minimal impact on the observed physics.}. 
As shown in \cref{app:cryostat_model}, the thermalization effect of the cryostat can be captured by adding the following cryostat-induced contributions to the normal self-energy in \cref{eq:normalElectronProp}
\begin{equation}
\begin{aligned}
    \Sigma^{\text{cryo},R}_\omega&=-i\eta_0,\\
    \Sigma^{\text{cryo},K}_\omega&=-2i\eta_0 F^{\text{cryo}}_\omega,\\
\end{aligned}
\label{eq:Sigma^cryo}
\end{equation}
where $F^{\text{cryo}}_\omega=\tanh\left(\omega/2T_0\right)$ and $\eta_0$ is the spectral broadening of the electrons induced by the cryostat, reflecting the strength of the cryostat-electron coupling.
The effect of the cryostat on the electron is thus implemented through the substitutions
\begin{equation}
    \begin{aligned}
        \Sigma^R_\omega&\to \Sigma^R_\omega+\Sigma^{\text{cryo},R}_\omega,\\\Sigma^K_\omega&\to \Sigma^K_\omega+\Sigma^{\text{cryo},K}_\omega.
    \end{aligned}
    \label{eq:cryo_sub}
\end{equation}

\section{Non-thermal contributions to the spectral gap and the superconducting phase transition}
\label{sec:nonThDeriv}
In this section, we will highlight the genuinely non-thermal effects contained in Eqs.~\eqref{eq:selfEnergy_EQs}. 
We will first sketch how they arise in the spectral gap within the superconducting phase, and then concentrate on properties at phase transition, deriving our main result in \cref{eq:gc}.

\subsection{Spectral gap}
\label{subsec:spectral_gap}
As discussed in the previous section, the gap that opens in the electron spectrum is given by $\Delta^R_{0}$ in the weak-coupling regime. 
This spectral gap is self-consistently determined from the $\Delta^R_\omega$ equation in Eqs.~\eqref{eq:selfEnergy_EQs}, evaluated at $\omega=0$. Consider now a regime where the spectral width $\eta$ of electrons at the FS is small compared to the characteristic energy scales of the bosons ($\omega_0$, $\kappa$) and the effective electron temperature $T_e$.
In this regime, the electron propagators are sharply peaked near the FS, whereas the boson propagators vary slowly with frequency. 
This qualitatively difference allows one to approximate the integrals in the equation for $\Delta^R_{0}$ by evaluating the boson propagators at $\omega=0$ and bringing them outside of the integral, obtaining
\begin{equation}
    \Delta^R_{0}=ig D^R_{0} \int_\epsilon \mathcal{F}^K_\epsilon +ig D^K_{0} \int_\epsilon \mathcal{F}^R_\epsilon.
    \label{eq:Delta0_wc}
    \end{equation}
The second term vanishes because the anomalous retarded propagator $\mathcal{F}^R$ contains the expectation value of an equal-time anti-commutator \begin{equation}\label{eq:zeroRetAnomProp}
\int_\epsilon\mathcal{F}^R_\epsilon=\mathcal{F}^R_{\tau=0}=-i \theta_{\tau=0} \langle \{\hat{c}_{\uparrow,\bm{k}_F,t},\hat{c}_{\downarrow,-\bm{k}_F,t}\}\rangle=0,
\end{equation}
where the relative-time variable $\tau=t-t'$ has been introduced. 

In the weak-coupling regime, the spectral gap therefore simplifies to 
\begin{equation}\label{eq:energy_gap_WC}
    \Delta^R_{0}=ig D^R_{0} \mathcal{F}^K_{\tau=0}. 
    \end{equation}
By Fourier transforming the anomalous Keldysh propagator defined in \cref{eq:anom_opDef}, one identifies $\mathcal{F}^K_{\tau=0}$ as $2i$ times the Cooper-pair expectation value $\langle\hat{c}_{\uparrow,\bm{k}_F,t}\hat{c}_{\downarrow,-\bm{k}_F,t}\rangle$.
Since they are linearly connected, either $\Delta^R_0$ or  
$\langle \hat{c}_{\uparrow,\bm{k}_F,t}\hat{c}_{\downarrow,-\bm{k}_F,t}\rangle$ can thus serve as an order parameter; 
throughout this work, we adopt $\Delta^R_0$.

Although our main results concern the phase transition, one can show that the spectral gap inside the superconducting phase also acquires a genuinely non-thermal correction. This can be seen by explicitly evaluating $\mathcal{F}^K_{\tau=0}=\int_\epsilon \mathcal{F}^K_\epsilon$ in \cref{eq:Delta0_wc} using \cref{eq:anomalousProp}, leading to (cf. \cref{app:GK_FK})
\begin{equation}\label{eq:FK_tau0_1}
\begin{aligned}
    \!\!\mathcal{F}^K_{\tau=0}
    &=\int_\epsilon \Big[2i F^{\Sigma}_\epsilon\Im \mathcal{F}^R_\epsilon-2i\left(F^\Delta_\epsilon-F^\Sigma_\epsilon\right)\\&\hspace{9mm}\times\abs{\mathcal{G}^R_\epsilon}^2\!\left(1-\abs{G^R_\epsilon}^2\abs{\Delta^R_\epsilon}^2\right)\Im \Delta^R_\epsilon \Big].
\end{aligned}
\end{equation}
Here the Keldysh components of the normal ($\Sigma^K_\epsilon$) and anomalous ($\Delta^K_\epsilon$) self-energies have been parametrized by generalized distributions $F^\Delta_\epsilon$ and $F^{\Sigma}_\epsilon$, analogously to the parametrization of the electron propagator in \cref{eq:electronFe} [cf.~\cref{eq:F^Sigma_param,eq:F^Delta_param}].
In thermal equilibrium, both generalized distributions reduce to $\tanh(\epsilon/2T)$, and only the first term in the integrand of \cref{eq:FK_tau0_1} contributes. 
In a NESS, however, $F^\Delta_\epsilon$ and $F^\Sigma_\epsilon$ generally differ, giving rise to a genuinely non-thermal correction to the spectral gap in \cref{eq:energy_gap_WC} through the second term in \cref{eq:FK_tau0_1}. 

\subsection{Critical coupling}
\label{subsec:critical_coupling}
We now consider how the genuinely non-thermal effects modify the onset of superconductivity. Building on the previous section, where we showed that the spectral properties of the electrons acquire non-thermal corrections in superconducting phase, we now analyze how these corrections enter the phase transition itself. At the critical transition point ($g=g_c$), the anomalous self-energy $\mathbbm{\Delta}_\omega$ becomes infinitesimal. Consequently, in Eqs.~\eqref{eq:selfEnergy_EQs} one can replace $\mathcal{G}_\omega \to G_\omega$, which implies $F^\Sigma_\omega\to F^e_\omega$ (cf.~\cref{app:GK_FK}). This renders $\mathcal{F}_\omega$ in \cref{eq:anomalousProp}, and therefore the anomalous self-energy equations in \eqref{eq:selfEnergy_EQs}, linear in $\mathbbm{\Delta}_\omega$. 
The expression for $\mathcal{F}^K_{\tau=0}$ in \cref{eq:FK_tau0_1} then simplifies to
\begin{equation}\label{eq:FK_tau0_3}
\begin{split}
    \mathcal{F}^K_{\tau=0}
    =-\int_\epsilon & 
    \Big\{2iF^e_{\epsilon}\Big\{\Im\left[\left(G^R_\epsilon\right)^2\Delta^R_\epsilon\right] \\
    &+|G^R_\epsilon|^2 \Im \Delta^R_\epsilon \Big\} +\abs{G^R_\epsilon}^2\Delta^K_\epsilon \Big\},
\end{split}
\end{equation}
Again exploiting that the electron spectrum remains sharply peaked around the FS, whereas the anomalous components $\Delta^R_\epsilon$ and $\Delta^K_\epsilon$ vary slowly, $\mathcal{F}^K_{\tau=0}$ can be approximated as
\begin{equation}\label{eq:anomFKt0_1}
\mathcal{F}^K_{\tau=0}\approx2i \mathcal{I}_K\Delta^R_0+i\mathcal{I}_R\Im\Delta^K_0,
\end{equation}
where
\begin{equation}\label{eq:IR_IK}
\begin{aligned}
    \mathcal{I}_{K}&=-\int_\epsilon F^e_\epsilon \,\text{Im}\left[\left(G^R_\epsilon\right)^2\right],\\
    \mathcal{I}_{R}&=-\int_\epsilon\vert G^R_\epsilon\vert^2.
\end{aligned}
\end{equation}
Furthermore, in the weak-coupling regime, the retarded electron propagator takes the approximate low-energy form
\begin{equation}
    G^R_\epsilon \approx \frac{1}{\omega+i\eta},
\end{equation}
with spectral width defined as
\begin{equation}\label{eq:eta}
    \eta=-\Im\Sigma^R_{\omega=0}.
\end{equation}
The electron distribution is likewise approximated by
\begin{equation}
F^e_\epsilon\approx\frac{\epsilon}{2T_e},
\label{eq:F^e_approx_T_eff}
\end{equation}
with the effective electron temperature defined as
\begin{equation}\label{eq:Te}
    T_e=\lim_{\omega\to0}\frac{1}{2 \partial_\omega F^e_\omega}=\lim_{\omega\to0}  \left[\partial_\omega \frac{\Im\Sigma^K_\omega}{\Im\Sigma^R_\omega}\right]^{-1},
\end{equation}
i.e.~from the slope of the electron distribution function $F^e_\omega$ around the FS.
These approximations allow for a closed-form solution of the integrals in \cref{eq:IR_IK}, yielding $\mathcal{I}_R=-1/2\eta$ and $\mathcal{I}_K=1/4T_e$.
Substituting back into $\mathcal{F}^K_{\tau=0}$ and using the result in \cref{eq:energy_gap_WC} leads to the critical coupling
\begin{equation}\label{eq:WCgc}
    g_c=\left[\frac{|D^R_0|}{2}\left(\frac{1}{T_e}-\frac{\anth}{\eta}\right)\right]^{-1},
\end{equation}
where $D^R_0=-\frac{\omega_0}{2(\omega_0^2+\kappa^2)}$ and we defined the non-thermal pairing parameter as
\begin{equation}
    \label{eq:alpha_nth}
\alpha_{\mathrm{nth}}=\frac{\Im\Delta^K_{\omega=0}}{\Delta^R_{\omega=0}} .
\end{equation}
With the definition $\bar{\omega}_0=\omega_0+\kappa^2/\omega_0$, \cref{eq:WCgc} corresponds exactly to our main result presented in \cref{eq:gc}.

Finally, let us briefly discuss the non-thermal pairing parameter of \cref{eq:alpha_nth}. At the phase transition the anomalous self-energies become infinitesimal. However, the ratio of the Keldysh component $\Delta^K_\omega$ to the retarded component, i.e.~the spectral gap, $\Delta^R_{\omega=0}$, is finite at and below the phase transition.
In thermal equilibrium, $\Delta^K_\omega$ is fixed by the thermal fluctuation-dissipation relation (FDR) $\Delta^K_\omega=\tanh(\omega/2T) (\Delta^R_\omega-\Delta^A_\omega)$ (see \cref{app:th_eq}) and becomes zero at $\omega=0$, implying $\alpha_{\mathrm{nth}}=0$ and $g_c=g_{c,\mathrm{th}}$, as anticipated in the discussion under \cref{eq:gc}. 
As soon as the system is brought outside of thermal equilibrium, thermal FDRs are broken and $\alpha_{\mathrm{nth}}$ becomes continuously non-zero. 
Its sign determines if it contributes to enhance ($\alpha_{\text{nth}}<0 \Rightarrow g_c<g_{c,\mathrm{th}}$) or decrease ($\alpha_{\text{nth}}>0 \Rightarrow g_c>g_{c,\mathrm{th}}$) the pairing with respect to thermal equilibrium at the temperature $T_e$.

\section{Non-thermal steady states: Two examples}
\label{sec:num_sol}
The expression for $g_c$ derived in \cref{eq:WCgc}, provides a compact analytic measure of how genuinely non-thermal effects reshape the superconducting instability. Yet the size and qualitative character of these corrections depend sensitively on the structure of the NESS. In order to get an understanding of how important the correction is and to connect the analytic weak-coupling estimates to fully self-consistent solutions, we considered two concrete examples in \cref{subsec:KR_examples}.
In this section we will examine these two examples in more detail.
We discuss the relevant physical quantities at the phase transition obtained both from self-consistent numerical solutions of Eqs.~\eqref{eq:selfEnergy_EQs} and analytically in the weak-coupling regime.
The numerical iterative method employed to solve Eqs.~\eqref{eq:selfEnergy_EQs} is described in \cref{app:num_methods}.

\subsection{Temperature bias}
\label{subsec:results_T-bias}
\begin{figure*}[tbp]
\centering
\includegraphics[width=0.85\textwidth]{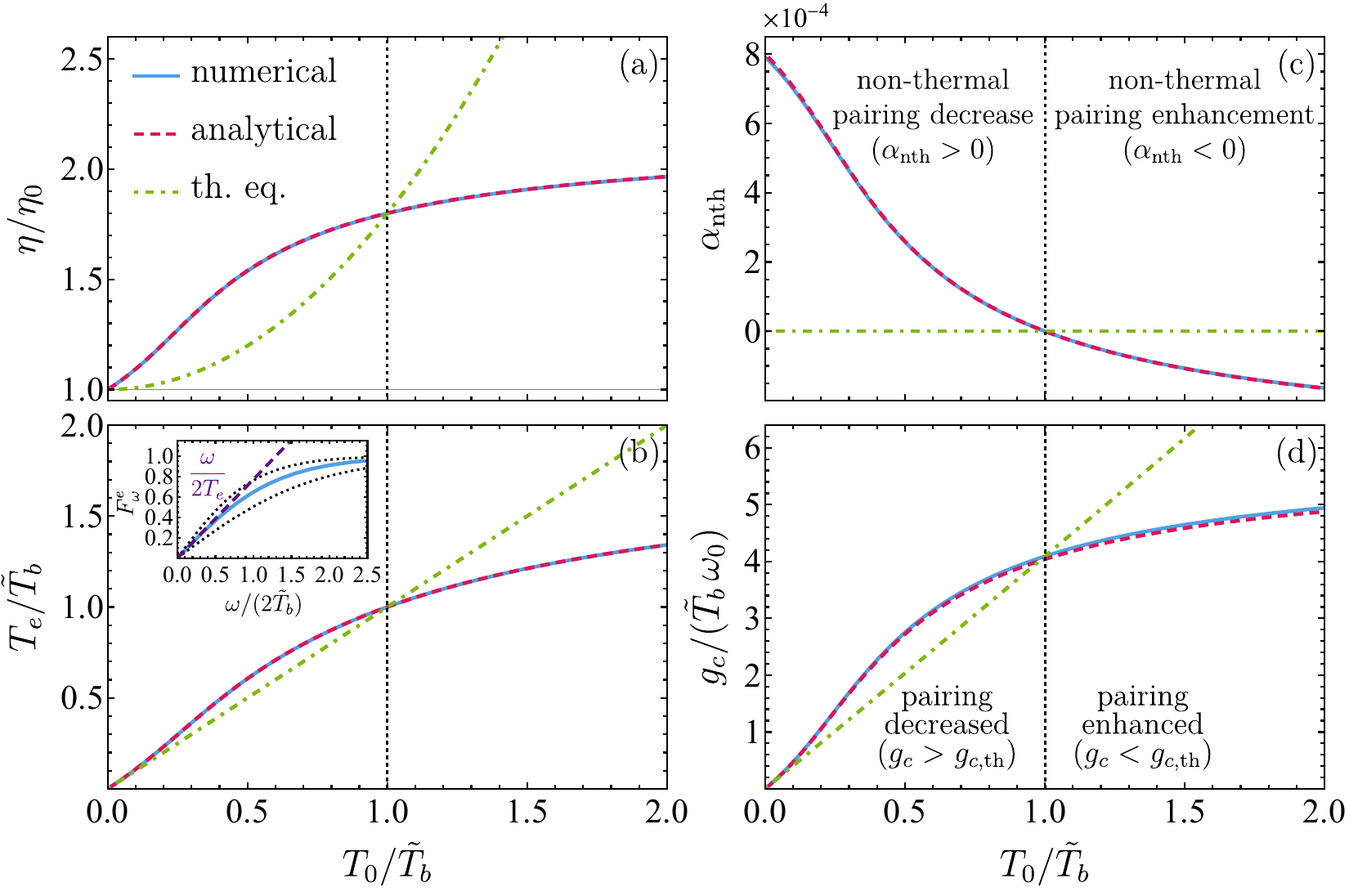}
\caption{Physical quantities appearing in \cref{eq:gc} for the temperature-bias example as a function of the cryostat temperature $T_0$:
(a) electron spectral width,
(b) effective electron temperature,
(c) non-thermal pairing parameter,
(d) critical coupling.
The NESS is realized by keeping $T_b=\tilde{T}_b$ fixed and varying $T_0$. The input parameters are identical to those in \pref{fig:KR}{a}. 
Blue solid lines are numerical results in the NESS, red dashed lines are the corresponding weak-coupling analytical expressions given in \cref{eq:eta_T-bias_wc,eq:Teff_analytic,eq:alpha_T-bias,eq:gc_closedform_T-bias_wc}, green dot-dashed lines are numerical results in thermal equilibrium ($T_b=T_0$ along the curve). The vertical black dotted line ($T_0=\tilde{T}_b$) separates the regions of decreased ($T_0<T_b$) and enhanced ($T_0>T_b$) pairing. 
Inset of (b): Numerical estimation of $T_e$ by fitting the electron distribution $F^e_\omega$ (blue solid line) with $\omega/2T_e$ (violet dashed line) for the point at $T_0/\tilde{T}_b=1.8$. The black dotted lines correspond to the Fermi distribution at temperatures $T_b$ (above) and $T_0$ (below).
}
\label{fig:T-bias_grid_results}
\end{figure*}
As a first concrete realization of a NESS, we consider a finite temperature bias between the two external environments in \pref{fig:setup}{b}. 
For the results in the NESS, we keep the boson temperature $T_b$ fixed at $\tilde{T}_b$ and vary the cryostat temperature $T_0$, thereby tuning the degree of non-thermality. 
Thermal equilibrium is restored when the bias is removed $(T_b=T_0)$, so the corresponding results are instead obtained by varying the two temperatures simultaneously.
All the results shown in this section are computed using the same parameters as in \pref{fig:KR}{a}, with a sufficiently small $\tilde{T}_b$ and a sufficiently large $\kappa$ to stay within the weak-coupling regime and justify neglecting the back-action of the electrons on the bosons.
The derivations of the analytical expressions for the various physical quantities can be found in \cref{app:T-bias_wc}.

In \cref{sec:key_results} we showed that genuinely non-thermal effects can enhance pairing when the bosons are cooler than the electrons.
To disentangle the underlying mechanisms, we now examine separately the behavior of the electron spectral width $\eta$, the effective electron temperature $T_e$ and the genuinely non-thermal pairing parameter $\anth$, clarifying how each quantity influences the critical coupling $g_c$ via \cref{eq:gc}. For each of these physical quantities we find a weak-coupling analytical expression which, when combined with \cref{eq:gc}, yields a closed analytical expression for $g_c$ expressed solely in terms of the external parameters ($T_0$, $T_b$, $\eta_0$, $\omega_0$ and $\kappa$).

Before considering the analytical approximations for the physical quantities, we point out that many of these can be expressed in terms of the constant
\begin{align}
    \gamma &= -\frac{\Im D^K_0}{2 T_b} = \frac{2 \omega_0 \kappa }{(\kappa^2+\omega_0^2)^2},
    \label{eq:gamma_def}
\end{align}
which has the units of inverse energy squared (see \cref{app:T-bias_wc}).
This quantity governs, for instance, the first-order broadening of the electron spectrum due to the interaction with bosons at $T_b$
\begin{equation}
    \eta^{(1)} = g_c \gamma \, T_b.
    \label{eq:eta^(1)_T-bias_wc}
\end{equation}
Including the cryostat contribution, the total spectral width becomes
\begin{equation}
    \eta \approx \eta_0 + g_c  \gamma \, T_b.
    \label{eq:eta_T-bias_wc}
\end{equation}

To confirm the validity of the analytical approximation in \cref{eq:eta_T-bias_wc}, we compare it to the numerical value obtained from $\eta=\eta_0-\Im\Sigma^R_0$ in  \pref{fig:T-bias_grid_results}{a}. 
In the figure one sees that the $T_0$ dependence of $\eta$ obtained numerically excellently agrees with the analytical result.
The thermal equilibrium spectral width is also shown for comparison. 
Interestingly, we see that for $T_0>T_b$ (cooler bosons) the electrons have a smaller spectral width than in thermal equilibrium, which boosts the genuinely non-thermal enhancement of pairing in this region due to its $\sim \anth/\eta$ dependence in \cref{eq:gc}.

Having established the behavior of the spectral width, we now consider the effective electron temperature $T_e$. 
Its weak-coupling behavior is captured by
\begin{equation}
    T_{e}\approx \frac{\eta_0+g_c \gamma \, T_b}{\eta_0+g_c \gamma \, T_0}T_0,
    \label{eq:Teff_analytic}
\end{equation}
from which the thermal equilibrium result $T_e =T_0$ immediately follows when $T_b=T_0$. In the NESS, however, $T_e$ acquires a non-trivial dependence on the self-consistently determined $g_c$, via \cref{eq:gc}, and the temperature bias.
Numerically, $T_e$ can be extracted by the slope of $F^e_\omega$ around the FS as shown in  \cref{eq:Te}. 
This corresponds to fitting the electron distribution with $\omega/2T_e$ close to the FS [see inset of \pref{fig:T-bias_grid_results}{b}]. The main panel of
\pref{fig:T-bias_grid_results}{b} shows that these numerical results again agree extremely well with the analytical expression. As a consistency check, it is observed that $T_e$ always lies between the temperatures of the two environments, $T_0$ and $T_b$.

The final self-consistent ingredient required to determine $g_c$ is the genuinely non-thermal pairing parameter $\anth$, defined in \cref{eq:alpha_nth}.
In the weak-coupling regime, its analytical approximation is found to be
\begin{equation}
    \anth 
    \approx \frac{g_c \gamma }{1+ \frac{g_c \gamma \, T_b}{\eta}}\frac{T_b-T_e}{T_e}
    \approx\frac{g_c \gamma \, \eta_0 }{\eta_0+2 g_c \gamma \, T_b }\frac{T_b-T_0}{T_0},
    \label{eq:alpha_T-bias}
\end{equation}
from which one can directly extract the constant $\mathcal{C}$ in \cref{eq:alpha_expansion}.
In thermal equilibrium, $T_b=T_e=T_0$, which immediately implies $\anth=0$. By contrast, in the NESS, 
$\anth$ becomes negative for $T_0 > T_e > T_b$, i.e.~when the bosons are cooler than the cryostat, as already discussed in \cref{subsec:KR_examples}.
Note that the sign-changing factor in \cref{eq:alpha_T-bias} exhibits the same functional dependence whether expressed in terms of $T_e$ or $T_0$. This follows directly from the monotonic dependence of $T_e$ on $T_0$, shown in \pref{fig:T-bias_grid_results}{b}, together with the fact that $T_e=T_0$ when $T_0=T_b$.

In \pref{fig:T-bias_grid_results}{c} we plot $\anth$ as a function of $T_0$, comparing the numerical solution from \cref{eq:alpha_nth} with the  analytical expression in \cref{eq:alpha_T-bias}. The two approaches show excellent agreement over the considered temperature range. Small deviations appear only at low temperatures, where $T_e \sim \eta$ and the assumption $\eta \ll T_e$ underlying the derivation of \cref{eq:alpha_T-bias} ceases to be valid.
As discussed in \cref{sec:key_results}, the sign of
$\anth$ determines whether the genuinely non-thermal contribution in \cref{eq:gc} enhances ($\anth<0$) or suppresses ($\anth>0$) pairing relative to the thermal-like contribution. Consistent with this interpretation, \pref{fig:T-bias_grid_results}{c} shows that pairing enhancement occurs when the boson are cooler than the cryostat ($T_0>T_b$), in agreement with the behavior observed in \pref{fig:KR}{a}.

In \pref{fig:KR}{a}, the critical coupling $g_c$ was presented as a function of the effective temperature $T_e$ to emphasize non-thermal effects. For completeness,  \pref{fig:T-bias_grid_results}{d} displays the same data as a function of the cryostat temperature $T_0$, which is the independent parameter. This representation makes clear that both the thermal-like renormalization of the temperature $T_e<T_0$ and the genuinely non-thermal contribution ($\anth<0$) act constructively, leading to an enhanced pairing instability when the bosons are cooled below the cryostat temperature.

The analytical weak-coupling expression for $g_c$, shown in \pref{fig:KR}{a} and \pref{fig:T-bias_grid_results}{d}, is obtained by combining \cref{eq:gc} with the analytical solutions for $\eta$, $T_e$ and $\anth$ given in \cref{eq:eta_T-bias_wc,eq:Teff_analytic,eq:alpha_T-bias}. This leads to 
\begin{equation}
    \begin{split}
        & g_c = \frac{2 T_b}{|D^R_0|}+\frac{1}{2} \left( \frac{2 T_b}{|D^R_0|} + \frac{\eta_0}{2\gamma \, T_0} \right) \\
        &\times \left( \! \sqrt{1+ \frac{4\eta_0(T_0-T_b)}{ \gamma \, T_0 |D^R_0|} \left( \frac{2 T_b}{|D^R_0|} + \frac{\eta_0}{2\gamma \, T_0 } \right)^{\!\!-2}} - 1 \! \right).
    \end{split}
    \label{eq:gc_closedform_T-bias_wc}
\end{equation}
At thermal equilibrium ($T_b=T_e=T_0$), the second term vanishes and one recovers the thermal result $g_c=2 T_e/|D^R_0|$ from \cref{eq:gc_th}.
As seen in \pref{fig:KR}{a} and \pref{fig:T-bias_grid_results}{d}, the weak-coupling approximation in \cref{eq:gc_closedform_T-bias_wc} reproduces the numerical results very well, thereby confirming the validity of the key result in \cref{eq:gc} \footnote{We remark that the small discrepancy of the weak-coupling solution at larger $T_0$ comes mostly from the weak-coupling approximation of the thermal contribution $g_{c,\mathrm{th}}$ of \cref{eq:gc_th}, rather than from the non-thermal contribution in \cref{eq:gc}. This is also discussed at the end of \cref{sec:inc}.}.

To summarize, we find that pairing is enhanced by cooling the boson mediating electron-electron interactions below the cryostat temperature. This enhancement arises from the combined action of a thermal-like reduction of the electronic temperature $T_e$, and a genuinely non-thermal contribution encoded in the finite value of $\anth$.

\subsection{Incoherent driving of bosons}\label{sec:inc}
\begin{figure*}[tbp]
\centering
\includegraphics[width=0.8\textwidth]{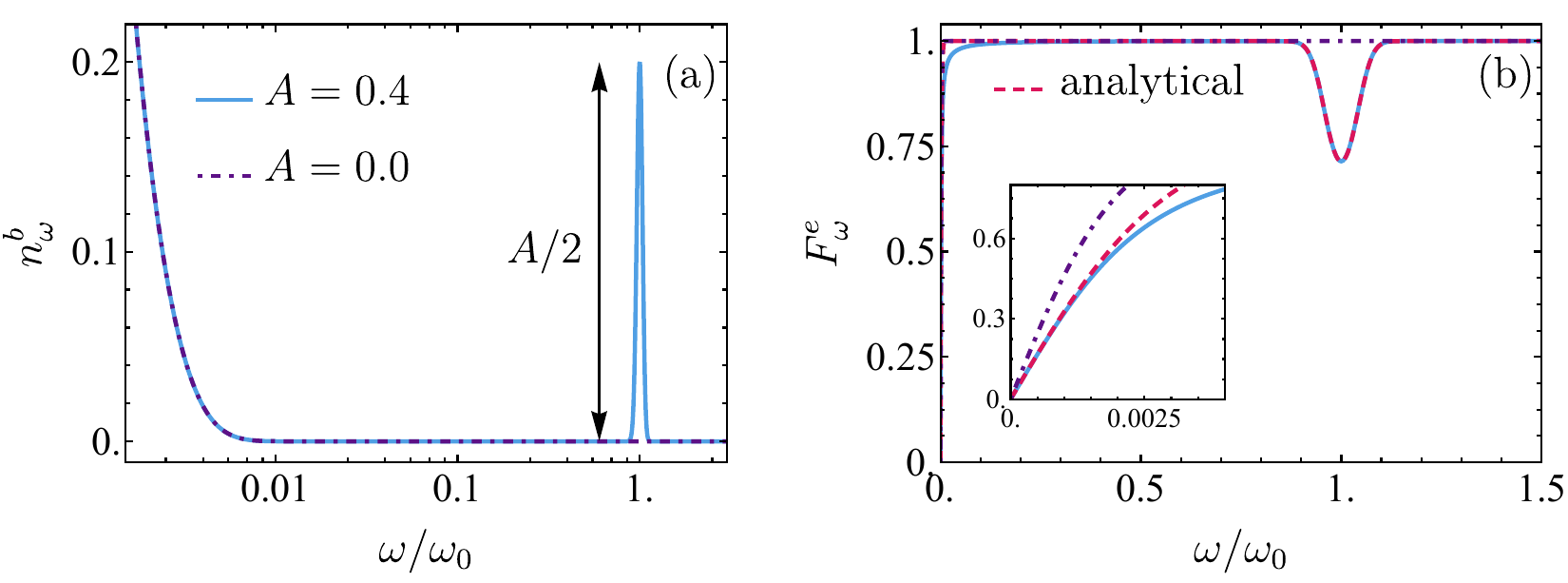}
\caption{
(a) Boson occupation function $n^b_\omega=(F^b_\omega-1)/2$ with $F^b_\omega$ given by \cref{eq:incFb} and (b) numerically computed electron distribution function $F^e_\omega$ when the bosons are incoherently driven with strength $A$. The $A=0.0$ (blue solid) lines represent the undriven/thermal case, while the $A=0.4$ (violet dot-dashed) lines denote the distributions for a finite drive strength. The additional ``analytical'' (red dashed) curve corresponds to the weak-coupling approximation in \cref{eq:ApproxFe_inc} for $A=0.4$, with the effective electron temperature $T_e$ determined via \cref{eq:incTe}. The inset shows a magnification of the behavior close to the FS, highlighting its thermal character.
}
\label{fig:inc_dist}
\end{figure*}
In the previous example, a NESS was generated by a finite temperature bias.
An alternative route to a NESS is to directly couple the bosons to a non-thermal environment, for instance by incoherently driving the bosons on top of their thermal background. 
With the amplitude of the incoherent drive acting as a tuning parameter for the degree of non-thermality, we can simplify the description by decoupling the cryostat in \pref{fig:setup}{b}.  Within our formalism, this corresponds to setting $\eta_0=0$ in \cref{eq:eta_T-bias_wc}, such that the steady-state properties of the electrons are entirely determined by the interactions with the bosons.

As in the previous section, we separately analyze the electron spectral width $\eta$, the effective electron temperature $T_e$
and the genuinely non-thermal pairing parameter $\anth$ as a function of the control parameter, here being the drive strength. For all three quantities, analytical  expressions can be derived in the weak-coupling regime.
Substituting these results into \cref{eq:gc}, we obtain a closed analytical expression for $g_c$. 
Details of the weak-coupling derivations are provided in \cref{app:incoherent_wc}.

The incoherent drive is modeled by augmenting the bosonic distribution with an additional non-thermal term (see \cref{app:bath_renormalization})
\begin{equation}\label{eq:incFb}
    F^{b}_\omega=F^{b,\mathrm{th}}_{T_b,\omega}+F^{b,d}_\omega,
\end{equation}
where $F^{b,\text{th}}_{T_b,\omega}$ is the thermal distribution defined in \cref{eq:cothDist}.
We parametrize the incoherent drive by a Gaussian 
\begin{equation}\label{eq:Fd_inc}
    F^{b,d}_\omega = A\left[\exp\left(-\frac{\left(\omega-\omega_d\right)^2}{v^2}\right)-\exp\left(-\frac{\left(\omega+\omega_d\right)^2}{v^2}\right)\right],
\end{equation}
where $A$ and $v$ control the strength and width of the incoherent drive, respectively, and the peak is centered at frequency $\omega_d$.
To maximize the electronic response, we choose $\omega_d=\omega_0$, i.e., resonant with the bosonic mode.
 
To remain within the weak-coupling regime, we consider a low-temperature  $T_b=10^{-3}\omega_0$. Together with a large bosonic spectral width $\kappa=0.25\omega_0$, this justifies neglecting the back-action of the electrons on the bosons.
The width of the drive is fixed to $v=0.05\omega_0$, which is much smaller than the driving frequency, ensuring that the thermal and driven contributions to $F^b_\omega$ are well separated in frequency. These parameters coincide with those used in \pref{fig:KR}{b}.

The resulting bosonic occupation function for finite drive strength is shown in \pref{fig:inc_dist}{a}, together with the undriven thermal occupation function. 
For $A=0$, the bosons are purely thermal, and the entire system thermalizes to the temperature $T_b$. 
This thermalization is a direct consequence of the decoupling from the cryostat, which leaves the boson as the sole source for determining the  electron distribution in the steady state.  

In the steady state, there exists a contribution to $F^e$ that is independent of the electron-boson coupling  
\begin{equation}\label{eq:Fe0_main}
    F^{e,(0)}_\omega=\frac{1}{F^b_\omega}=\frac{1}{F^{b,\mathrm{th}}_{T_b,\omega}+F^{b,d}_\omega}.
\end{equation}
Although the time required to reach this contribution diverges for $g_c\to0$, it is always there in the steady state. 
For $A=0$, this expression reduces to $F^e_\omega=F^{e,(0)}_\omega=1/\coth(\omega/2T_b)= \tanh(\omega/2T_b)$, as expected in thermal equilibrium. 

For finite drive strength $(A\neq0)$, the electronic distribution acquires a non-thermal component, related to $1/F^{b,d}_\omega$, that is non-perturbative in the coupling.
The numerical result for $F^e_\omega$ is shown in \pref{fig:inc_dist}{b}.
In the weak-coupling regime, the behavior of the electron distribution near $\omega_d$ closely mirrors the inverse behavior of the driven bosonic distribution shown in \pref{fig:inc_dist}{a}.
By contrast, as highlighted in the inset of \pref{fig:inc_dist}{b}, the distribution near the FS deviates from the zeroth-order result in \cref{eq:Fe0_main}. 
This deviation reflects the combined effect of interactions and driving, which leads to an increase of the effective electron temperature $T_e$. To capture this behavior, we refine our weak-coupling approximation for the electron distribution \eqref{eq:Fe0_main} by replacing the the boson temperature $T_b$ with the effective electron temperature $T_e$ 
\begin{equation}\label{eq:ApproxFe_inc}
    F^e_\omega\approx\frac{1}{F^{b,\mathrm{th}}_{T_e,\omega}+F^{b,d}_\omega}.
\end{equation}
As demonstrated in \pref{fig:inc_dist}{b}, this ansatz provides a good description of the electron distribution both near the drive frequency $\omega_d$ and close to the FS. 

\begin{figure*}[tbp]
\centering
\includegraphics[width=0.85\textwidth]{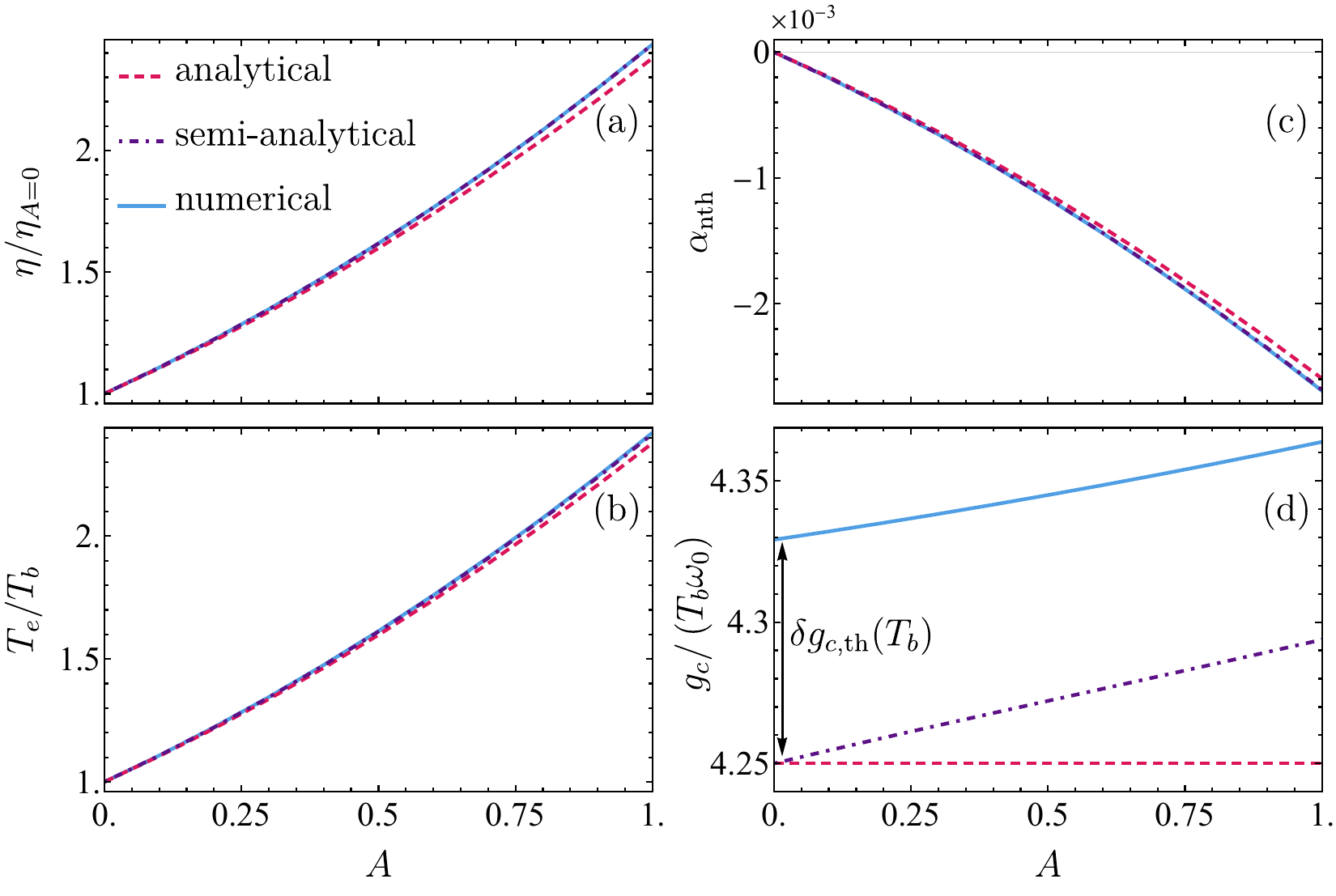}
\caption{
Physical quantities appearing in \cref{eq:gc} for the setting with incoherent driving of the bosons as a function of the driving strength: 
(a) electron spectral width,
(b) effective electron temperature,
(c) non-thermal pairing parameter,
(d) critical coupling [zoom of the plot in \pref{fig:KR}{b}].
Curves denoted with ``numerical'' (blue solid) are obtained via the self-consistent numerical solution of \cref{eq:selfEnergy_EQs}, curves denoted with ``analytical'' (red dashed) are the corresponding weak-coupling approximations obtained via \cref{eq:inceta,eq:incTe,eq:Ieta,eq:alpha_inc,eq:inc_gc} and curves denoted with ``semi-analytical'' (violet dot-dashed) are computed with the weak-coupling approximations \cref{eq:gc,eq:inceta,eq:incTe,eq:Ieta,eq:alpha_inc}, but with the numerically determined $g_c$. The difference between the numerical and analytical thermal equilibrium critical coupling is denoted as $\delta g_{c,\mathrm{th}}(T_b)$ in panel (d).  The input parameters are identical to those used in \pref{fig:KR}{b}.
}
\label{fig:inc_square}
\end{figure*}

To obtain a reliable self-consistent approximation for the effective electronic temperature $T_e$, it is first necessary to understand how incoherent driving modifies the electronic spectrum and, in particular, the spectral width $\eta$ at the FS.
The spectral width extracted from the numerical determination of $\Im\Sigma^R_0$ from Eqs.~\eqref{eq:selfEnergy_EQs} is shown in \pref{fig:inc_square}{a}. 
Strong incoherent driving of the bosons is seen to significantly increase the electronic spectral width compared to the undriven case. 
This demonstrates that incoherent driving of the bosons substantially modifies the thermal-like properties of the electrons and, 
as evident from \cref{eq:gc}, an increased spectral width reduces the effectiveness of the genuinely non-thermal contribution to the critical coupling.

To understand the mechanism behind the increase of $\eta$, we now consider its weak-coupling approximation. 
In contrast to the electronic distribution, modifications to the spectrum start appearing at linear order in $g_c$. Specifically, the first-order contribution to the imaginary part of the retarded self-energy takes the form 
\begin{equation}\label{eq:incSigmaR}
    \Im \Sigma^{R,(1)}_\omega=\frac{g_c\Im D^K_{\omega}}{2},
\end{equation}
where $D^K_\omega$ is given in \cref{eq:D_omega_RAK} and contains the driven bosonic distribution $F^b_\omega$ defined in \cref{eq:incFb}.
Since we consider $T_b\ll\omega_d$, the incoherent drive is energetically well-separated from the low-energy thermal sector.  This separation allows us to decompose $\Im\Sigma^{R,(1)}_\omega$ into two physically distinct contributions. 
In the vicinity of the FS, $\Im \Sigma^{R,(1)}_\omega$ is governed entirely by the thermal background at temperature $T_b$, adding $g_c\gamma T_b$ to $\eta$. This contribution is identical to the thermal broadening obtained in the temperature-bias example [cf.~\cref{eq:eta^(1)_T-bias_wc}].
At energies close to the drive frequency $\omega_d$, however, the incoherent drive creates spectral weight that was not present in the temperature-bias example.
Owing to the convolution structure of the self-energies in Eqs.~\eqref{eq:selfEnergy_EQs}, this spectral weight makes the electrons at the FS in the NESS sensitive to the behavior of the electron distributions near $\omega_d$. 
As a result an additional contribution arises in $\eta$, such that its weak-coupling approximation becomes
\begin{equation}\label{eq:inceta}
    \eta\approx g_c\left(\gamma T_b+g_cI_\eta\right),
\end{equation}
where the effect of the drive is encoded in
\begin{equation}\label{eq:Ieta}
    I_\eta=2\int_\epsilon\left[2\,\mathrm{sign}(\epsilon)F^{b,d}_\epsilon+\left(F^{b,d}_\epsilon\right)^2\right]\frac{\left(\Im D^R_{\epsilon}\right)^2}{\epsilon^2},
\end{equation}
which is strictly positive. 
Anticipating the closed analytical expression for $g_c$ in \cref{eq:inc_gc}, the approximation for $\eta$ in \cref{eq:inceta} is compared to the numerical result in \pref{fig:inc_square}{a}.
We find a good qualitative agreement over the full range of drive strengths, with the analytical expression slightly underestimating $\eta$ at larger $A$. 
This underestimate can be mitigated by inserting the numerically determined value of $g_c$ into \cref{eq:inceta}, as shown by the semi-analytical curve in \pref{fig:inc_square}{a}. This observation confirms that the  functional form of \cref{eq:inceta} correctly captures the dominant mechanism underlying the increase of $\eta$.

Having established a weak-coupling description of the modification of the electron spectrum via \cref{eq:incSigmaR}, we now turn to the effective electron temperature $T_e$. 
As already indicated in \pref{fig:inc_dist}{b} and shown explicitly in \pref{fig:inc_square}{b}, $T_e$ increases monotonically with the drive strength $A$. 
This heating arises due to the interacting nature of the system and leads to an increase of the thermal contribution $g_{c,\mathrm{th}}$ defined in \cref{eq:gc_th}.
To derive a weak-coupling approximation for $T_e$, we note that, since the bosons near the FS remain thermal with a temperature $T_b$, the heating of the electrons must originate from the spectral weight in $\Im G^R_\omega$ at energies $\omega\sim\omega_d$.
This spectral weight is generated by  $\Im\Sigma^{R,(1)}_\omega$ in \cref{eq:incSigmaR}, highlighting the essential role of the spectral modifications in determining $T_e$.
Exploiting the energy separation between the thermal and driven sectors, one finds the estimate
\begin{equation}\label{eq:incTe}
    T_e\approx \frac{\eta}{g_c\gamma}=T_b\left(1+\frac{g_c I_\eta}{\gamma T_b}\right).
\end{equation}
As for $\eta$, using the analytical approximation \eqref{eq:inc_gc} for $g_c$ in \cref{eq:incTe} leads to a slight underestimate of $T_e$ observed in \pref{fig:inc_square}{b}, whereas substituting  the numerical value of $g_c$ yields excellent agreement. 
\cref{eq:incTe} makes explicit that the heating of the electrons is governed by the same quantity $I_\eta$ that controls the increase of the spectral width. 

With the electronic properties near the FS under control, the final ingredient required to estimate the critical coupling in \cref{eq:gc} is the genuinely non-thermal pairing parameter $\anth$.
Within a weak-coupling treatment, its evaluation requires an appropriate ansatz for the frequency dependence of the anomalous self-energy near $\omega_d$. 
At lowest order, this dependence is well approximated by that of the bosonic propagators (cf.~\cref{fig:inc_anom}), leading to
\begin{equation}\label{eq:alpha_inc}
    \anth\approx\frac{g_c}{1+\frac{g_c \gamma T_b}{\eta}}\left[\gamma \left(\frac{T_b}{T_e}-1\right)-\frac{2I_\eta}{\abs{D^R_0}}\right].
\end{equation}
This weak-coupling approximation and the numerical result for $\anth$ are compared in \pref{fig:inc_square}{c}. As for $\eta$ and $T_e$, the analytical approximation slightly underestimates the magnitude of $\anth$, an effect that is again remedied by inserting the numerically determined value of $g_c$, as done in the semi-analytical curve.

At thermal equilibrium $(A=0)$, $\anth$ vanishes identically, since $T_e=T_b$ and $I_\eta=0$ as a consequence of $F^{b,d}_\omega\vert_{A=0}=0$.
The first term in the parenthesis of \cref{eq:alpha_inc} is identical to the temperature-bias expression for $\anth$ in \cref{eq:alpha_T-bias}. Because the incoherent drive increases $T_e$, this term leads to a negative contribution to $\anth$, corresponding to an enhancement of the genuinely non-thermal pairing effect.
The second term in \cref{eq:alpha_inc}, is also negative and proportional to $I_\eta$, which is positive and increases with the drive strength. Remarkably, this shows that the same mechanism responsible for the increase of $T_e$ and $\eta$ also enhances the genuinely non-thermal pairing.

Having obtained analytical expressions for $\eta$, $T_e$ and $\anth$ as functions of $g_c$, we can now substitute them into the weak-coupling approximation for $g_c$ in \cref{eq:gc} and solve for $g_c$. This yields 
\begin{equation}\label{eq:inc_gc}
    g_c\approx g_{c,\mathrm{th}}(T_b),
\end{equation}
indicating that, to leading order, the critical coupling is essentially unaffected by the incoherent drive.
A detailed comparison with the numerical result in \pref{fig:inc_square}{d} shows that the analytical approximation slightly underestimates the thermal value, leading to an offset $\delta g_{c,\mathrm{th}}(T_b)=g_{c,A=0}-g_{c,\mathrm{th}}(T_b)\approx 0.02\, g_{c,\mathrm{th}}(T_b)$.
By using the ``semi-analytical'' results for $\eta$, $T_e$ and $\anth$ in \cref{eq:gc} the offset is unaffected, but the monotonic increase with $A$, observed for the numerical result in \pref{fig:inc_square}{d}, is captured.
Including high-order corrections in  $g_{c,\text{th}}(T_e)$ would reduce the offset $\delta g_{c\mathrm{th}}(T_b)$ and capture the small increase of $g_c$ with $A$, but we do not pursue this extension here, since the correction would be anyway subleading [$\delta g_{c,\mathrm{th}}(T_b)\ll g_{c,\mathrm{th}}(T_b)$]
\footnote{More refined weak-coupling treatments at thermal equilibrium have been developed by accounting for the frequency dependence of $\Delta^R_\omega$ arising from the bosonic propagator (cf.~\cref{fig:inc_anom}) in \cref{eq:selfEnergy_EQs} \cite{Rademaker2016,Marsiglio2018,MirabiMarsiglio2020}.
}.
It is nevertheless notable that the dominant discrepancy arises entirely from the thermal contribution $g_{c,\text{th}}$, while the non-thermal contributions are captured with high accuracy.

In summary, when the incoherent drive is well separated from low-energy thermal modes, 
the critical coupling is only weakly sensitive to the specific form of the drive distribution, rendering this result applicable to a broad class of incoherent driving protocols. In this 
regime, the electrons near the FS heat up and additional occupation near the drive frequency is generated, yet these effects are largely compensated by the enhanced genuinely non-thermal pairing.
As a consequence, the critical coupling is almost unperturbed, despite the system being driven far from thermal equilibrium and exhibiting a substantially increased electron temperature.

\section{Conclusions}
\label{sec:conclusion}
We have developed a non-thermal extension of Eliashberg theory to describe superconductivity in steady states where the bosonic pairing mediator and the electrons are not in mutual thermal equilibrium. 
Within this framework, we derived an analytical expression for the critical coupling that depends on three self-consistently determined quantities: the effective electron temperature, the quasiparticle spectral width, and a genuinely non-thermal pairing contribution. Applying this formalism to two representative non-thermal examples in the weak-coupling regime, we obtained self-consistent analytic approximations for these quantities and found good agreement with fully numerical solutions. 
These results highlight the analytically tractable nature of the theory and demonstrated that, away from thermal-equilibrium, the retarded nature of the boson-mediated interaction is essential even in the weak-coupling regime. 

More specifically, we showed that bosons colder than the electrons enhance pairing due to a combination of a thermal-like cooling and genuinely non-thermal enhancement of the pairing strength.
By contrast, incoherent driving of the bosons at energies much larger than the 
temperature leaves the critical coupling nearly unchanged. In this case the heating of the electrons is largely compensated by an enhancement of the genuinely non-thermal pairing contribution. Nevertheless, 
the incoherent drive pushes the system into a strongly non-thermal steady state, which is simultaneously hotter and more strongly paired than the corresponding thermal equilibrium state.

Our work demonstrates that non-thermal states of the boson mediators provide a new mechanism to modify superconductivity and motivate further exploration of regimes beyond weak coupling, steady states, finite-momentum interactions, and dynamics inside the superconducting phase.

\section*{Acknowledgements}
We would like to thank C.~Hooley, D.~Gole\v{z} and V.~Plastovets for useful discussions. C.J.~gratefully acknowledge financial support from Provincia Autonoma di Trento (PAT). The data that support the findings of this article are openly available \cite{ZenodoRep}.

\section*{Summary of the appendices}
The appendices are organized as follows. \cref{app:pathIntegral} details the derivation of NESS-Eliashberg equations using a real-time Keldysh path-integral formalism. \cref{app:th_eq} shows how the thermal equilibrium results are recovered by the NESS-Eliashberg equations and how they relate to Matsubara formalism. \cref{app:causal_structure_MF} clarifies the causal structure of anomalous and normal propagators at the mean-field level in Keldysh formalism and how the choice of the gauge for the anomalous self-energy can lead to simplifications. \cref{app:GK_FK} presents useful expressions for the retarded and Keldysh components of the normal and anomalous propagators, both in the superconducting phase and at phase transition.
\cref{app:boson_self-energy} discusses why the feedback of the electrons on the bosons can be neglected in the weak-coupling regime. \cref{app:QuadInt,app:cryostat_model} respectively presents how to describe the effect of the boson environment and the cryostat. \cref{app:num_methods} illustrates the numerical method employed to solve the NESS-Eliashberg equations at the phase transition. In \cref{app:T-bias_wc,app:incoherent_wc} weak-coupling analytical expressions for relevant physical quantities are derived for the temperature-bias and the incoherently driven settings, respectively. Finally, in \cref{app:quasi-classical_approx} the implementation of the NESS-Eliashberg equations for smooth momentum-dependent electron-electron interactions and the subsequent momentum averaging within the quasi-classical approximation are discussed.

\appendix
\section{Path-integral derivation of NESS-Eliashberg equations}
\label{app:pathIntegral}
The first step in deriving the NESS-Eliashberg equations, is to construct a steady-state real-time partition function based on the Hamiltonian \eqref{eq:H_sys}. 
This can be done using the Schwinger-Keldysh path-integral formulation \cite{KamenevBook,Chakraborty2021}. After having integrated out the $P$-quadrature of the bosonic field and having included the boson environment
self-energy (see \cref{app:QuadInt}), the partition function for the Bose-Fermi system becomes
\begin{equation}
    \begin{split}
    &\mathcal{Z}= \prod_\sigma \int \mathcal{D}\left[\bar{\psi}_\sigma,\psi_\sigma\right] \int \mathcal{D}\left[X\right] \\
    &\times \exp\left\{i \left(S_{e,0 }\left[\bar{\psi},\psi\right]+S_{b,0 }\left[X\right]+S_{e-b}\left[\bar{\psi},\psi,X\right]\right)\right\},
    \end{split}
    \label{eq:Z}
\end{equation}
where
\begin{equation}
    \begin{split}
    &S_{e,0}\left[\bar{\psi}, \psi\right]    \\
    &= \int_{x,x'} \sum_{\sigma} \begin{pmatrix}
    \bar{\psi}^{c}_\sigma \\
    \bar{\psi}^{q}_\sigma
    \end{pmatrix}^{T}_{x}
    \begin{pmatrix}
    0 & [G_0^{-1}]^{A} \\
    [G_0^{-1}]^{R} & [G_0^{-1}]^{K}
    \end{pmatrix}_{x-x'}
    \begin{pmatrix}
    \psi^{c}_\sigma \\
    \psi^{q}_\sigma
    \end{pmatrix}_{x'}\label{eq:free_fermion_action_kt}
    \end{split}
\end{equation}
is the free action for the electrons written in the space-time coordinates $x=(\bm{r},t)$ with the shorthand notation $\int_x=\sum_{\bm{r}}  \int dt$, while
\begin{equation}
    \begin{split}
    &S_{b,0 }\left[X\right] \\
    &= \frac{1}{2} \int_{t,t'} \begin{pmatrix}
    X^{c} \\
    X^{q}
    \end{pmatrix}^{T}_{t}
    \begin{pmatrix}
    0 & [D^{-1}]^{A} \\
    [D^{-1}]^{R} & [D^{-1}]^{K}
    \end{pmatrix}_{t-t'}
    \begin{pmatrix}
    X^{c} \\
    X^{q}
    \end{pmatrix}_{t'}
    \end{split}
\label{eq:free_boson_action_t}
\end{equation}
is the free action for the real bosons, and
\begin{equation}
    \begin{split}
        &S_{e-b}\left[\bar{\psi}, \psi,X\right] =- 2
        \left[4\left(1+\kappa^2/\omega_0^2\right)\right]^{1/4}
        g_0 \\
        &\times \int_x \sum_{\sigma} \begin{pmatrix} X^{c} & 
        X^{q}
        \end{pmatrix}_{t}
        \begin{pmatrix}
        \bar{\psi}^{c}_\sigma \psi^{q}_\sigma + \bar{\psi}^{q}_\sigma \psi^{c}_\sigma \\
         \bar{\psi}^{c}_\sigma \psi^{c}_\sigma + \bar{\psi}^{q}_\sigma \psi^{q}_\sigma
        \end{pmatrix}_{x}
    \end{split}
    \label{eq:boson_fermion_int_action}
\end{equation}
is the action for the electron-boson interaction vertex. 
Here $\bar{\psi}_{\sigma,x}$ and $\psi_{\sigma,x}$ are the Grassmann fermionic fields associated with the electronic operators $\hat{c}^\dag_{\sigma \bm{r}}$ and $\hat{c}_{\sigma \bm{r}}$, and $X(t)=\left[a^*(t)+a(t)\right]\left[4\left(1+\kappa^2/\omega_0^2\right)\right]^{-1/4}$ \footnote{The scaling constant is chosen to have a particularly simple form of the propagator, as shown in \cref{eq:ret_boson}.} is the real quadrature of the bosonic field, where the complex fields $a^*(t)$, $a(t)$ are associated to the bosonic operators $\hat{a}^\dag$, $\hat{a}$. 
The classical ($c$) and quantum ($q$) components of the fields are related to the fields on the forward ($+$) and backward ($-$) time contours through the Keldysh rotation \cite{KamenevBook}
\begin{equation}
    \begin{split}
    &\psi^{(c/q)}_{\sigma,x}=\left(\psi^{+}_{\sigma,x} \pm \psi^{-}_{\sigma,x}\right)/\sqrt{2}, \\
    &\bar{\psi}^{(c/q)}_{\sigma,x}=\left(\bar{\psi}^{+}_{\sigma,x} \pm \bar{\psi}^{-}_{\sigma,x}\right)/\sqrt{2}
    \end{split}
    \label{eq:Keldysh_rot_fermions}
\end{equation}
for the fermionic fields and
\begin{equation}
    X^{(c/q)}_t=\left(X^{+}_t\pm X^{-}_t\right)/2
    \label{eq:Keldysh_rot_bosons}
\end{equation}
for the real bosonic field.  
Note that we do not employ the Larkin-Ovchinnikov rotation for fermions, but instead use the same transformations as for the bosons. 
Consequently, the fermionic and bosonic propagators will have the same Keldysh structure. 
Inverting the matrix structures in \cref{eq:free_fermion_action_kt,eq:free_boson_action_t}, one obtains a free electron propagator with the retarded, advanced and Keldysh (RAK) matrix structure
\begin{equation}
\label{eq:G0_k}
    [G_0]_{x-x'}=\begin{pmatrix}
        [G_0]^K_{x-x'} & [G_0]^R_{x-x'} \\
        [G_0]^A_{x-x'} & 0
    \end{pmatrix},
\end{equation}
and a real boson propagator with the RAK matrix structure
\begin{equation}
    D_{t-t'}=\begin{pmatrix}
        D^K_{t-t'} & D^R_{t-t'} \\
        D^A_{t-t'} & 0
    \end{pmatrix}.
\end{equation}
The Fourier transformed boson propagators are defined in \cref{eq:D_omega_RAK}, while the Fourier transformed free electron propagators are given by 
\begin{equation}\label{eq:G0_omega_RAK}
    \begin{aligned}
        \left[G_0\right]^{R(A)}_{\bf{k},\omega} &=  \frac{1}{\omega -\xi_{\bm{k}}\pm i 0^+},\\
        \left[G_0\right]^K_{\bf{k},\omega}  &= 2i F_{0,\omega} \Im \left[G_0\right]^{R}_{\bf{k},\omega} ,
    \end{aligned}    
\end{equation}
with $F_{0,\omega}$ being the initial distribution function of the electrons \footnote{Note that the initial distribution of the electrons will actually never enter the steady-state results, since the memory of it will be lost due to the interaction with the cryostat and/or the bosons.}.

\subsection{Derivation of NESS-Eliashberg equations}
\label{sec:derivation}
To derive the NESS-Eliashberg equations, we generalize the approach introduced for the thermal equilibrium case in Ref.~\cite{Protter2021} to a NESS. 
The steps of the procedure are the following: (i) we integrate out the bosons exactly, obtaining a purely fermionic action with an effective boson-mediated interaction vertex, (ii) we perform a double Hubbard-Stratonovich (HS) transformation in the Cooper and exchange channels, introducing the four-component auxiliary fields $\Delta$ and $\mathcal{S}_\sigma$, which are respectively related to the anomalous and normal self-energies, (iii) we reorganize the electron propagator in a Nambu structure and integrate out the electron fields, obtaining an effective action for the auxiliary fields, and finally (iv) we perform a saddle-point (or mean-field) approximation, obtaining the NESS-Eliashberg equations for the auxiliary fields as the stationary path of their effective action.

\subsection{Integrating out the bosons and deriving the effective interaction for the electrons}
\label{subsec:integrating_out_bosons}
 By performing the Gaussian functional integral of the $X$-field in \cref{eq:Z}, one finds an effective action $S_e$ for the electron fields $(\bar{\psi},\psi)$ (see for example the supplemental material of Ref.~\cite{Chakraborty2021}). 
 The action $S_e=S_{e,0 }+S_\mathrm{int}$ includes an effective boson-mediated interaction term $S_\mathrm{int}$ which can be expressed as
\begin{equation}
    \begin{split}
    &S_\mathrm{int}\left[\bar{\psi},\psi\right]
    = -\frac{g}{2}  \sum_{\sigma, \sigma'}  \int_{x,x'} \\
    &\Big[ D^{R}_{t-t'} \big(\bar{\psi}^{c}_{\sigma, x} \bar{\psi}^{c}_{\sigma', x'} \psi^{c}_{\sigma', x'} \psi^{q}_{\sigma, x} 
    +\bar{\psi}^{c}_{\sigma, x} \bar{\psi}^{q}_{\sigma', x'} \psi^{q}_{\sigma', x'} \psi^{q}_{\sigma, x} \\
    &\quad \quad+ \bar{\psi}^{q}_{\sigma, x} \bar{\psi}^{c}_{\sigma', x'} \psi^{c}_{\sigma', x'} \psi^{c}_{\sigma, x} +\bar{\psi}^{q}_{\sigma, x} \bar{\psi}^{q}_{\sigma', x'} \psi^{q}_{\sigma', x'} \psi^{c}_{\sigma, x}\big) \\
    &+D^{A}_{t-t'} \big(\bar{\psi}^{c}_{\sigma, x} \bar{\psi}^{c}_{\sigma', x'} \psi^{q}_{\sigma', x'} \psi^{c}_{\sigma, x} 
    +\bar{\psi}^{c}_{\sigma, x} \bar{\psi}^{q}_{\sigma', x'} \psi^{c}_{\sigma', x'} \psi^{c}_{\sigma, x} \\
    &\quad \quad + \bar{\psi}^{q}_{\sigma, x} \bar{\psi}^{c}_{\sigma', x'} \psi^{q}_{\sigma', x'} \psi^{q}_{\sigma, x} +\bar{\psi}^{q}_{\sigma, x} \bar{\psi}^{q}_{\sigma', x'} \psi^{c}_{\sigma', x'} \psi^{q}_{\sigma, x}\big) \\
        &+D^{K}_{t-t'} \big(\bar{\psi}^{c}_{\sigma, x} \bar{\psi}^{c}_{\sigma', x'} \psi^{q}_{\sigma', x'} \psi^{q}_{\sigma, x} 
    +\bar{\psi}^{c}_{\sigma, x} \bar{\psi}^{q}_{\sigma', x'} \psi^{c}_{\sigma', x'} \psi^{q}_{\sigma, x} \\
    &\quad \quad + \bar{\psi}^{q}_{\sigma, x} \bar{\psi}^{c}_{\sigma', x'} \psi^{q}_{\sigma', x'} \psi^{c}_{\sigma, x} +\bar{\psi}^{q}_{\sigma, x} \bar{\psi}^{q}_{\sigma', x'} \psi^{c}_{\sigma', x'} \psi^{c}_{\sigma, x}\big)  
    \Big],
    \end{split}
    \label{eq:fermionic_vertices}
\end{equation}
where we defined the coupling $g=8
g_0^2\sqrt{1+\kappa^2/\omega_0^2}$. 
We can now separate \cref{eq:fermionic_vertices} in two terms, which either describe interactions between electrons with the same or opposite spin:
\begin{equation}
S_\mathrm{int}\left[\bar{\psi},\psi\right] = \sum_\sigma S^{\sigma \sigma}_\mathrm{int}\left[\bar{\psi},\psi\right]+S^{\uparrow \downarrow}_\mathrm{int}\left[\bar{\psi},\psi\right].
\label{eq:Sint}
\end{equation}
The opposite-spin interaction term of \cref{eq:Sint} can be reorganized in the matrix form
\begin{equation}
    S^{\uparrow \downarrow}_{\mathrm{int}}\left[\bar{\psi},\psi\right] =-g \int_{x,x'} \big(\bar{\rho}^{\uparrow \downarrow}_{x, x'}\big)^{T} \, \mathds{D}_{t-t'} \,  \rho^{\uparrow \downarrow}_{x, x'},
    \label{eq:Sint_updown}
\end{equation}
where we defined the (symmetric) $4\times 4$ matrix
\begin{equation}
    \mathds{D}_{t-t'}=
     \begin{pmatrix}
         0 & D^{A}_{t-t'} & D^{R}_{t-t'} & D^{K}_{t-t'} \\
         D^{A}_{t-t'} &  0 & D^{K}_{t-t'} & D^{R}_{t-t'} \\
        D^{R}_{t-t'} &  D^{K}_{t-t'} & 0 &  D^{A}_{t-t'} \\
        D^{K}_{t-t'} &  D^{R}_{t-t'} &  D^{A}_{t-t'} & 0 
    \end{pmatrix} 
    \label{eq:D_matrix}
\end{equation}
and the four-dimensional electron bilinears 
\begin{equation}
    \begin{split}
        \rho^{\uparrow \downarrow}_{x, x'}=
        \begin{pmatrix}
            \psi^{c}_{\downarrow, x'} \psi^{c}_{\uparrow, x} \\
            \psi^{q}_{\downarrow, x'} \psi^{c}_{\uparrow, x} \\
            \psi^{c}_{\downarrow, x'} \psi^{q}_{\uparrow, x} \\
            \psi^{q}_{\downarrow, x'} \psi^{q}_{\uparrow, x}
        \end{pmatrix}, \quad
        \bar{\rho}^{\uparrow \downarrow}_{x, x'}=
        \begin{pmatrix}
            \bar{\psi}^{c}_{\uparrow, x} \bar{\psi}^{c}_{\downarrow, x'} \\ \bar{\psi}^{c}_{\uparrow, x} \bar{\psi}^{q}_{\downarrow, x'} \\
            \bar{\psi}^{q}_{\uparrow, x} \bar{\psi}^{c}_{\downarrow, x'} \\
            \bar{\psi}^{q}_{\uparrow, x} \bar{\psi}^{q}_{\downarrow, x'} 
        \end{pmatrix}.  
        \label{eq:rho_updown}
    \end{split}
\end{equation}

Analogously, the same-spin interaction term in \cref{eq:Sint} can be written as
\begin{equation}
    S^{\sigma \sigma}_\mathrm{int}\left[\bar{\psi},\psi\right] =\frac{g}{2} \int_{x,x'}\big(\bar{\rho}^{\sigma}_{x, x'}\big)^{T}  \, \mathds{D}_{t-t'} \, \rho^{\sigma}_{x, x'},
    \label{eq:Sint_sigmasigma}
\end{equation}
where the electron bilinears are defined as
\begin{equation}
    \rho^{\sigma}_{x,x'}=
    \begin{pmatrix}
        \bar{\psi}^{c}_{\sigma, x'} \psi^{c}_{\sigma, x} \\
        \bar{\psi}^{q}_{\sigma, x'} \psi^{c}_{\sigma, x} \\
        \bar{\psi}^{c}_{\sigma, x'} \psi^{q}_{\sigma, x} \\
        \bar{\psi}^{q}_{\sigma, x'} \psi^{q}_{\sigma, x}
    \end{pmatrix}, \quad
    \bar{\rho}^{\sigma}_{x,x'}=
    \begin{pmatrix}
        \bar{\psi}^{c}_{\sigma, x} \psi^{c}_{\sigma, x'} \\
        \bar{\psi}^{c}_{\sigma, x} \psi^{q}_{\sigma, x'} \\
        \bar{\psi}^{q}_{\sigma, x} \psi^{c}_{\sigma, x'} \\
        \bar{\psi}^{q}_{\sigma, x} \psi^{q}_{\sigma, x'} 
    \end{pmatrix}.
    \label{eq:rho_sigmasigma}
\end{equation}

Before introducing auxiliary fields through the HS transformation, the symmetric matrix \eqref{eq:D_matrix} must diagonalized
\begin{equation}
    g\mathds{D}_{t-t'}=U^T \Lambda_{t-t'} U,
    \label{eq:D_diag_trans}
\end{equation}
where $\Lambda_{t-t'}$ is a diagonal matrix 
with the entries
\begin{equation}\label{eq:D_EWs}
    \begin{split}
    &(\lambda_{t-t'})_1= g\left[D^A_{t-t'}-D^R_{t-t'}-D^K_{t-t'}\right], \\
        &(\lambda_{t-t'})_2= g\left[-D^A_{t-t'}+D^R_{t-t'}-D^K_{t-t'}\right], \\
        &(\lambda_{t-t'})_3= g\left[-D^A_{t-t'}-D^R_{t-t'}+D^K_{t-t'}\right], \\
        &(\lambda_{t-t'})_4= g\left[D^A_{t-t'}+D^R_{t-t'}+D^K_{t-t'}\right],
    \end{split}
\end{equation}
and $U$ is a time-independent orthogonal ($U^{-1}=U^{T}$) matrix given by 
\begin{equation}
    U=\frac{1}{2}
    \begin{pmatrix}
        -1 & -1 & 1 & 1 \\
        -1 & 1 & -1 & 1 \\
        1 & -1 & -1 & 1 \\
        1 & 1 & 1 & 1 \\
    \end{pmatrix}.
    \label{eq:U_trans_matrix}
\end{equation}
Note that the specific form of $\Lambda_{t-t'}$ and $U$ are not necessary for the following derivation, and are reported here only for completeness.
With the transformation \eqref{eq:D_diag_trans} one can then rewrite the two terms of \cref{eq:Sint} as
\begin{equation}
    \begin{split}
        &S^{\uparrow \downarrow}_\mathrm{int}\left[\bar{\psi},\psi\right] =-\int_{x,x'} \big(U \bar{\rho}^{\uparrow \downarrow}_{x, x'}\big)^{T}  \Lambda_{t-t'}   \big(U \rho^{\uparrow \downarrow}_{x, x'}\big) \\ &=-\int_{x,x'} \sum_{n} \big(\btilde{\rho}^{\uparrow \downarrow}_{x, x'}\big)_{n} \big(\lambda_{t-t'}\big)_{n}  \big(\tilde{\rho}^{\uparrow \downarrow}_{x, x'}\big)_{n},
    \end{split}
    \label{eq:Sint_updown_diag}
\end{equation}
and
\begin{equation}
    \begin{split}
        &S^{\sigma \sigma}_\mathrm{int}\left[\bar{\psi},\psi\right] =\frac{1}{2}\int_{x,x'} \big(U \bar{\rho}^{\sigma}_{x x'}\big)^{T}  \Lambda_{t-t'}   \big(U \rho^{\sigma}_{x x'}\big) \\ &=\frac{1}{2}\int_{x,x'} \sum_n \big(\btilde{\rho}^{\sigma}_{x, x'}\big)_{n} \big(\lambda_{t-t'}\big)_{n}  \big(\tilde{\rho}^{\sigma}_{x, x'}\big)_{n},
    \end{split}
    \label{eq:Sint_sigmasigma_diag}
\end{equation}
where the transformed electron bilinears have been denoted with the tilde symbol ($\tilde{\rho}=U\rho$).
\subsection{Hubbard-Stratonovich transformation in the Cooper and exchange channels}
\label{subsec:Hubbard-Stratonovic}
To make the quartic interaction terms tractable one performs a HS transformation \cite{KamenevBook,Altland&Simons} in the Cooper channel for the opposite-spin interaction of \cref{eq:Sint_updown} and in the exchange channel for the same-spin interaction of \cref{eq:Sint_sigmasigma}.
This is done by introducing two four-component auxiliary complex fields $\Delta$ and $\mathcal{S}^\sigma$ through the Gaussian identities
\begin{equation}
\begin{split}
    1=\prod_{n}& \int \mathcal{D}\left[\bar{\Delta}_{n},\Delta_{n}\right] \\
    & \times\exp \left\{ \int_{x,x'} \, \left[i \frac{\big(\bar{\Delta}_{x,x'}\big)_n \big(\Delta_{x,x'}\big)_n}{\big(\lambda_{t-t'}\big)_n}  \right]\right\},
\end{split}
\label{eq:naive_HST_Delta}
\end{equation}
and
\begin{equation}
\begin{split}
    1= \prod_{n}& \int \mathcal{D}\left[\bar{\mathcal{S}}_{n}^{\sigma},\mathcal{S}_{n}^{\sigma}\right] \\
    &\times\exp \left\{ \int_{x,x'} \left[-2i \frac{\big(\bar{\mathcal{S}}^{\sigma}_{x,x'}\big)_n \big(\mathcal{S}^\sigma_{x,x'}\big)_n}{\big(\lambda_{t-t'}\big)_n} \right]\right\}.
\end{split}
\label{eq:naive_HST_phi}
\end{equation}
For these relations to be true, the Gaussian functional integrals over the auxiliary fields need to be convergent. Strictly speaking, this is the case only if $\Im\big(\lambda_{t-t'}\big)_n<0$ in the first integral and $\Im\big(\lambda_{t-t'}\big)_n>0$ in the second integral. However, from \cref{eq:D_EWs} one sees that  in general $\Im\big(\lambda_{t-t'}\big)_n$ is given by $\pm \Im D^{K}_{t-t'}$, which is sign-changing as a function of $(t-t')$ \footnote{The symmetry of the real boson forces $D^R_t$ to be real and $D^K_t$ to be imaginary.}. To mitigate this sign problem we follow a similar approach  as the one used in Ref.~\cite{Dalal2023} for the derivation of Eliashberg equations in thermal equilibrium for repulsive interactions.

We start by introducing two ``modified'' four-component complex fields $\Gamma$ and $\zeta^\sigma$ satisfying the following Gaussian identities
\begin{equation}
\begin{split}
    &1= \prod_{n} \int \mathcal{D}\left[\Gamma^*_n,\Gamma_n\right]\\
    &\times\exp \left\{ \int_{x,x'} \left[i \frac{\big[\big(v_{t-t'}\big)_n\big]^2}{\big(\lambda_{t-t'}\big)_n} \big(\tilde{\Gamma}_{x,x'}\big)^*_n \big(\Gamma_{x,x'}\big)_n \right]\right\},
\end{split}
\label{eq:Gauss_id_Delta_tilde}
\end{equation}
and
\begin{equation}
\begin{split}
    &1= \prod_{n} \int \mathcal{D}\left[\zeta_{n}^{\sigma *},\zeta_{n}^{\sigma}\right] \\
    &\times\exp \left\{ \int_{x,x'} \, \left[-i \frac{2\big[\big(\tilde{v}_{t-t'}\big)_n\big]^2}{\big(\lambda_{t-t'}\big)_n} \big(\zeta^{\sigma}_{x,x'}\big)^*_n \big(\zeta^\sigma_{x,x'}\big)_n \right]\right\},
\end{split}
\label{eq:Gauss_id_S_tilde}
\end{equation}
where the * denotes complex conjugation and we have introduced the factors
\begin{equation}
    \begin{split}
        \big(v_{t-t'}\big)_n&=\frac{1}{2}\left\{ 1 -  \, \mathrm{sign}\big[ \Im\big(\lambda_{t-t'}\big)_n \big] \right\}\\
        &+ \frac{i}{2}\left\{ 1 +  \mathrm{sign}\big[ \Im\big(\lambda_{t-t'}\big)_n \big] \right\}, \\
        \big(\tilde{v}_{t-t'}\big)_n&=\frac{1}{2}\left\{ 1 +  \mathrm{sign}\big[ \Im\big(\lambda_{t-t'}\big)_n \big] \right\} \\
        &+ \frac{i}{2}\left\{ 1 -  \mathrm{sign}\big[ \Im\big(\lambda_{t-t'}\big)_n \big] \right\}.
    \end{split}
    \label{eq:v_def}
\end{equation}
From the definitions in \cref{eq:v_def} it follows that
$\big[\big(v_{t-t'}\big)_n\big]^2=-\operatorname{sign}\big[\Im\big(\lambda_{t-t'}\big)_n\big]$ and $\big[\big(\tilde{v}_{t-t'}\big)_n\big]^2=\operatorname{sign}\big[\Im\big(\lambda_{t-t'}\big)_n\big]$, ensuring the convergence of the Gaussian integrals of \cref{eq:Gauss_id_Delta_tilde,eq:Gauss_id_S_tilde}. 

We now multiply the interaction term $\exp(iS_{\mathrm{int}}^{\uparrow\downarrow}[\bar{\psi},\psi])$ in the partition function by the identity \eqref{eq:Gauss_id_Delta_tilde} and perform a change of variable
\begin{equation}
    \begin{split}
        &\big(\tilde{\Delta}_{x,x'}\big)_n = \big(\Gamma_{x,x'}\big)_n -\frac{\big(\lambda_{t-t'}\big)_n}{\big(v_{t-t'}\big)_n} \, \big(\tilde{\rho}^{\uparrow \downarrow}_{x, x'}\big)_{n}, \\
        &\big(\btilde{\Delta}_{x,x'}\big)_n = \big(\Gamma_{x,x'}\big)^*_n -\frac{\big(\lambda_{t-t'}\big)_n}{\big(v_{t-t'}\big)_n} \,\big(\btilde{\rho}^{\uparrow \downarrow}_{x, x'}\big)_{n},
    \end{split}
\end{equation}
to obtain
\begin{equation}
    \begin{split}
        e^{iS_{\mathrm{int}}^{\uparrow\downarrow}\left[\bar{\psi},\psi\right]} &=\prod_{n}  \int \mathcal{D}\left[\btilde{\Delta}_n,\tilde{\Delta}_n\right] \\
        &\quad \times e^{i \left(S^{\uparrow\downarrow}_{\mathrm{HS}, 0}\left[\btilde{\Delta},\tilde{\Delta}\right]+ S^{\uparrow\downarrow}_{\mathrm{HS}, \mathrm{int}}\left[\bar{\psi},\psi,\btilde{\Delta},\tilde{\Delta}\right]\right)},
    \end{split}
    \label{eq:Sint_updown_HST_tilde}
\end{equation}
where
\begin{equation}
    \begin{split}
        S^{\uparrow\downarrow}_{\mathrm{HS}, 0}\left[\btilde{\Delta},\tilde{\Delta}\right]=
         \int_{x,x'}\frac{\big[\big(v_{t-t'}\big)_n\big]^2}{\big(\lambda_{t-t'}\big)_n} \big(\btilde{\Delta}_{x,x'}\big)_n \big(\tilde{\Delta}_{x,x'}\big)_n,
    \end{split}
\end{equation}
and
\begin{equation}
    \begin{split}
        &S^{\uparrow\downarrow}_{\mathrm{HS}, \mathrm{int}}\left[\bar{\psi},\psi,\btilde{\Delta},\tilde{\Delta}\right]= \int_{x,x'} \big(v_{t-t'}\big)_n \\ &\quad\times\Big[\big(\btilde{\Delta}_{x,x'}\big)_n \big(\btilde{\rho}^{\uparrow \downarrow}_{x, x'}\big)_{n}
        + \big(\btilde{\rho}^{\uparrow \downarrow}_{x, x'}\big)_{n}  \big(\tilde{\Delta}_{x,x'}\big)_n \Big] .
    \end{split}
\end{equation}
We note that, as discussed in \cite{Dalal2023} $\btilde{\Delta}$ and $\tilde{\Delta}$ are not necessarily connected by complex conjugation. 

Equivalently, one can multiply the interaction term $\exp(i S_{\mathrm{int}}^{\sigma \sigma}[\bar{\psi},\psi])$ by the identity \eqref{eq:Gauss_id_S_tilde} and perform the change of variables
\begin{equation}
    \begin{split}
        &\tilde{\mathcal{S}}^{\sigma}_{n}(x,x') = \big(\zeta^{\sigma}_{x,x'}\big)_n - \frac{\big(\lambda_{t-t'}\big)_n}{2 \big(\tilde{v}_{t-t'}\big)_n} \big(\btilde{\rho}^{\sigma}_{x, x'}\big)_{n}, \\
        &\big(\btilde{\mathcal{S}}^{\sigma}_{x,x'}\big)_n = \big(\zeta^{\sigma}_{x,x'}\big)^*_n - \frac{\big(\lambda_{t-t'}\big)_n}{2 \big(\tilde{v}_{t-t'}\big)_n} \big(\btilde{\rho}^{\sigma}_{x, x'}\big)_{n},
    \end{split} 
\end{equation}
to obtain
\begin{equation}
    \begin{split}
        e^{iS_{\mathrm{int}}^{\sigma \sigma}\left[\bar{\psi},\psi\right]}&=\prod_{n}  \int \mathcal{D}\left[\btilde{\mathcal{S}}^{\sigma}_n,\tilde{\mathcal{S}}^{\sigma}_n\right]  \\
        &\quad\times
        e^{i \left(S^{\sigma \sigma}_{\mathrm{HS}, 0}\left[\btilde{\mathcal{S}},\tilde{\mathcal{S}}\right]
        +S^{\sigma \sigma}_{\mathrm{HS}, \mathrm{int}}\left[\bar{\psi},\psi,\btilde{\mathcal{S}},\tilde{\mathcal{S}}\right]\right)},
    \end{split}
    \label{eq:Sint_sigmasigma_HST_tilde}
\end{equation}
where
\begin{equation}
    \begin{split}
        S^{\sigma \sigma}_{\mathrm{HS}, 0}&\left[\btilde{\mathcal{S}},\tilde{\mathcal{S}}\right] \\ 
        &= -\int_{x,x'}\frac{2\big[\big(\tilde{v}_{t-t'}\big)_n\big]^2}{\big(\lambda_{t-t'}\big)_n} \big(\btilde{\mathcal{S}}^{\sigma}_{x,x'}\big)_n \big(\tilde{\mathcal{S}}^\sigma_{x,x'}\big)_n,
    \end{split}
\end{equation}
and
\begin{equation}
    \begin{split}
        &S^{\sigma \sigma}_{\mathrm{HS}, \mathrm{int}}\left[\bar{\psi},\psi,\btilde{\mathcal{S}},\tilde{\mathcal{S}}\right]= - \int_{x,x'}\big(\tilde{v}_{t-t'}\big)_n \\
        &\quad\times\Big[\big(\btilde{\mathcal{S}}^{\sigma}_{x,x'}\big)_n \big(\btilde{\rho}^{\sigma}_{x, x'}\big)_{n}
        +  \big(\bar{\btilde{\rho}}^{\sigma}_{x, x'}\big)_{n}  \big(\tilde{\mathcal{S}}^{\sigma}_{x,x'}\big)_n \Big]. 
    \end{split}
\end{equation}

The quartic interaction terms \eqref{eq:Sint_updown_diag} and \eqref{eq:Sint_sigmasigma_diag} for the electrons have then been replaced with a linear coupling of electron bilinears ($\tilde{\rho}$) with the auxiliary fields ($\tilde{\Delta}$, $\tilde{\mathcal{S}}$), plus the free quadratic action for the auxiliary fields. We can now use the transformation matrix of \cref{eq:U_trans_matrix} to express the fermion bilinears $\tilde{\rho}$ back in terms of the bilinears $\rho$ in the original Keldysh basis [\cref{eq:rho_updown,eq:rho_sigmasigma}] as
\begin{equation}
    \tilde{\rho}^{\uparrow \downarrow}_{n} = \sum_{m} U_{n m} \rho^{\uparrow \downarrow}_{m},\quad
    \btilde{\rho}^{\uparrow \downarrow}_{n} = \sum_{m} U_{n m} \bar{\rho}^{\uparrow \downarrow}_{m},
\end{equation}
and
\begin{equation}
    \tilde{\rho}^{\sigma}_{n}  = \sum_{m} U_{n m} \rho^{\sigma}_{m},\quad
    \btilde{\rho}^{\sigma}_{n} = \sum_{m} U_{n m} \bar{\rho}^{\sigma}_{m}.
\end{equation}
Moreover, one can now redefine the auxiliary fields as 
\begin{equation}
    \Delta_{n}=\sum_{m} U_{n m} v_{m} \tilde{\Delta}_{m}, \quad
    \bar{\Delta}_{n}=\sum_{m} U_{n m} v_{m} \btilde{\Delta}_{m},
    \label{eq:Delta_rotation}
\end{equation}
and
\begin{equation}
    \mathcal{S}^{\sigma}_{n}=\sum_{m} U_{n m} \tilde{v}_{m} \tilde{\mathcal{S}}^{\sigma}_{m}, \quad
    \bar{\mathcal{S}}^{\sigma}_{n}=\sum_{m} U_{n m} \tilde{v}_{m} \btilde{\mathcal{S}}^{\sigma}_{m},
    \label{eq:S_rotation}
\end{equation}
where the space-time indices $(x,x')$ have been dropped for simplicity. The introduction of the factors $v_m$ and $\tilde{v}_m$ in this redefinition allows us to go back to auxiliary fields similar to the ones introduced in the ``naive'' formulation of the HS transformations in \cref{eq:naive_HST_Delta,eq:naive_HST_phi}, with the advantage that now the Gaussian integrals are convergent, because the measure of the integrals $\int \mathcal{D}[\bar\Delta,\Delta]$ and $\int\mathcal{D}[\bar{\mathcal{S}},\mathcal{S}]$ keeps memory of the rotations of \cref{eq:Delta_rotation,eq:S_rotation}.
With these transformations plus \cref{eq:D_diag_trans}, one can then rewrite \cref{eq:Sint_updown_HST_tilde} as
\begin{equation}
    \label{eq:Sint_gap}
    \begin{split}
        &e^{iS_{\mathrm{int}}^{\uparrow\downarrow}\left[\bar{\psi},\psi\right]}\\
        &= \int \mathcal{D}\left[\bar{\Delta},\Delta\right] e^{i \left(S^{\uparrow\downarrow}_{\mathrm{HS}, 0}[\bar{\Delta},\Delta]+ S^{\uparrow\downarrow}_{\mathrm{HS}, \mathrm{int}}\left[\bar{\psi},\psi,\bar{\Delta},\Delta\right] \right)}
    \end{split}
\end{equation}
where we used the shorthand notation  $\int\mathcal{D}\left[\bar{\Delta},{\Delta}\right]=\prod_{n} \int\mathcal{D}\left[\bar{\Delta}_n,\Delta_n\right]$
and we defined
\begin{equation}\label{eq:DeltaBareAction}
    S^{\uparrow\downarrow}_{\mathrm{HS}, 0}\left[\bar{\Delta},\Delta\right]=\frac{1}{g}\int_{x,x'}  \big(\bar{\Delta}_{x,x'})^T  \mathds{D}_{t-t'}^{-1}  \Delta_{x,x'},
\end{equation}
and
\begin{equation}\label{eq:DeltaIntAction}
    \begin{split}
        S^{\uparrow\downarrow}_{\mathrm{HS}, \mathrm{int}}&\left[\bar{\psi},\psi,\bar{\Delta},\Delta\right]=\\
        &\int_{x,x'} \left[ (\bar{\Delta}_{x,x'})^T  \rho^{\uparrow \downarrow}_{x,x'} + (\bar{\rho}^{\uparrow \downarrow}_{x,x'})^T   \Delta_{x,x'} \right],
    \end{split}
\end{equation}
where we reorganized the fields $\Delta_{n}$ (and their barred counterparts) in the vector form $\Delta^T=(\Delta_1,\Delta_2,\Delta_3,\Delta_4)$.
Equivalently, one can rewrite \cref{eq:Sint_sigmasigma_HST_tilde} as
\begin{equation}\label{eq:Sint_SE}
    \begin{split}
        &e^{iS_{\mathrm{int}}^{\sigma \sigma}\left[\bar{\psi},\psi\right]} \\
        &= \int \mathcal{D}\left[\bar{\mathcal{S}}^{\sigma},\mathcal{S}^{\sigma}\right] e^{i \left(S^{\sigma \sigma}_{\mathrm{HS}, 0}\left[\bar{\mathcal{S}},\mathcal{S}\right]+ S^{\sigma \sigma}_{\mathrm{HS},\mathrm{int}}\left[\bar{\psi},\psi,\bar{\mathcal{S}},\mathcal{S}\right] \right)} 
    \end{split}
\end{equation}
where we used the shorthand notation $\int\mathcal{D}\left[\bar{\mathcal{S}}^{\sigma},\mathcal{S}^{\sigma}\right]=\prod_{n} \int\mathcal{D}\left[\bar{\mathcal{S}}^{\sigma}_n,\mathcal{S}^{\sigma}_n\right]$ and we defined
\begin{equation}\label{eq:SEBareAction}
    S^{\sigma \sigma}_{\mathrm{HS}, 0}\left[\bar{\mathcal{S}},\mathcal{S}\right]=-\frac{2}{g} \int_{x,x'} \big(\bar{\mathcal{S}}^{\sigma}_{x,x'})^T  \mathds{D}_{t-t'}^{-1}  \mathcal{S}^{\sigma}_{x,x'},
\end{equation}
\begin{equation}\label{eq:SEintAction}
    \begin{split}   
     S^{\sigma \sigma}_{\mathrm{HS},\mathrm{int}}&\left[\bar{\psi},\psi,\bar{\mathcal{S}},\mathcal{S}\right]=\\
     &-\int_{x,x'} \left[(\bar{\mathcal{S}}^{\sigma}_{x,x'})^T  \rho^{\sigma}_{x,x'} +  (\bar{\rho}^{\sigma}_{x,x'})^T   \mathcal{S}^{\sigma}_{x,x'} \right],
    \end{split} 
\end{equation}
where we reorganized the fields $\mathcal{S}^{\sigma}_{n}$ (and their barred counterparts) in the vector form $(\mathcal{S}^{\sigma})^T=(\mathcal{S}^{\sigma}_{1},\mathcal{S}^{\sigma}_{2},\mathcal{S}^{\sigma}_{3},\mathcal{S}^{\sigma}_{4})$.

\subsection{Nambu structure and integrating out the electrons}
\label{subsec:integrating_out_electrons}
The next step is integrating out the fermionic fields $(\bar{\psi},\psi)$ to obtain an effective action for the auxiliary fields $(\bar{\Delta},\Delta,\bar{\mathcal{S}},\mathcal{S})$.
To this extent, it is convenient to transform the action back into momentum-frequency space $k=(\bm{k},\omega)$. 
This relies on the system being translationally invariant in space and time, so that the auxiliary fields and the propagators are only functions of the difference $(x-x')$. 
The resulting action is
\begin{equation}
    \begin{split}
        &S^{\uparrow\downarrow}_{\mathrm{HS},0}\left[\bar{\Delta},\Delta\right]= \frac{1}{g} \sum_{k,k'} \bar{\Delta}_{k}^T  \mathds{D}^{-1}_{k-k'}  \Delta_{k'}, \\
        &S^{\uparrow\downarrow}_{\mathrm{HS},\mathrm{int}}\left[\bar{\psi},\psi,\bar{\Delta},\Delta\right]= \frac{1}{g} \sum_{k} \left[ (\bar{\rho}^{\uparrow\downarrow}_{k, -k})^T   \Delta_{k} + (\bar{\Delta}_{k})^T  \rho^{\uparrow\downarrow}_{k, -k}    \right], \\ 
    \end{split}
    \label{eq:S_HS_updown}
\end{equation}
for the $\Delta$-fields and
\begin{equation}
    \begin{split}
        &S^{\sigma\sigma}_{\mathrm{HS},0}\left[\bar{\mathcal{S}},\mathcal{S}\right]=
        -\frac{2}{g} \sum_{k,k'} (\bar{\mathcal{S}}^{\sigma}_{k})^T  \mathds{D}^{-1}_{k-k'}  \mathcal{S}^\sigma_{k'}, \\
        &S^{\sigma \sigma}_{\mathrm{HS},\mathrm{int}}\left[\bar{\psi},\psi,\bar{\mathcal{S}},\mathcal{S}\right]= -\frac{1}{g} \sum_{k} \left[ (\bar{\rho}^{\sigma}_{k,k} )^T   \mathcal{S}^{\sigma}_{k} + (\bar{\mathcal{S}}^{\sigma}_{k})^T  \rho^{\sigma}_{k,k}     \right], 
    \end{split}
    \label{eq:S_HS_sigmasigma}
\end{equation}
for the $\mathcal{S}^\sigma$-fields, where we have used the shorthand notations $\sum_{k}= \sum_{\bm{k}} \int \frac{d\omega}{2 \pi}$ and $\mathds{D}^{-1}_{k-k'}=\mathds{D}^{-1}_{\omega-\omega'} \, \delta_{\bm{k},\bm{k}'}$.
Finally, the free part of the electron action in \cref{eq:free_fermion_action_kt} can be expressed in the momentum-frequency basis as
\begin{equation}
    S_{e,0 }\left[\bar{\psi}, \psi\right] = \sum_{k,\sigma} \begin{pmatrix} \bar{\psi}^{c}_\sigma & \bar{\psi}^{q}_\sigma \end{pmatrix}_{k}
    \begin{pmatrix}
    0 & [G_0^{-1}]^{A} \\
    [G_0^{-1}]^{R} & [G_0^{-1}]^{K}
    \end{pmatrix}_{k}
    \begin{pmatrix}
    \psi^{c}_\sigma \\
    \psi^{q}_\sigma
    \end{pmatrix}_{k}.
    \label{eq:free_fermion_action}
\end{equation}
To integrate out the electrons, it is necessary to reorganize the $\psi$-dependent terms in a Nambu structure \cite{Altland&Simons},which is achieved by introducing the Keldysh-Nambu spinors
\begin{equation}
    \Psi_{k}=
    \begin{pmatrix}
        \psi^{c}_{\uparrow k} \\
        \psi^{q}_{\uparrow k}\\
        {\bar{\psi}}^{c}_{\downarrow \, -k} \\
        {\bar{\psi}}^{q}_{\downarrow \, -k}
    \end{pmatrix}, \quad
    \bar{\Psi}_{k}=
    \begin{pmatrix}
        \bar{\psi}^{c}_{\uparrow k} \\
        \bar{\psi}^{q}_{\uparrow k}\\
        \psi^{\,c}_{\downarrow \, -k} \\
        \psi^{\,q}_{\downarrow \, -k}
    \end{pmatrix},
    \label{eq:Keldysh-Nambu}
\end{equation}
and reorganize the auxiliary field vectors in the $2\times2$ matrix forms
\begin{equation}
    \Delta_{k}=
    \begin{pmatrix}
        \big(\Delta_k)_1\\
        \big(\Delta_k)_2\\
        \big(\Delta_k)_3\\
        \big(\Delta_k\big)_4
    \end{pmatrix} \quad \to \quad
    \mathbbm{\Delta}_k=
    \begin{pmatrix}
        \big(\Delta_k)_1 &
        \big(\Delta_k)_2\\
        \big(\Delta_k)_3 &
        \big(\Delta_k)_4
    \end{pmatrix}, 
    \label{eq:Delta_matrix}
\end{equation}
and
\begin{equation}
    \mathcal{S}^{\sigma}_{k}=
    \begin{pmatrix}
        \big(\mathcal{S}^{\sigma}_k\big)_1\\
        \big(\mathcal{S}^{\sigma}_k\big)_2\\
        \big(\mathcal{S}^{\sigma}_k\big)_3\\
        \big(\mathcal{S}^{\sigma}_k\big)_4
    \end{pmatrix} \quad \to \quad
    \mathbb{S}^{\sigma}_k=
    \begin{pmatrix}
        \big(\mathcal{S}^{\sigma}_k\big)_1 &
        \big(\mathcal{S}^{\sigma}_k\big)_2\\
        \big(\mathcal{S}^{\sigma}_k\big)_3 &
        \big(\mathcal{S}^{\sigma}_k\big)_4
    \end{pmatrix},    
    \label{eq:S_matrix}
\end{equation}
and equivalently for the barred fields. The three $\psi$-dependent terms can then be written in the form
\begin{equation}
    \begin{split}
        &S_{e,\text{HS}}\left[\bar{\psi}, \psi,\bar{\Delta},\Delta,\bar{\mathcal{S}},\mathcal{S}\right]\equiv S_{e,0 }[\bar{\psi}, \psi]+ S^{\uparrow\downarrow}_{\mathrm{HS},\mathrm{int}}\left[\bar{\psi},\psi,\bar{\Delta},\Delta\right] \\&\quad+ \sum_\sigma S^{\sigma \sigma}_{\mathrm{HS},\mathrm{int}}\left[\bar{\psi},\psi,\bar{\mathcal{S}},\mathcal{S}\right]
        =\sum_k \bar{\Psi}_{k}^T  \mathbb{G}^{-1}_{k} \Psi_{k},
    \end{split}
\end{equation}
with the inverse Keldysh-Nambu propagator defined as the following block matrix
\begin{equation}
    \mathbb{G}^{-1}_{k}=
    \begin{pmatrix}
         G^{\uparrow \, -1}_{k} & \mathbbm{\Delta}_{k} \\
         \bar{\mathbbm{\Delta}}^T_{k} & -G^{\downarrow \, -T}_{-k}
    \end{pmatrix},
    \label{eq:G-1_Nambu}
\end{equation}
where we used the shorthand notation
\begin{equation}
    G^{\sigma}_{k}=\left(\left[G_{0}^{-1}\right]_k-\mathbbm{\Sigma}^{\sigma}_{k}\right)^{-1}
    \label{eq:Dyson_normalG}
\end{equation}
for the normal-state propagator and we have defined
\begin{equation}
    \mathbbm{\Sigma}^{\sigma}_k= \mathbb{S}^{\sigma}_{k}+\bar{\mathbb{S}}^{\sigma  T}_{k},
    \label{eq:Sigma_def}
\end{equation}
anticipating that this symmetrized combination of the auxiliary fields $(\bar{\mathbb{S}}^{\sigma}_{k},\mathbb{S}^{\sigma}_{k})$ will assume the role of the normal electron self-energy.
With this interpretation of $\mathbbm{\Sigma}^{\sigma}_k$, \cref{eq:Dyson_normalG} simply becomes the Dyson equation for the normal-state propagator. From the structure of \cref{eq:G-1_Nambu}, one sees that $-\mathbbm{\Delta}_{k}$ can instead be interpreted as an anomalous self-energy.

Proceeding with the Gaussian integration of the electron fields $(\bar{\Psi},\Psi)$, one obtains
\begin{equation}
    \begin{split}
        &e^{i S_{\mathrm{HS}}\left[\bar{\Delta},\Delta,\bar{\mathcal{S}},\mathcal{S}\right]}=\int \mathcal{D}\left[\bar{\psi},\psi\right] \, e^{i S_{e,\text{HS}}\left[\bar{\psi}, \psi,\bar{\Delta},\Delta,\bar{\mathcal{S}},\mathcal{S}\right] }
        \\
        &= \det(-i\mathbb{G}^{-1}) =e^{\ln\left[\det\left(-i\mathbb{G}^{-1}\right)\right]}=e^{\operatorname{Tr}\left[\ln\left(-i\mathbb{G}^{-1}\right)\right]},
    \end{split}
    \label{eq:S_HS_final}
\end{equation}
where the trace operator ($\operatorname{Tr}$) sums both over the diagonal terms in the Keldysh-Nambu structure and sums/integrates over $k=(\bm{k},\omega)$. The partition function then reads
\begin{equation}
    \begin{split}
    \mathcal{Z}=& \int  \mathcal{D}\left[\bar{\Delta},\Delta\right] \, \prod_\sigma \int \mathcal{D}\left[\bar{\mathcal{S}}^\sigma,\mathcal{S}^\sigma\right] \\
    & \times e^{i \left(S^{\uparrow\downarrow}_{\mathrm{HS},0}\left[\bar{\Delta},\Delta\right]+ \sum_\sigma S^{\sigma \sigma}_{\mathrm{HS},0}\left[\bar{\mathcal{S}},\mathcal{S}\right]+ S_{\mathrm{HS}}\left[\bar{\Delta},\Delta,\bar{\mathcal{S}},\mathcal{S}\right]\right) }.
    \end{split}
    \label{eq:HS_partition_function}
\end{equation}

\subsection{Saddle-point equations for the auxiliary fields}
\label{subsec:saddle-point_equations}
The final step of our derivation is to perform a saddle-point (or mean-field) approximation, i.e.~approximating our auxiliary fields with their non-fluctuating values that extremize the action in \cref{eq:HS_partition_function}. This is done by setting to zero the functional derivatives of the action in \cref{eq:HS_partition_function} with respect to the auxiliary fields.

Let us start with the saddle-point equation for the normal self-energy $\mathbbm{\Sigma}^{\sigma}_{k}=\mathbb{S}^{\sigma}_{k}+\bar{\mathbb{S}}^{\sigma T}_{k}$. As the system has balanced spin populations ($\mu=\mu_\uparrow=\mu_\downarrow$), one has $\mathbbm{\Sigma}^{\uparrow}_{k}=\mathbbm{\Sigma}^{\downarrow}_{k}$ making it sufficient to only derive the equations for $\mathbbm{\Sigma}^{\downarrow}_{k}$. 

Taking the functional derivative of the free action $S^{\downarrow\downarrow}_{\mathrm{HS},0}$ defined in \cref{eq:S_HS_sigmasigma} with respect to the fields $\bar{\mathcal{S}}^{\downarrow}_{n}$ and $\mathcal{S}^{\downarrow}_{n}$, one obtains
\begin{equation}
    \begin{split}
        \frac{\delta S^{\downarrow \downarrow}_{\mathrm{HS},0}\left[\bar{\mathcal{S}},\mathcal{S}\right]}{\delta \big(\bar{\mathcal{S}}^{\downarrow}_{k}\big)_n} 
        &=-\frac{2}{g}\sum_{k'} \sum_{m}  \big(\mathds{D}^{-1}_{k-k'}\big)_{n m}  \big(\mathcal{S}^{\downarrow}_{k'}\big)_m, \\
        \frac{\delta S^{\downarrow \downarrow}_{\mathrm{HS},0}\left[\bar{\mathcal{S}},\mathcal{S}\right]}{\delta \big(\mathcal{S}^{\downarrow}_{k}\big)_n} 
        &=-\frac{2}{g}\sum_{k'} \sum_{m}  \big(\bar{\mathcal{S}}^{\downarrow}_{k'}\big)_m \big(\mathds{D}^{-1}_{k'-k}\big)_{m n}  .
    \end{split}
    \label{eq:derivative_S_HS_sigmasigma}
\end{equation}
The functional derivatives of $S_{\mathrm{HS}}$ defined in \cref{eq:S_HS_final} are instead given by
\begin{equation}
    \begin{split}
        &\frac{\delta S_{\mathrm{HS}}\left[\bar{\Delta},\Delta,\bar{\mathcal{S}},\mathcal{S}\right]}{\delta \big(\bar{\mathcal{S}}^{\downarrow}_{k}\big)_n} \\
        &\quad= -i \frac{\delta \operatorname{Tr}\left[\ln(-i\mathbb{G}^{-1}_{k'})\right]}{\delta \big(\bar{\mathcal{S}}^{\downarrow}_{k}\big)_n}=-i \operatorname{Tr}\left[ \mathbb{G}_{k'}  \frac{\delta\mathbb{G}^{-1}_{k'}}{\delta \big(\bar{\mathcal{S}}^{\downarrow}_{k}\big)_n} \right], 
    \end{split}
    \label{eq:derivative_Trlog_barS}
\end{equation}
and
\begin{equation}
    \begin{split}
        &\frac{\delta S_{\mathrm{HS}}\left[\bar{\Delta},\Delta,\bar{\mathcal{S}},\mathcal{S}\right]}{\delta \big(\mathcal{S}^{\downarrow}_{k}\big)_n} \\
        &\quad= -i \frac{\delta \operatorname{Tr}\left[\ln(-i\mathbb{G}^{-1}_{k'})\right]}{\delta \big(\mathcal{S}^{\downarrow}_{k}\big)_n}=-i \operatorname{Tr}\left[ \mathbb{G}_{k'}  \frac{\delta\mathbb{G}^{-1}_{k'}}{\delta \big(\mathcal{S}^{\downarrow}_{k}\big)_n} \right].
    \end{split}
    \label{eq:derivative_Trlog_S}
\end{equation}
To further proceed, one needs then to compute the full Keldysh-Nambu propagator $\mathbb{G}_{k}$ by inverting the matrix $\mathbb{G}^{-1}_{k}$ defined in \cref{eq:G-1_Nambu}. By employing the Schur complement formula for the inversion of a block matrix \cite{HornJohnsonBook}, one obtains
\begin{equation}
    \mathbb{G}_{k}=
    \begin{pmatrix}
        \mathcal{G}^{\uparrow}_{k} & \mathcal{F}_{k} \\
        \bar{\mathcal{F}}_{k} & \bar{\mathcal{G}}^{\downarrow}_{k}
    \end{pmatrix}
    \label{eq:G_Nambu}
\end{equation}
where $\mathcal{G}$ and $\mathcal{F}$ are respectively the normal and anomalous electron propagators in the superconducting phase, and they are given by
\begin{equation}
    \begin{split}
        &\mathcal{G}^{\uparrow}_{k} =\big(G^{\uparrow \,-1}_{k}+\mathbbm{\Delta}_{k} G^{\downarrow  \, T}_{-k} \bar{\mathbbm{\Delta}}^{T}_{k}\big)^{-1}, \\
        &\bar{\mathcal{G}}^{\downarrow}_{k}=-\big(G^{\downarrow \,-T}_{-k}+\bar{\mathbbm{\Delta}}^T_{k} G^{\uparrow}_{k} \mathbbm{\Delta}_{k}\big)^{-1}, \\
        &\mathcal{F}_{k}=G^{\uparrow}_{k} \mathbbm{\Delta}_{k} \big(G^{\downarrow \,-T}_{-k}+\bar{\mathbbm{\Delta}}^T_{k} G^{\uparrow}_{k} \mathbbm{\Delta}_{k}\big)^{-1}=-G^{\uparrow}_{k} \mathbbm{\Delta}_{k} \bar{\mathcal{G}}^{\downarrow}_{k},\\
        &\bar{\mathcal{F}}_{k}=\big(G^{\downarrow \,-T}_{-k}\!+\bar{\mathbbm{\Delta}}^T_{k} G^{\uparrow}_{k} \mathbbm{\Delta}_{k}\big)^{-1} \bar{\mathbbm{\Delta}}^T_{k} G^{\uparrow}_{k}=-\bar{\mathcal{G}}^{\downarrow}_{k}\bar{\mathbbm{\Delta}}^T_{k} G^{\uparrow}_{k}.
    \end{split} 
    \label{eq:G_Nambu_elements}
\end{equation}
At the mean-field level, it can be shown (see \cref{app:causal_structure_MF}) that the upper and lower matrix blocks are connected by the relations $\bar{\mathcal{F}}_{k}=-\sigma_z \mathcal{F}^{\dag}_{k}\sigma_z$ ($\sigma_z$ being the third Pauli matrix) and $\bar{\mathcal{G}}^{\downarrow}_k=-\mathcal{G}^{\downarrow T}_{-k}$.

The functional derivatives of $\mathbb{G}^{-1}_{k}$ are instead given by
\begin{equation}
    \begin{split}
        \frac{\delta\mathbb{G}^{-1}_{k'}}{\delta \big(\bar{\mathcal{S}}^{\downarrow}_{k}\big)_n} = 
        \begin{pmatrix}
            \mathbbm{0} & \mathbbm{0} \\
            \mathbbm{0} & \mathbb{I}_{n} \delta_{k+k'}
        \end{pmatrix}, \\
        \frac{\delta\mathbb{G}^{-1}_{k'}}{\delta \big(\mathcal{S}^{\downarrow}_{k}\big)_n} = 
        \begin{pmatrix}
            \mathbbm{0} & \mathbbm{0} \\
            \mathbbm{0} & \mathbb{I}_{n}^{\, T} \delta_{k+k'}
        \end{pmatrix},   
    \end{split}
     \label{eq:dG-1_dS}
\end{equation}
where $\delta_{k+k'}=\delta_{\bm{k},-\bm{k}'}\delta_{\omega+\omega'}$ and the matrix $\mathbb{I}_{n}$ is defined as 
\begin{equation}
    \mathbb{I}_{n}=
    \begin{pmatrix}
        \delta_{n,1} & \delta_{n,2}\\
        \delta_{n,3} & \delta_{n,4}\\
    \end{pmatrix},
    \label{eq:I_n}
\end{equation}
i.e.~a matrix with a $1$ in the position $n$ and the rest of the elements zero. Plugging \cref{eq:G_Nambu,eq:dG-1_dS} in \cref{eq:derivative_Trlog_S}, one obtains
\begin{equation}
    \begin{split}
        \frac{\delta S_{\mathrm{HS}}\left[\bar{\Delta},\Delta,\bar{\mathcal{S}},\mathcal{S}\right]}{\delta \big(\bar{\mathcal{S}}^{\downarrow}_{k}\big)_n}
        &=i\operatorname{Tr}\left( \mathcal{G}^{\downarrow \, T}_{-k'} \mathbb{I}_{n} \delta_{k+k'}\right)=i \big(\mathcal{G}^{\downarrow}_{k}\big)_{n},
        \\
        \frac{\delta S_{\mathrm{HS}}\left[\bar{\Delta},\Delta,\bar{\mathcal{S}},\mathcal{S}\right]}{\delta \big(\mathcal{S}^{\downarrow}_{k}\big)_n} 
        &=i\operatorname{Tr}\left( \mathcal{G}^{\downarrow \, T}_{-k'} \mathbb{I}_{n}^{\, T} \delta_{k+k'}\right)=i \big(\mathcal{G}^{\downarrow  \, T}_{k}\big)_{n},
    \end{split}
\end{equation}
where the $n$-th element of the matrix $\mathcal{G}^{\downarrow}_k=(G^{\uparrow \,-1}_{k}+\mathbbm{\Delta}_{k} G^{\downarrow  \, T}_{-k} \bar{\mathbbm{\Delta}}^{T}_{k})^{-1}$ is defined as in the matrix structure of \cref{eq:I_n}. 
The saddle-point equations for $\bar{\mathcal{S}}^{\downarrow}_n$ and $\mathcal{S}^{\downarrow}_n$ are then
\begin{equation}
    \begin{split}
        &\frac{2}{g}\sum_{k'} \sum_{m}  \big(\mathds{D}^{-1}_{k-k'}\big)_{n m}  \big(\mathcal{S}^{\downarrow}_{k'}\big)_m = i \big(\mathcal{G}^{\downarrow}_{k}\big)_{n},
        \\
        &\frac{2}{g}\sum_{k'} \sum_{m}  \big(\bar{\mathcal{S}}^{\downarrow}_{k'}\big)_m \big(\mathds{D}^{-1}_{k'-k}\big)_{m n} = i\big(\mathcal{G}^{\downarrow \,T}_{k}\big)_{n}.
    \end{split}
\end{equation}
Taking the transpose of the second equation and summing the two equation together, one obtains the saddle-point equations for the self-energy $\mathbbm{\Sigma}^{\downarrow}_{k}=\mathbb{S}^{\downarrow}_{k}+\bar{\mathbb{S}}^{\downarrow \, T}_{k}$:
\begin{equation}
    \frac{1}{g}\sum_{k'} \sum_{m}  \big(\mathds{D}^{-1}_{k-k'}\big)_{n m}  \big(\Sigma^{\downarrow}_{k'}\big)_{m} = i\big(\mathcal{G}^{\downarrow}_{k}\big)_{n}.
\end{equation}
Finally, using the inversion relation for the matrix $\mathds{D}_{k-k'}$
\begin{equation}
    \sum_{k'} \sum_{n'}  \big(\mathds{D}_{k-k'}\big)_{n n'} \big(\mathds{D}^{-1}_{k'-k''}\big)_{n' m} = \delta_{n m} \delta_{k-k''},
    \label{eq:inversion_D}
\end{equation}
and dropping the spin-indices, the saddle-point equations for the normal self-energy become
\begin{equation}
     \Sigma_{n}(k) = i g \sum_{k'} \sum_{m}  \big(\mathds{D}_{k-k'}\big)_{n m} \big(\mathcal{G}_{k'}\big)_{m},
     \label{eq:4comp_Sigma_eq}
\end{equation}
where the structure of a Fock self-energy is now evident [cf.~\pref{fig:SE}{a}]. 

Due to the convolution structure  of \cref{eq:4comp_Sigma_eq}, $\mathbbm{\Sigma}_k$ inherits the causal structure of $\mathds{D}_{k-k'}$ and $\mathcal{G}_{k}$ [see \cref{eq:G_operatorial}], resulting in $\Sigma_{1}=0$.
The remaining matrix elements can be identified as the advanced ($\Sigma_2\to\Sigma^A$), retarded ($\Sigma_3 \to \Sigma^R$)
and Keldysh ($\Sigma_4 \to \Sigma^K$) components of the normal self-energy.
The matrix 
$\mathbbm{\Sigma}_k$ of \cref{eq:Sigma_def} is then simplified to the form
\begin{equation}
    \mathbbm{\Sigma}_k=
    \begin{pmatrix}
        0 &
        \Sigma^A_k\\
        \Sigma^R_k &
        \Sigma^K_k
    \end{pmatrix}.
    \label{eq:Sigma_matrix_RAK}
\end{equation}
Using this RAK notation, and recalling the definition for $(\mathds{D}_{k-k'})_{nm}=(\mathds{D}_{\omega-\omega'})_{nm} \delta_{\bm{k},\bm{k}'}$ as the Fourier transform of \cref{eq:D_matrix}, \cref{eq:4comp_Sigma_eq} becomes
\begin{equation}
    \begin{pmatrix}
         \Sigma^{A}_{\bm{k}, \omega} \\     \Sigma^{R}_{\bm{k}, \omega} \\ \Sigma^{K}_{\bm{k}, \omega}   
    \end{pmatrix}
    = i g \int_{\omega'} \\
    \begin{pmatrix}
        D^{A} \mathcal{G}^{K} + D^{K} \mathcal{G}^{A} \\
        D^{R} \mathcal{G}^{K} + D^{K} \mathcal{G}^{R} \\
        D^{K} \mathcal{G}^{K} + D^{R} \mathcal{G}^{R}+ D^{A} \mathcal{G}^{A} 
    \end{pmatrix},
    \label{eq:RAK_Sigma_eqs}
\end{equation}
where we used the shorthand notations $\int_{\omega'}=\int \frac{d\omega'}{2\pi}$ and $D^{\alpha}\mathcal{G}^{\beta}=D^{\alpha}_{\omega-\omega'}\mathcal{G}^{\beta}_{\bm{k}, \omega'}$. 

Next, we consider the saddle-point equation for the $\Delta$-fields. Taking the functional derivative of the free action term $S^{\uparrow\downarrow}_{\mathrm{HS},0}$ defined in \cref{eq:S_HS_updown} with respect to $\bar{\Delta}_{n}$, one obtains
\begin{equation}
    \frac{\delta S^{\uparrow\downarrow}_{\mathrm{HS},0}\left[\bar{\Delta},\Delta\right]}{\delta \bar{\Delta}_{n}(k)}=\frac{1}{g}\sum_{k'} \sum_{m}  \big(\mathds{D}^{-1}_{k-k'}\big)_{n m}  \big(\Delta_{k'}\big)_m,
    \label{eq:derivative_S_HS_updown}
\end{equation}
The functional derivative of $S_{\mathrm{HS}}$ defined in \cref{eq:S_HS_final} is instead given by 
\begin{equation}
    \begin{split}
    &\frac{\delta S_{\mathrm{HS}}\left[\bar{\Delta},\Delta,\bar{\mathcal{S}},\mathcal{S}\right]}{\delta \big(\bar{\Delta}_{k}\big)_n}
    \\
    &=-i \frac{\delta \operatorname{Tr}\left[\ln\left(-i\mathbb{G}^{-1}_{k'}\right)\right]}{\delta \big(\bar{\Delta}_{k}\big)_n}=-i \operatorname{Tr}\left[ \mathbb{G}_{k'}  \frac{\delta\mathbb{G}^{-1}_{k'}}{\delta \big(\bar{\Delta}_{k}\big)_n} \right],
    \end{split}
    \label{eq:derivative_Trlog_Delta}
\end{equation}
where $\mathbb{G}_{k}$ is defined in \cref{eq:G_Nambu}, while the functional derivative of $\mathbb{G}^{-1}_{k}$ is given by
\begin{equation}
     \frac{\delta\mathbb{G}^{-1}_{k'}}{\delta \big(\bar{\Delta}_{k}\big)_n} = 
     \begin{pmatrix}
         \mathbbm{0} & \mathbbm{0} \\
         \mathbb{I}_{n}^{\,T}  \delta_{k-k'} & \mathbbm{0}
     \end{pmatrix},
     \label{eq:dG-1_dbDelta}
\end{equation}
where we used the same notations of \cref{eq:dG-1_dS}.
Plugging \cref{eq:G_Nambu,eq:dG-1_dbDelta} in \cref{eq:derivative_Trlog_Delta} one obtains then
\begin{equation}
    \begin{split}
        &\frac{\delta S_{\mathrm{HS}}\left[\bar{\Delta},\Delta,\bar{\mathcal{S}},\mathcal{S}\right]}{\delta \big(\bar{\Delta}_{k}\big)_n} \\
        &\quad= -i \operatorname{Tr}\left(\mathcal{F}_{k'} \, \mathbb{I}_{n}^{T}\delta_{k-k'}\right)=-i\big(\mathcal{F}_{k}\big)_{n},
    \end{split}
    \label{eq:derivative_Trlog_Delta_final}
\end{equation}
where the $n$-th element of the matrix $\mathcal{F}_k=G^{\uparrow}_{k} \mathbbm{\Delta}_{k} \mathcal{G}^{\downarrow \, T}_{-k}$ is defined as in the matrix structure of \cref{eq:I_n}. The saddle-point equations for the anomalous self-energy are then obtained by setting the sum of \cref{eq:derivative_S_HS_updown,eq:derivative_Trlog_Delta_final} to zero:
\begin{equation}
    \begin{split}
        \frac{1}{g}\sum_{k'} \sum_{m}&  \big(\mathds{D}^{-1}_{k-k'}\big)_{n m}  \big(\Delta_{k'}\big)_m= i \big(\mathcal{F}_k\big)_n.
    \end{split}
\end{equation}
Using again the inversion relation for the matrix $\mathds{D}_{k-k'}$ of \cref{eq:inversion_D}
one obtains the equations for the anomalous self-energy
\begin{equation}
    \begin{split}
        \big(\Delta_{k}\big)_n=& i g \sum_{k'} \sum_{m}  (\mathds{D}_{k-k'})_{n m} (\mathcal{F}_{k'})_m,
    \end{split}
     \label{eq:4comp_Delta_eq}
\end{equation}
Analogously to the normal self-energy of \cref{eq:Sigma_matrix_RAK}, the mean-field structure of $\Delta$ inherits the causal structure of the boson propagator and $\mathcal{F}$ [see \cref{eq:F_operatorial}].
As a result the component $\Delta_1$ of the anomalous self-energy is zero, and the remaining matrix elements can be identified as the advanced ($\Delta_2 \to \Delta^{A}$), retarded ($\Delta_3 \to \Delta^{R}$) and Keldysh ($\Delta_4 \to \Delta^{K}$) components of the anomalous self-energy. 
The matrix $\mathbbm{\Delta}_k$ of \cref{eq:Delta_matrix} then takes the form
\begin{equation}
    \mathbbm{\Delta}_k=
    \begin{pmatrix}
        0 &
        \Delta^A_k\\
        \Delta^R_k &
        \Delta^K_k
    \end{pmatrix}.
    \label{eq:Delta_matrix_RAK}
\end{equation}
Using this RAK notation for $\mathbbm{\Delta}_k$ and $\mathcal{F}_k$, \cref{eq:4comp_Delta_eq} becomes
\begin{equation}\label{eq:Delta_withFs} 
    \begin{pmatrix}
         \Delta^{A}_{\bm{k}, \omega} \\     \Delta^{R}_{\bm{k}, \omega} \\ \Delta^{K}_{\bm{k}, \omega}   
    \end{pmatrix}
    = i g \int_{\omega'} \\
    \begin{pmatrix}
        D^{A} \mathcal{F}^{K} + D^{K} \mathcal{F}^{A} \\
        D^{R} \mathcal{F}^{K} + D^{K} \mathcal{F}^{R} \\
        D^{K} \mathcal{F}^{K} + D^{R} \mathcal{F}^{R}+ D^{A} \mathcal{F}^{A} 
    \end{pmatrix},
\end{equation}
or, by dropping the spin indices and writing the structure of $\mathcal{F}_k=-G_k \mathbbm\Delta_k\bar{\mathcal{G}}_{k}=G_k \mathbbm\Delta_k\mathcal{G}^T_{-k}$ explicitly:
\begin{widetext}
\begin{equation}
\begin{split}
&\begin{pmatrix}
     \Delta^{A}_{\bm{k}, \omega} \\ \Delta^{R}_{\bm{k}, \omega} \\ \Delta^{K}_{\bm{k}, \omega}   
\end{pmatrix}
= i g \int_{\omega'} \begin{pmatrix}
    D^{K} G^{A} \mathcal{G}^{R}_{-}+
    D^{A} G^{K} \mathcal{G}^{R}_{-} &
    D^{A} G^{R} \mathcal{G}^{K}_{-} &
    D^{A} G^{R} \mathcal{G}^{R}_{-}\\
    D^{R} G^{K} \mathcal{G}^{R}_{-} & D^{K} G^{R} \mathcal{G}^{A}_{-}+D^{R} G^{R} \mathcal{G}^{K}_{-} & D^{R} G^{R} \mathcal{G}^{R}_{-}  \\
    D^{A} G^{A} \mathcal{G}^{R}_{-}+
    D^{K} G^{K} \mathcal{G}^{R}_{-} &
    D^{R} G^{R} \mathcal{G}^{A}_{-}+
    D^{K} G^{R} \mathcal{G}^{K}_{-} & D^{K} G^{R} \mathcal{G}^{R}_{-}
\end{pmatrix}
\begin{pmatrix}
     \Delta^{A}_{\bm{k}, \omega'} \\ \Delta^{R}_{\bm{k}, \omega'} \\ \Delta^{K}_{\bm{k}, \omega'}   
\end{pmatrix},
\label{eq:RAK_Delta_eqs}
\end{split}
\end{equation}
\end{widetext}
where we used the shorthand notations $D^{\alpha}\mathcal{F}^{\beta}=D^{\alpha}_{\omega-\omega'}\mathcal{F}^{\beta}_{\bm{k}, \omega'}$ and $D^{\alpha} G^{\beta} \mathcal{G}^{\gamma}_{-}=D^{\alpha}_{\omega-\omega'} G^{\beta}_{\bm{k},\omega'} \mathcal{G}^{\gamma}_{-\bm{k},-\omega'}$. 

The NESS-Eliashberg \cref{eq:RAK_Sigma_eqs,eq:RAK_Delta_eqs}, together with the Dyson equation for the normal-state propagator $G$ of \cref{eq:Dyson_normalG} and the definition of the full normal propagator $\mathcal{G}$ in \cref{eq:G_Nambu_elements}, form a close set of equations.
In the normal phase or at the phase transition, one can simply replace $\mathcal{G}\to G$ in \cref{eq:RAK_Sigma_eqs,eq:RAK_Delta_eqs}. 
After this replacement, \cref{eq:RAK_Delta_eqs} leads to the linearized equations for the anomalous self-energy, which admit a non-zero solution only at the phase transition. 
Both \cref{eq:RAK_Sigma_eqs,eq:RAK_Delta_eqs} can equivalently be rewritten in terms of distributions and retarded propagators, as shown in \cref{app:distr_repr_and_th_eq}.

At thermal equilibrium, the Keldysh components $\Sigma^K$  and $\Delta^K$ satisfy thermal fluctuation-dissipation relations and \cref{eq:RAK_Sigma_eqs,eq:RAK_Delta_eqs} are largely simplified, as discussed in \cref{app:th_eq}. In the same appendix, it is also shown that these simplified thermal equilibrium equations are completely equivalent to the real-frequency analytic continuation of Eliashberg equations in Matsubara frequencies.

\section{Connection to thermal equilibrium} 
\label{app:th_eq}
In this appendix, we show how the NESS-Eliashberg \cref{eq:RAK_Sigma_eqs,eq:RAK_Delta_eqs} are connected to thermal equilibrium by expressing them in terms of distributions and relating them to the Eliashberg equations in the Matsubara formalism. 
For simplicity, we will work here with the linearized equations at the phase transition and at $|\bm{k}|=k_F$, dropping the parametric momentum dependence of propagators and self-energies.

\subsection{NESS-Eliashberg equations in terms of distributions and thermal equilibrium condition}
\label{app:distr_repr_and_th_eq}
Our starting point is the linearized version ($\mathcal{G}\to G$) of the self-energy equations \eqref{eq:RAK_Sigma_eqs} at $|\bm{k}|=k_F$.
In the steady state, one can express the Keldysh components of the electron and boson propagator in terms of their distributions
i.e.
\begin{align}
    G^K_\omega&=2iF^e_\omega \, \Im G^R_\omega, 
    \label{eq:GK_def_app}\\
    D^K_\omega&=2iF^b_\omega \, \Im D^R_\omega.
    \label{eq:DK_def_app}
\end{align}
Combining \cref{eq:GK_def_app} with the Keldysh component of the Dyson equation, i.e.~$G^{K}_{\omega}=G^{R}_{\omega} \Sigma^{K}_{\omega} G^{A}_{\omega}$, one also has
\begin{equation}
    \Sigma^K_\omega =2 i F^e_\omega \Im \Sigma^R_\omega,
    \label{eq:SigmaK_def_app}
\end{equation}
so, at the phase transition,  $F^\Sigma_\omega=F^e_\omega$.
 Replacing \cref{eq:GK_def_app,eq:DK_def_app} in the equation for $\Sigma^R_\omega$, one obtains
\begin{equation}
    \begin{split}
        \Sigma^R_\omega=& -2g_c \int_\epsilon \Big[D^R_{\omega-\epsilon} F^e_{\epsilon} \Im G^R_{\epsilon}  \\
        &+ F^b_{\omega-\epsilon}  G^R_{\epsilon}\Im D^R_{\omega-\epsilon}
        \Big].
        \label{eq:SigmaR_dist_eq}
    \end{split}
\end{equation}
Similarly for $\Sigma^K_\omega$ one finds
\begin{equation}
    \begin{split}
        \Sigma^K_\omega = - 4i g_c \int_\epsilon \left(F^b_{\omega-\epsilon} F^e_{\epsilon}+1\right) \Im D^R_{\omega-\epsilon} \Im G^R_{\epsilon}.
    \end{split}
    \label{eq:SigmaK_dist_eq}
\end{equation}
Replacing this expression and the imaginary part of \cref{eq:SigmaR_dist_eq} in \cref{eq:SigmaK_def_app}, one finally obtains the kinetic equation for the distribution $F^e_\omega$
\begin{equation}
    \begin{split}
        &\int_\epsilon \left[ F^b_{\omega-\epsilon} F^e_{\epsilon}+1 -F^e_{\omega} \left(F^e_{\epsilon}+F^b_{\omega-\epsilon}\right)\right] \\
        &\times\Im\left(D^R_{\omega-\epsilon}\right)\, \Im\left(G^R_{\epsilon}\right)=0.
    \end{split}
    \label{eq:Fe_kin_eq}
\end{equation}

Let us consider now the linearized equations for the anomalous self-energy \eqref{eq:RAK_Delta_eqs} at $\vert\bm{k}\vert=k_F$. First of all, one can express the Keldysh component of the anomalous self-energy in terms of its generalized distribution defined through
\begin{equation}
    \Delta^K_\omega= 2 i F^\Delta_\omega \Im\Delta^R_\omega.
    \label{eq:DeltaK_def_app}
\end{equation}
Replacing this expression together with \cref{eq:GK_def_app,eq:DK_def_app} in the equation for $\Delta^R_\omega$ and using the particle-hole symmetry at $\vert\bm{k}\vert=k_F$, i.e.~$G^A_{-\omega}=-G^R_{\omega}$ and $F^e_{-\omega}=-F^e_\omega$, one obtains
\begin{equation}
    \begin{split}
        \Delta^R_{\omega}&= 2 g_c \int_\epsilon \bigg\{ F^b_{\omega-\epsilon} \left(G^R_{\epsilon}\right)^2 \Delta^R_{\epsilon}\Im D^R_{\omega-\epsilon} \\
        &+D^R_{\omega-\epsilon} \left\{F^e_{\epsilon} \Im\left[\left(G^R_{\epsilon}\right)^2 \Delta^R_{\epsilon}\right]\right. \\
        &+ \left.\left(F^\Delta_{\epsilon}-F^e_{\epsilon}\right) \abs{G^R_{\epsilon}}^2 \Im \Delta^R_{\epsilon}\right\} \Big\}. 
        \label{eq:DeltaR_dist_eq}
    \end{split}
\end{equation}
Considering now the equation for $\Delta^K_\omega$, one can rewrite it by adding the zero integral
\begin{equation}
    \begin{split}
        0 =-i g_c \int_\epsilon  &\left(D^R_{\omega-\epsilon} G^A_{\epsilon} G^R_{-\epsilon} \Delta^A_{\epsilon}\right. \\
        &\left.+D^A_{\omega-\epsilon} G^R_{\epsilon} G^A_{-\epsilon} \Delta^R_{\epsilon}\right) 
    \end{split}
\end{equation}
to its r.h.s. Using again \cref{eq:GK_def_app,eq:DK_def_app,eq:DeltaK_def_app} one obtains
\begin{equation}
    \begin{split}
        F^\Delta_\omega \Im \Delta^R_\omega =&2 g_c \int_\epsilon \Im\left(D^R_{\omega-\epsilon}\right)  \\
        &\times\left\{ \left( F^b_{\omega-\epsilon}F^e_{\epsilon}+1\right) \Im\left[\left(G^R_{\epsilon}\right)^2 \Delta^R_{\epsilon}\right]\right. \\
        &+\left. F^b_{\omega-\epsilon} \left( F^\Delta_{\epsilon}-F^e_{\epsilon}\right) \abs{G^R_{\epsilon}}^2 \Im\Delta^R_{\epsilon} \right\}.
    \end{split}
    \label{eq:F^Delta_ImDelta}
\end{equation}
Replacing the imaginary part of \cref{eq:DeltaR_dist_eq} on the l.h.s.~of this equation and bringing everything on the r.h.s., one finally arrives at the equation for $F^\Delta_\omega$
\begin{equation}
    \begin{split}
        0=& \int_\epsilon  \Im(D^R_{\omega-\epsilon}) \Big\{ \big[ F^b_{\omega-\epsilon}F^e_{\epsilon}+1\\
        &-F^\Delta_{\omega} \left(F^e_{\epsilon}+F^b_{\omega-\epsilon}\right) \big] \Im\left[\left(G^R_{\epsilon}\right)^2 \Delta^R_{\epsilon}\right]  \\
        & +\left( F^b_{\omega-\epsilon}-F^\Delta_{\omega} \right) \left( F^\Delta_{\epsilon}-F^e_{\epsilon}\right) \abs{G^R_{\epsilon}}^2 \Im \Delta^R_{\epsilon} \Big\}.
    \end{split} 
    \label{eq:FDelta_kin_eq}
\end{equation}
\cref{eq:SigmaR_dist_eq,eq:Fe_kin_eq,eq:DeltaR_dist_eq,eq:FDelta_kin_eq} form a set of self-consistent equations equivalent to  \cref{eq:RAK_Sigma_eqs,eq:RAK_Delta_eqs}.

Assuming that bosons and electrons are at thermal equilibrium at the temperature $T$, one can show that \cref{eq:Fe_kin_eq,eq:FDelta_kin_eq} are self-consistently solved by the thermal distributions \footnote{These distributions correspond to the thermal Bose-Einstein $n^{b}_\omega=(F^{b}_\omega-1)/2=(e^{\omega/T}-1)^{-1}$ and Fermi-Dirac $n^{e}_\omega=(1-F^{e}_\omega)/2=(e^{\omega/T}+1)^{-1}$ occupations for bosons and electrons, respectively.}
\begin{equation}
    F^{b}_\omega= \coth \left( \frac{\omega}{2T} \right), \quad F^e_\omega=F^\Delta_\omega= \tanh \left( \frac{\omega}{2T} \right) .
    \label{eq:Fb_Fe_FDelta_theq}
\end{equation}
This can be proved by replacing \cref{eq:Fb_Fe_FDelta_theq} in \cref{eq:Fe_kin_eq,eq:FDelta_kin_eq} and showing that the terms in the square brackets vanish because of the relation \cite{KamenevBook}
\begin{equation}
    \begin{split}
    &1 + \coth \left( \frac{\omega-\epsilon}{2T}\right) \tanh \left( \frac{\epsilon}{2T} \right)\\
    &= \tanh \left( \frac{\omega}{2T} \right) \left[ \tanh \left( \frac{\epsilon}{2T} \right)+ \coth \left( \frac{\omega-\epsilon}{2T} \right)\right] ,  
    \end{split}
    \label{eq:magic_formula}
\end{equation}
and that the last term in \cref{eq:FDelta_kin_eq} vanishes because $F^\Delta_\omega=F^e_\omega$. Note that, for the same reason, also the last term on the r.h.s.~of \cref{eq:DeltaR_dist_eq} becomes zero.

At thermal equilibrium, one has to self-consistently solve only the \cref{eq:SigmaR_dist_eq,eq:DeltaR_dist_eq} for $\Sigma^R_\omega$ and $\Delta^R_\omega$, as the distributions are already fixed by the thermal fluctuation-dissipation relations in  \cref{eq:Fb_Fe_FDelta_theq}.

\subsection{Analytic continuation of Matsubara Eliashberg equations at thermal equilibrium}
\label{app:analytical_continuation}
The Eliashberg \cref{eq:SigmaR_dist_eq,eq:DeltaR_dist_eq} for $\Sigma^R_\omega$ and $\Delta^R_\omega$ at thermal equilibrium could be equivalently obtained as the analytic continuation to real frequencies of Eliashberg equations in Matsubara formalism, which is the one typically employed in thermal equilibrium calculations.
Within this formalism, the frequency dependence of propagators and self-energies is expressed in terms of discrete fermionic Matsubara frequencies $\omega_{n}=(2n+1)\pi T$ ($n$ integer) on the imaginary frequency axis, rather than in terms of real frequencies $\omega$ \cite{Fetter&Walecka}. The Eliashberg equations in Matsubara frequencies for our model can be derived following Ref.~\cite{Protter2021} and in the linearized case at $|\bm{k}|=k_F$ read \footnote{The factor of $2$ in \cref{eq:Sigma_Matsubara,eq:Delta_Matsubara} comes from a different definition of the free boson Hamiltonian in \cref{eq:H_0b} with respect to Ref.~\cite{Protter2021}.}
\begin{align}
    \Sigma_{i \omega_n} &= - 2 g_c  T \sum_{\omega_{m}} D_{i\omega_n-i\omega_{m}} G_{i\omega_{m}},
    \label{eq:Sigma_Matsubara}
    \\
    \Delta_{i \omega_n}&= - 2 g_c  T  \sum_{\omega_{m}} D_{i\omega_n-i\omega_{m}}  G_{i \omega_{m}} G_{-i \omega_{m}}\Delta_{i \omega_{m}},
    \label{eq:Delta_Matsubara}
\end{align}
where the dressed fermion propagator is defined through the Dyson equation
\begin{equation}
    G_{\bm{k},i \omega_n}=\frac{1}{i \omega_{n} -\xi_{\bm{k}}-\Sigma_{\bm{k},i \omega_n}},
    \label{eq:G_Matsubara}
\end{equation}
and the boson propagator is
\begin{equation}
   D_{i\Omega_{n}}= \frac{1}{2} \frac{\omega_0}{\left[i\Omega_{n}+i \kappa \,\mathrm{sgn}\,\Omega_{n}\right]^2-\omega_0^2},
   \label{eq:D_Matsubara}
\end{equation}
with $\Omega_n=2\pi n T$ ($n$ integer) is a bosonic Matsubara frequency. Note that the $\mathrm{sgn}\,\Omega_{n}$ factor in \cref{eq:D_Matsubara} is necessary to ensure the proper causality structure $\Pi^{\mathrm{env}}_{-i\Omega_n}=\left[\Pi^{\mathrm{env}}_{i\Omega_n}\right]^*$ of the boson environment self-energy in Matsubara formalism. 

\cref{eq:Sigma_Matsubara,eq:Delta_Matsubara} can be analytically continued exactly from the positive Matsubara axis ($\omega_n>0$) to frequencies $\omega+i0^+$ just above the real axis to obtain the Eliashberg equations for the retarded components $\Delta^{R}_\omega$ and $\Sigma^{R}_\omega$ \cite{Marsiglio1988,Combescot1995}.
To do so, we express the propagators \eqref{eq:D_Matsubara} and \eqref{eq:G_Matsubara}  through their spectral representations~\cite{Fetter&Walecka}
\begin{equation}
    \begin{split}
        D_{i\Omega_{n}}&=
        -2 \int_\Omega
         \frac{\Im D^R_{\Omega}}{i\Omega_{n}-\Omega} \\
        &=- 2 \int_\Omega  \Im D^R_{\Omega} \, \frac{\Omega}{(i\Omega_{n})^2-\Omega^2},
    \label{eq:D_spectral_rep}
    \end{split}
\end{equation}
where the symmetry $\Im D^R_{-\Omega}=-\Im D^R_{\Omega}$ has been used in the second line,
and 
\begin{equation}
    G_{i\omega_{n}}= -2\int_\epsilon  \frac{\Im G^R_{\epsilon}}{i\omega_{n}-\epsilon}.
    \label{eq:G_spectral_rep}
\end{equation}
Moreover, one can also express the function $G_{i \omega_{m}} G_{-i \omega_{m}}\Delta_{i \omega_{m}}$ in \cref{eq:Delta_Matsubara} through its spectral representation
\begin{equation}
    G_{i \omega_{m}} G_{-i \omega_{m}}\Delta_{i \omega_{m}}=-2 \int_\varepsilon \frac{\Im\left(G^R_{\epsilon} G^A_{-\epsilon}\Delta^R_{\epsilon}\right)}{i\omega_n -\epsilon}.
\end{equation}
We can now replace these spectral representations in \cref{eq:Sigma_Matsubara,eq:Delta_Matsubara}, obtaining
\begin{equation}
    \begin{split}
        \Sigma_{i \omega_n} = &-
        8 g_c \int_\Omega \int_\epsilon
        \, \Im D^R_{\Omega} \, \Im G^R_{\epsilon} \\
        &\times T \sum_{\omega_{m}} \frac{\Omega}{\left(i\omega_n-i\omega_{m}\right)^2-\Omega^2} \frac{1}{i \omega_m -\epsilon},
    \end{split}
\end{equation}
and
\begin{equation}
    \begin{split}
        \Delta_{i \omega_n} &=-8 g_c \int_\Omega \int_\epsilon
        \, \Im D^R_{\Omega} \, \Im\left(G^R_{\epsilon} G^A_{-\epsilon}\Delta^R_{\epsilon}\right) \\
        &\times T \sum_{\omega_{m}} \frac{\Omega}{\left(i\omega_n-i\omega_{m}\right)^2-\Omega^2} \frac{1}{i \omega_m -\epsilon}.
    \end{split}
\end{equation}
The Matsubara frequency sum in both equations can be computed by transforming it in a contour integration \cite{Fetter&Walecka}
\begin{equation}
    \begin{split}   
        T& \sum_{\omega_{m}} \frac{\Omega}{\left(i\omega_n-i\omega_{m}\right)^2-\Omega^2} \frac{1}{i \omega_m -\epsilon} \\
        =&-\frac{1}{2} \tanh \left(\frac{\epsilon}{2T} \right) \frac{\Omega}{\left(\epsilon-i\omega_n\right)^2-\Omega^2} -\frac{1}{4} \coth \left( \frac{\Omega}{2T} \right)\\
        &\times \left( \frac{1}{i \omega_n-\Omega-\epsilon}+\frac{1}{i \omega_n+\Omega-\epsilon} \right),
    \end{split}    
\end{equation}
where we have used the property $\tanh \left(i \omega_n-z\right)=-\coth(z)$ to obtain the $\coth$ in the last term. Note that this last replacement is essential to obtain the correct causal structure of the functions when the analytic continuation is performed, i.e.~to obtain functions that are bound and analytic in the upper complex plane \cite{Marsiglio1988}. We can now perform the analytic continuation $i\omega_n \to \omega +i 0^+$ of the external Matsubara frequency in the previous expressions, obtaining
\begin{equation}
    \begin{split}
         &\Sigma^R_{\omega} = 
         8 g_c \int_\Omega \int_\epsilon
        \, \Im D^R_{\Omega} \, \Im G^R_{\epsilon} \\
        &\times\bigg[ \frac{1}{2} \tanh \left(\frac{\epsilon}{2T} \right) \frac{\Omega}{\left(\omega -\epsilon+i0^{+}\right)^2-\Omega^2}+\frac{1}{4} \coth \left( \frac{\Omega}{2T} \right)\\
        &\times\left( \frac{1}{\omega -\Omega-\epsilon+i0^{+}}+\frac{1}{\omega +\Omega-\epsilon+i0^{+}} \right) \bigg] ,
    \end{split}
\end{equation}
and
\begin{equation}
    \begin{split}
        &\Delta^R_{\omega}  =
        8 g_c \int_\Omega \int_\epsilon
        \, \Im D^R_{\Omega} \, \Im\left(G^R_{\epsilon} G^A_{-\epsilon}\Delta^R_{\epsilon}\right) \\
        &\bigg[ \frac{1}{2} \tanh \left(\frac{\epsilon}{2T} \right) \frac{\Omega}{(\omega -\epsilon+i0^{+})^2-\Omega^2} +\frac{1}{4} \coth \left( \frac{\Omega}{2T} \right)\\
        &\times \left( \frac{1}{\omega-\Omega-\epsilon+i0^{+}}+\frac{1}{\omega+\Omega-\epsilon+i0^{+}} \right) \bigg] .
    \end{split}
\end{equation}
Performing the change of variable $\Omega\to -\Omega$ in the second term in round brackets, one can replace
\begin{equation}
    \begin{split}
        &\left( \frac{1}{\omega -\Omega-\epsilon+i0^{+}}+\frac{1}{\omega +\Omega-\epsilon+i0^{+}} \right) \\
        & \to \frac{2}{\omega -\Omega-\epsilon+i0^{+}},
    \end{split}
\end{equation}
as $\Im\left(D^R_\Omega\right) \coth \left( \frac{\Omega}{2T} \right)$ is symmetric in $\Omega$.
Moreover, one can identify in the previous expressions the spectral representations for the retarded functions \cite{Fetter&Walecka}
\begin{align}
    &D^R_{\omega-\epsilon} = -2 \int_\Omega
    \Im D^R_{\Omega} \, \frac{\Omega}{\left(\omega-\epsilon+i0^{+}\right)^2-\Omega^2},\\
    &G^R_{\omega+\Omega}= -2\int_\epsilon  \frac{\Im G^R_{\epsilon}}{\omega+\Omega-\epsilon+i0^{+}},
\end{align}
and
\begin{equation}
    \begin{split}
        & G^R_{\omega+\Omega} G^A_{-\omega-\Omega}\Delta^R_{\omega+\Omega} =-2 \int_\epsilon \frac{\Im\left(G^R_{\epsilon} G^A_{-\epsilon}\Delta^R_{\epsilon}\right)}{\omega +\Omega-\epsilon+i0^{+}},
    \end{split}
\end{equation}
to obtain
\begin{equation}
    \begin{split}
        \Sigma^R_{\omega}  = &- 2g_c  \int_\epsilon  \tanh \left(\frac{\epsilon}{2T}\right)
        \, D^R_{\omega-\epsilon} \, \Im G^R_{\epsilon} \\
        & - 2 g_c  \int_\Omega  \coth \left( \frac{\Omega}{2T} \right)G^R_{\omega+\Omega} \, \Im D^R_{\Omega},
    \end{split}
    \label{eq:SigmaR_theq}
\end{equation}
and
\begin{equation}
    \begin{split}
        &\Delta^R_{\omega}  = - 2 g_c  \int_\epsilon  \tanh \left(\frac{\epsilon}{2T}\right)
        \, D^R_{\omega-\epsilon} \Im\left(G^R_{ \epsilon} G^A_{-\epsilon}\Delta^R_{ \epsilon}\right)   \\
        & - 2 g_c  \int_\Omega  \coth \left( \frac{\Omega}{2T} \right) G^R_{\omega+\Omega} G^A_{-\omega-\Omega}\Delta^R_{\omega+\Omega}\Im D^R_{\Omega}.
    \end{split}
    \label{eq:DeltaR_theq}
\end{equation}
Finally, by performing the change of variables $\epsilon=\Omega+\omega$ in the second integrals and using the relation $G^A_{-\omega}=-G^R_{\omega}$, one finds that \cref{eq:DeltaR_theq,eq:SigmaR_theq} are equivalent to \cref{eq:SigmaR_dist_eq,eq:DeltaR_dist_eq} at thermal equilibrium, i.e.~obtained by setting  $F^e_\omega=F^\Delta_\omega=\tanh(\frac{\omega}{2T})$ and $F^b_\omega = \coth (\frac{\omega}{2T})$.

\section{Causal structure of mean-field theory}
\label{app:causal_structure_MF}
In this appendix we will consider the causal structure of our theory at the mean-field level.
The mean-field level means that we neglect the quantum fluctuations of the normal and anomalous self-energies and they are therefore described fully by \cref{eq:selfEnergy_EQs}. 
The fermionic partition function then collapses to 
\begin{equation}
    \mathcal{Z}= \int  \mathcal{D}[\bar{\psi},\psi] e^{i\sum_k \bar{\Psi}_{k}^T  \mathbb{G}^{-1}_{k} \Psi_{k}},
\end{equation}
with the Keldysh-Nambu spinors defined in \cref{eq:Keldysh-Nambu} and the inverse Keldysh-Nambu propagator defined in \cref{eq:G-1_Nambu}. 
We can now derive the casual structure of the sixteen elements in $\mathbb{G}$ by relating them to two-point correlation functions of the electrons.
Due to the quadratic nature of the mean-field action, this can be straightforwardly obtained by adding linear source fields and taking functional derivatives thereof \cite{KamenevBook}. 

\subsection{Anomalous propagators and self-energies}
\label{app:anomalous_prop}
The eight different anomalous terms in $\mathbb{G}$ are connected to the fermionic expectation values
\begin{align*}
\mathcal{F}^{(1)}_k&=-i\langle \psi_{\uparrow,k}^c\psi_{\downarrow,-k}^c\rangle,   &     \bar{\mathcal{F}}^{(1)}_k&=-i\langle \bar{\psi}_{\downarrow,-k}^c \bar{\psi}_{\uparrow,k}^c\rangle,\\  
\mathcal{F}^{(2)}_k&=-i\langle \psi_{\uparrow,k}^c\psi_{\downarrow,-k}^q\rangle,   &     \bar{\mathcal{F}}^{(2)}_k&=-i\langle \bar{\psi}_{\downarrow,-k}^c \bar{\psi}_{\uparrow,k}^q\rangle,\\
\mathcal{F}^{(3)}_k&=-i\langle \psi_{\uparrow,k}^q\psi_{\downarrow,-k}^c\rangle,   &     \bar{\mathcal{F}}^{(3)}_k&=-i\langle \bar{\psi}_{\downarrow,-k}^q \bar{\psi}_{\uparrow,k}^c\rangle,\\
\mathcal{F}^{(4)}_k&=-i\langle \psi_{\uparrow,k}^q\psi_{\downarrow,-k}^q\rangle,   &     \bar{\mathcal{F}}^{(4)}_k&=-i\langle \bar{\psi}_{\downarrow,-k}^q \bar{\psi}_{\uparrow,k}^q\rangle,
\end{align*}
where the numbering follows the structure in \cref{eq:Delta_matrix}.
To find the explicit causal structure of the these objects we map the fields from the Keldysh contour back into time-ordered operators. 
To this extent, the first step is to Fourier transform to real-space.
To ensure consistency with the definition used for the $\bar{\Delta}$ fields, we define the transformation of $\bar{\mathcal{F}}$ with the same sign as $\bar{\Delta}$. 
Due to the temporal and spatial translational invariance of the system, the real-space form depends only on the difference $x-x'$ and assumes the form
\begin{equation}
    \mathcal{F}^{(n)}_{x-x'}=-i\langle\psi_{\uparrow,x}^{\alpha_n}\psi_{\downarrow,x'}^{\beta_n}\rangle,\quad   \bar{\mathcal{F}}^{(n)}_{x-x'}=-i\langle\bar{\psi}_{\downarrow,x'}^{\alpha_n}\bar{\psi}_{\uparrow,x}^{\beta_n}\rangle,
\end{equation}
where $\alpha_n,\beta_n\in\{c,q\}$. 
The next steps for mapping back to operators is to undo the Keldysh rotation \eqref{eq:Keldysh_rot_fermions} and using the appropriate contour ordering, followed by writing out the time-ordering \cite{Sieberer2016}. 
To keep the presentation self-contained, this procedure is shown for one of the expectation values
\begin{equation}\label{eq:fieldtoOp}
    \begin{aligned}
\left<\psi^c_{\uparrow,x}\psi^q_{\downarrow,x'}\right>&=\frac{1}{2}\left<\left(\psi_{\uparrow,x}^++\psi_{\uparrow,x}^-\right)\left(\psi_{\downarrow,x'}^+-\psi_{\downarrow,x'}^-\right)\right>,\\
        &=\frac{1}{2}\bigg(\left<\psi_{\uparrow,x}^+\psi_{\downarrow,x'}^+\right>-\left<\psi_{\uparrow,x}^-\psi_{\downarrow,x'}^-\right>\\&\hspace{10mm}+\left<\psi_{\uparrow,x}^-\psi_{\downarrow,x'}^+\right>-\left<\psi_{\uparrow,x}^+\psi_{\downarrow,x'}^-\right>\bigg),\\
        &=\frac{1}{2}\Bigg(\mathcal{T}\left<\hat{c}_{\uparrow,x}\hat{c}_{\downarrow,x'}\right>-\mathcal{T}^-\left<\hat{c}_{\uparrow,x}\hat{c}_{\downarrow,x'}\right>\\&\hspace{10mm}+\left<\hat{c}_{\uparrow,x}\hat{c}_{\downarrow,x'}\right>+\left<\hat{c}_{\downarrow,x'}\hat{c}_{\uparrow,x}\right>\Bigg),\\
        &=\frac{1}{2}\bigg[\theta_{t-t'}\bigg(\left<\hat{c}_{\uparrow,x}\hat{c}_{\downarrow,x'}\right>+\left<\hat{c}_{\downarrow,x'}\hat{c}_{\uparrow,x}\right>\\&\hspace{24mm}+\left<\hat{c}_{\uparrow,x}\hat{c}_{\downarrow,x'}\right>+\left<\hat{c}_{\downarrow,x'}\hat{c}_{\uparrow,x}\right>\bigg)\\&\hspace{8mm}-\theta_{t'-t}\bigg(\left<\hat{c}_{\downarrow,x'}\hat{c}_{\uparrow,x}\right>+\left<\hat{c}_{\uparrow,x}\hat{c}_{\downarrow,x'}\right>\\&\hspace{24mm}-\left<\hat{c}_{\uparrow,x}\hat{c}_{\downarrow,x'}\right>-\left<\hat{c}_{\downarrow,x'}\hat{c}_{\uparrow,x}\right>\bigg)\bigg],
        \\
       &=\theta_{t-t'}\left<\left\{\hat{c}_{\uparrow,x},\hat{c}_{\downarrow,x'}\right\}\right>=i\mathcal{F}^R_{x-x'},
    \end{aligned}
\end{equation}

where $\left\{\cdot,\cdot\right\}$ is the anticommutator and the hat has been used to highlight the operator nature.
In the first line the Keldysh rotation has been undone. To get from the second to the third line we have used the contour ordering to write the fields as operators. 
This  is done by knowing that all fields on the $+$ ($-$) branch must be (anti-)time-ordered \cite{KamenevBook}, otherwise the fields on the $+$ branch must be evaluated before the ones on the $-$ branch.
Going to the fourth line, the time-ordering have been written explicitly, which highlights the cancellations leading to the final result. 
As a last step the result have been identified with the retarded anomalous propagators defined in \cref{eq:anom_opDef}. 
Following the same procedure for the other seven expectation values, the causal structure of all the different anomalous propagators is given by
\begin{equation}
    \begin{aligned}\label{eq:F_operatorial}
    \mathcal{F}^{(1)}_{x-x'}&=-i\left<\left[\hat{c}_{\uparrow,x},\hat{c}_{\downarrow,x'}\right]\right>=\mathcal{F}^{K}_{x-x'},\\
    \mathcal{F}^{(2)}_{x-x'}&=-i\theta_{t-t'}\left<\left\{\hat{c}_{\uparrow,x},\hat{c}_{\downarrow,x'}\right\}\right>=\mathcal{F}^R_{x-x'},\\
    \mathcal{F}^{(3)}_{x-x'}&=i\theta_{t'-t}\left<\left\{\hat{c}_{\uparrow,x},\hat{c}_{\downarrow,x'}\right\}\right>=\mathcal{F}^A_{x-x'}, \\
    \mathcal{F}^{(4)}_{x-x'}&=0, \\ 
    \bar{\mathcal{F}}^{(1)}_{x-x'}&=-i\left<\left[\hat{c}^\dagger_{\downarrow,x'},\hat{c}^\dagger_{\uparrow,x},\right]\right>=-\left(\mathcal{F}^{K}_{x-x'}\right)^*,\\
     \bar{\mathcal{F}}^{(2)}_{x-x'}&=-i\theta_{t'-t}\left<\left\{\hat{c}^\dagger_{\downarrow,x'},\hat{c}^\dagger_{\uparrow,x}\right\}\right>=\left(\mathcal{F}^A_{x-x'}\right)^*,\\
    \bar{\mathcal{F}}^{(3)}_{x-x'}&=i\theta_{t-t'}\left<\left\{\hat{c}^\dagger_{\downarrow,x'},\hat{c}^\dagger_{\uparrow,x}\right\}\right>=\left(\mathcal{F}^R_{x-x'}\right)^*, \\
    \bar{\mathcal{F}}^{(4)}_{x-x'}&=0.
\end{aligned}
\end{equation}
Transforming back to energy-momentum space, the relation between the two off-diagonal $2\times2$ blocks in $\mathbb{G}$ is thus found to be 
\begin{equation}\label{eq:F_Fbar_relations}
    \bar{\mathcal{F}}_k=-\sigma_z \mathcal{F}_k^\dag \sigma_z.
\end{equation}
With this relation we can connect $\Delta$ to $\bar{\Delta}$.
To do this we follow the same  procedure used for deriving the equations for $\Delta$ to derive the equations for $\bar{\Delta}$. 
The result of the almost identical derivations is
\begin{equation}\label{eq:barDelta}
    \begin{pmatrix}
         \bar{\Delta}^{A}_{\bm{k}, \omega} \\     \bar{\Delta}^{R}_{\bm{k}, \omega} \\ \bar{\Delta}^{K}_{\bm{k}, \omega}   
    \end{pmatrix}
    = i g \int_{\omega'} \\
    \begin{pmatrix}
         \mathcal{F}^{A*}D^{K}-\mathcal{F}^{K*}D^{A} \\
 \mathcal{F}^{R*}D^{K}          -\mathcal{F}^{K*}D^{R}\\
      \mathcal{F}^{R*}D^{R}+  \mathcal{F}^{A*}D^{A}-\mathcal{F}^{K*}D^{K} 
    \end{pmatrix},
\end{equation}
where we have used the notation $\mathcal{F^{\alpha*}}D^\beta=\left(\mathcal{F^{\alpha}}_{\bm{k},\omega'}\right)^*D^\beta_{\omega'-\omega}$ and used the relation between $\mathcal{F}$ and $\bar{\mathcal{F}}$ in \cref{eq:F_Fbar_relations}. 
As the $D$'s are propagators of a real field, they obey the symmetries
\begin{equation}
    D^R_{-\omega}=D^A_\omega\text{ and } D^K_{-\omega}=D^K_\omega.
\end{equation}
Using these relations and comparing to \cref{eq:Delta_withFs} one finds the connection
between $\Delta$ and $\bar{\Delta}$
\begin{equation}
\bar{\Delta}^R_{k}=\left(\Delta^R_k\right)^*,\quad \bar{\Delta}^A_{k}=\left(\Delta^A_k\right)^*,\quad \bar{\Delta}^K_{k}=-\left(\Delta^K_k\right)^*. 
\label{eq:barDelta_relations}
\end{equation}
As observed for the anomalous propagators in \cref{eq:F_Fbar_relations} we see that $\bar{\Delta}$ is connected to $\Delta$ through complex conjugation, but with an additional change of sign for the Keldysh component.  

From \cref{eq:Delta_withFs} and \cref{eq:barDelta}, it follows directly that if
\begin{equation}\label{eq:casualF}
\mathcal{F}^R_k=\left(\mathcal{F}^A_k\right)^* \text{ and  } \mathcal{F}^K_k=-\left(\mathcal{F}^K_k\right)^*,
\end{equation}
then it is also true that 
\begin{equation}\label{eq:causal_Delta}
    \Delta^R_k=\left(\Delta^A_k\right)^* \text{ and  } \Delta^K_k=-\left(\Delta^K_k\right)^*,
\end{equation}
and vice versa.
However, due to anomalous nature of $\mathcal{F}$, the causal connections in \cref{eq:casualF} are not generically obeyed. To satisfy the first equality in \cref{eq:casualF}, one has always the gauge freedom to choose a global phase for $\mathcal{F}^R_k$ (or equivalently for $\Delta^R_k$) such that the retarded and advanced propagators are connected through complex conjugation.
In thermal equilibrium, $\mathcal{F}^K$ is then connected to $\mathcal{F}^R_k$ through thermal FDRs [similarly to \cref{eq:DeltaK_def_app} for $\Delta^K$], so its anti-Hermicity shown in the second equality of \cref{eq:casualF} is also straightforwardly verified.
When the system is not in thermal equilibrium, instead, $\mathcal{F}^K$ is not generically anti-Hermitian. We can show this by expressing it via \cref{eq:anomalousProp} in the form
\begin{equation}\label{eq:FK_alt}
    \begin{split}
    \mathcal{F}^K_k&=G^{ R}_k\Delta^K_k\mathcal{G}^{R}_{-k}+G^{ K}_k\Delta^A_k\mathcal{G}^{ R}_{-k}+G^{ R}_k\Delta^R_k\mathcal{G}^{K}_{-k}\\
    &=\mathcal{G}^{R}_k\Delta^K_kG^{R}_{-k}+\mathcal{G}^{K}_k\Delta^A_kG^{R}_{-k}+\mathcal{G}^{R}_k\Delta^R_kG^{K}_{-k}.
    \end{split}
\end{equation}
Assuming the causal connections of $\Delta^{R,A,K}$ in \cref{eq:causal_Delta} to be true, and assuming that $\mathcal{G}^{R,A,K}$ (and equivalently $G^{R,A,K}$) follow the standard causal connections $\mathcal{G}^R_k=(\mathcal{G}^A_k)^*$ and $\mathcal{G}^K_k=-(\mathcal{G}^K_k)^*$ \cite{KamenevBook}, the conjugate of \cref{eq:FK_alt} gives
\begin{equation}
\left(\mathcal{F}^K_k\right)^*=-\mathcal{G}^{A}_k\Delta^K_kG^{A}_{-k}-\mathcal{G}^{K}_k\Delta^R_kG^{A}_{-k}-\mathcal{G}^{A}_k\Delta^A_kG^{K}_{-k}.
\end{equation}
We see then that $\left(\mathcal{F}^K_k\right)^*=-\mathcal{F}^K_k$ does not hold in general. It holds, however, if we fix the momentum to $\abs{\bm{k}}=k_F$. The proof is based on the fact that, due to the momentum structure of our interaction \cref{eq:H_bf}, the propagators $\mathcal{G}^{R,A,K}$ (and equivalently $G^{R,A,K}$) satisfy the particle-hole symmetry
\begin{equation}
    \mathcal{G}^{R}_{-\bm{k}_F,-\omega}=-\mathcal{G}^{A}_{\bm{k}_F,\omega} \ \text{ and } \    \mathcal{G}^{K}_{-\bm{k}_F,\omega}=-\mathcal{G}^{K}_{\bm{k}_F,\omega}.
    \label{eq:ph_symmetry}
\end{equation}
Using this symmetry one finds 
\begin{equation}
\begin{aligned}
    \left(\mathcal{F}^K_{k_F,\omega}\right)^*=&-\mathcal{G}^{R}_{-\bm{k}_F,\omega}\Delta^K_{\bm{k}_F,\omega}G^{R}_{\bm{k}_F,\omega}-\mathcal{G}^{K}_{-\bm{k}_F,-\omega}\Delta^R_{\bm{k}_F,\omega}G^{R}_{\bm{k}_F,\omega}\\&-\mathcal{G}^{R}_{-\bm{k}_F,-\omega}\Delta^A_{\bm{k}_F,\omega}G^{K}_{\bm{k}_F,\omega},
\end{aligned}
\end{equation}
which is equivalent to $-\mathcal{F}^K_{\bm{k}_F,\omega}$ in \cref{eq:FK_alt}.
So we find that, even when the system is in a NESS, as long as it obeys particle-hole symmetry as in \cref{eq:ph_symmetry}, one can always choose a gauge where $\mathcal{F}^K$ is anti-Hermitian, i.e.~purely imaginary in momentum-frequency space.
In this case $\Delta^{R,A,K}$ inherit the causal structure of $\mathcal{F}^{R,A,K}$.
\subsection{Normal propagators}
Let us consider now the eight different normal terms in $\mathbb{G}$. For this purpose, we reintroduce the spin index $\sigma=\uparrow,\downarrow$ to properly identify them. Similarly to the anomalous terms, we find that they are connected to the fermionic expectation values
\begin{align*}
\mathcal{G}^{\uparrow (1)}_k&=-i\langle \psi_{\uparrow,k}^c\bar{\psi}_{\uparrow,k}^c\rangle,   &     \bar{\mathcal{G}}^{\downarrow(1)}_k&=-i\langle \bar{\psi}_{\downarrow,-k}^c \psi_{\downarrow,-k}^c\rangle,\\  
\mathcal{G}^{\uparrow (2)}_k&=-i\langle \psi_{\uparrow,k}^c\bar{\psi}_{\uparrow,k}^q\rangle,   &     \bar{\mathcal{G}}^{\downarrow (2)}_k&=-i\langle \bar{\psi}_{\downarrow,-k}^c \psi_{\downarrow,-k}^q\rangle,\\ \mathcal{G}^{\uparrow (3)}_k&=-i\langle \psi_{\uparrow,k}^q\bar{\psi}_{\uparrow,k}^c\rangle,   &     \bar{\mathcal{G}}^{\downarrow (3)}_k&=-i\langle \bar{\psi}_{\downarrow,-k}^q \psi_{\downarrow,-k}^c\rangle,\\ \mathcal{G}^{\uparrow (4)}_k&=-i\langle \psi_{\uparrow,k}^q\bar{\psi}_{\uparrow,k}^q\rangle,   &     \bar{\mathcal{G}}^{\downarrow (4)}_k&=-i\langle \bar{\psi}_{\downarrow,-k}^q \psi_{\downarrow,-k}^q\rangle.\\ 
\end{align*}
Using the anticommutation relations of fermionic fields, it is immediately evident that the relation between the $2\times2$ diagonal blocks in $\mathbb{G}$ is
\begin{equation}
\label{eq:G_Gbar_relation}
    \bar{\mathcal{G}}^\sigma_k=-\mathcal{G}^{\sigma \, T}_{-k}.
\end{equation}
We can then just work on the $\mathcal{G}^\sigma_k$ block, as the properties of the $\bar{\mathcal{G}}^\sigma_k$ block immediately follow.
To find the causal structure of $\mathcal{G}^\sigma_k$, we start by Fourier transforming it to real space, obtaining
\begin{equation}
    \mathcal{G}^{\sigma (n)}_{x-x'}=-i\langle \psi_{\sigma,x}^{\alpha_n}\bar{\psi}_{\sigma,x'}^{\beta_n}\rangle.
\end{equation}
We can now map the fields from the Keldysh contour back into time-ordered operators with a procedure analogous to \cref{eq:fieldtoOp} and obtain
\begin{equation}\
    \begin{aligned}
        \label{eq:G_operatorial}
        \mathcal{G}^{\sigma (1)}_{x-x'}&=-i\langle [ \hat{c}_{\sigma,x},\hat{c}^\dag_{\sigma,x'} ] \rangle=\mathcal{G}^{\sigma K}_{x-x'},\\
        \mathcal{G}^{\sigma (2)}_{x-x'}&=-i\theta_{t-t'}\langle \{ \hat{c}_{\sigma,x},\hat{c}^\dag_{\sigma,x'} \} \rangle=\mathcal{G}^{\sigma R}_{x-x'},\\
        \mathcal{G}^{\sigma (3)}_k&=i\theta_{t'-t}\langle \{ \hat{c}_{\sigma,x},\hat{c}^\dag_{\sigma,x'} \} \rangle=\mathcal{G}^{\sigma A}_{x-x'},\\
        \mathcal{G}^{\sigma (4)}_{x-x'}&=0.\\
    \end{aligned}
\end{equation}
This result can then be used to derive the causal structure of the normal self-energy, as discussed above \cref{eq:Sigma_matrix_RAK}.

\section{RAK components of electron propagators}
\label{app:GK_FK}
In this appendix we present some useful expressions for the retarded and Keldysh components of the normal ($\mathcal{G}$)  and anomalous ($\mathcal{F}$) electron propagators written in terms of the normal-state propagator ($G$) and the anomalous self-energy ($\mathbbm{\Delta}$).

Within the superconducting phase, the electron propagators are given by \cref{eq:G_Nambu_elements}:
\begin{align}
    \mathcal{G}_k&=\left(G^{-1}_{k}+\mathbbm{\Delta}_{k} G^{ T}_{-k} \bar{\mathbbm{\Delta}}^{T}_{k}\right)^{-1}, \\
    \mathcal{F}_k&=G_k \mathbbm{\Delta}_k \mathcal{G}_{-k}^T=\mathcal{G}_k \mathbbm{\Delta}_k G_{-k}^T,
\end{align}
where $G_k=\left[(\omega -\xi_{\bm{k}})\sigma_x - \mathbbm{\Sigma}_k\right]^{-1}$ [cf.~\cref{eq:normalElectronProp}].
By expliciting the RAK matrix structures as in \cref{eq:Full_G,eq:Causal_self-energy,eq:barDelta_def}, one can explicitly compute the RAK components of the normal and anomalous propagator, obtaining
\begin{align}
    \mathcal{G}^R_k&=\frac{G^R_k}{1+G^R_k G^A_{-k}\bar{\Delta}^A_k\Delta^R_k},
    \label{eq:mathcalGR_app}\\
    \mathcal{F}^R_k&=\frac{G^R_k G^A_{-k} \Delta^R_{k}}{1+G^R_k G^A_{-k}\bar{\Delta}^A_k\Delta^R_k}, \label{eq:mathcalFR_app}
\end{align}
for the retarded components and
\begin{widetext}
\begin{align}
    \mathcal{G}^K_k&=\frac{G^K_k-G^A_k G^R_k(G^R_{-k}\bar{\Delta}^R_k \Delta^K_k+G^A_{-k}\bar{\Delta}^K_k \Delta^R_k+G^K_{-k}\bar{\Delta}^R_k \Delta^R_k)}{(1+G^A_k G^R_{-k}\bar{\Delta}^R_k \Delta^A_k)(1+G^R_k G^A_{-k}\bar{\Delta}^A_k \Delta^R_k)},  \label{eq:mathcalG_Keldysh} \\
    \mathcal{F}^K_k&=\frac{G^R_{-k}G^K_{k}\Delta^A_k +G^K_{-k}G^R_{k}\Delta^R_k+G^R_{-k}G^R_{k}\Delta^K_k-G^R_{-k}G^A_{k}G^R_k G^A_{-k} \Delta^A_k  \bar{\Delta}^K_k \Delta^R_k   }{(1+G^A_k G^R_{-k}\bar{\Delta}^R_k \Delta^A_k)(1+G^R_k G^A_{-k}\bar{\Delta}^A_k \Delta^R_k)},
    \label{eq:mathcalF_Keldysh}
\end{align}
\end{widetext}
for the Keldysh components. We remind that $\bar{\Delta}_\omega$ and $\Delta_\omega$ are related  by complex conjugation relations as in \cref{eq:barDelta_relations}. 

To derive the expression for $\mathcal{F}^K_\omega$ used in of \cref{eq:FK_tau0_1}, one can set $|\bm{k}|=k_F$ in \cref{eq:mathcalF_Keldysh} (dropping the momentum dependence) and express $\Delta^K_\omega$ and $\Sigma^K_\omega$ in terms of distributions 
\begin{align}
    \Sigma^K_\omega&=2iF^{\Sigma}_\omega \Im\Sigma^R_\omega, \label{eq:F^Sigma_param} \\
    \Delta^K_\omega&=2iF^\Delta_\omega \Im\Delta^R_\omega. \label{eq:F^Delta_param}
\end{align}
It is easy to show through \cref{eq:normalElectronProp} that $F^\Sigma_\omega$ is also the distribution function for the normal-state propagator $G$, that is
\begin{equation}
    \label{eq:G^K_F^Sigma_param}
    G^K_\omega=2i F^{\Sigma}_\omega \Im  G^R_\omega .   
\end{equation}
On the other hand, in the superconducting phase the electron distribution $F^e_\omega$ for the full normal propagator $\mathcal{G}$, defined through the parametrization $\mathcal{G}^K_\omega=2i F^e_\omega \Im \mathcal{G}^R_\omega$,
is in general different from $F^\Sigma_\omega$.
Being at $|\bm{k}|=k_F$, one can further use the particle-hole symmetry $G^R_{-\omega}=-G^A_\omega$ [cf.~\cref{eq:ph_symmetry}] and the relation $(\Delta^R_\omega)^*=\Delta^A_\omega$ [cf.~\eqref{eq:causal_Delta}] to obtain
\begin{equation}
    \begin{split}
        &\mathcal{F}^K_\omega =2i F^{\Sigma}_\omega \frac{-\Im\left[\left(G^R_\omega\right)^2 \Delta^R_\omega\right]-\abs{G^R_\omega}^4 \abs{\Delta^R_\omega}^2 \Im \Delta^R_\omega }{\abs{1-\left(G^R_\omega\right)^2 (\Delta^R_\omega)^2}^2}\\
        &-2i\left(F^\Delta_\omega-F^{\Sigma}_\omega\right) \frac{\abs{G^R_\omega}^2 \Im\left(\Delta^R_\omega\right)+\abs{G^R_\omega}^4 \abs{\Delta^R_\omega}^2 \Im\left(\Delta^R_\omega\right) }{\abs{1-\left(G^R_\omega\right)^2 \left(\Delta^R_\omega\right)^2}^2},
    \end{split}
    \end{equation}
where we have separated a ``thermal-like'' contribution proportional to $F^{\Sigma}_\omega$ and a ``non-thermal'' contribution proportional $F^{\Delta}_\omega-F^{\Sigma}_\omega$. From this expression it is directly evident that $\mathcal{F}^K_\omega$ is purely imaginary.
Further rewriting of this expression in terms of \cref{eq:mathcalGR_app,eq:mathcalFR_app} at $\abs{\bm{k}}=k_F$ finally results in the expression
\begin{equation}
    \label{eq:F^K}
    \begin{split}
        \mathcal{F}^K_\omega &=2i F^{\Sigma}_\omega \Im\left(\mathcal{F}^R_\omega\right) 
        -2i\left(F^\Delta_\omega-F^{\Sigma}_\omega\right)\Im\Delta^R_\omega \\
        &\times\abs{\mathcal{G}^R_\omega}^2\left(1-\abs{G^R_\omega}^2 \abs{\Delta^R_\omega}^2\right),
    \end{split}
\end{equation}
which corresponds to the integrand of \cref{eq:FK_tau0_1}.

At the superconducting phase transition, one has that $\mathcal{G}_\omega\to G_\omega$ and $F^\Sigma_\omega \to F^e_\omega$,  and this allows to simplify \cref{eq:mathcalFR_app,eq:mathcalF_Keldysh} for the anomalous propagator as
\begin{equation}
    \mathcal{F}^R_k=G^R_k G^A_{-k} \Delta^R_{k}, \label{eq:mathcalFR_trans_app}
\end{equation}
for the retarded component and
\begin{equation}
    \mathcal{F}^K_k=G^R_{-k}G^K_{k}\Delta^A_k +G^K_{-k}G^R_{k}\Delta^R_k+G^R_{-k}G^R_{k}\Delta^K_k,
    \label{eq:mathcalF_Keldysh_trans}
\end{equation}
for the Keldysh component. Setting again $\abs{\bm{k}}=k_F$ one can use the parametrization \cref{eq:G^K_F^Sigma_param} with $F^\Sigma_\omega \to F^e_\omega$ and the particle-hole symmetry to rewrite \cref{eq:mathcalF_Keldysh_trans} as
\begin{equation}
    \begin{split}
    \mathcal{F}^K_\omega=&
    -2iF^e_{\omega}\left\{\Im\left[\left(G^R_\omega\right)^2\Delta^R_\omega\right] +|G^R_\omega|^2 \Im \Delta^R_\omega \right\} \\
    &+\abs{G^R_\omega}^2\Delta^K_\omega,
    \end{split}
\end{equation}
which corresponds to the integrand of \cref{eq:FK_tau0_3}.

\section{Electron back-action onto the bosons}
\label{app:boson_self-energy}
In this appendix, we give a justification of why there is always a weak-coupling regime where the boson self-energy $\Pi(\omega)$ due to electron-boson interaction can be neglected. For simplicity, we suppose here to be at the critical coupling $g_c$ of the phase transition, so that we can consider only normal-state propagators $G$ for the electrons.

Starting from the partition function for the boson-electron system of \cref{eq:Z}, one can keep only the one-loop contributions to the boson self-energy to obtain \cite{KamenevBook}
\begin{equation}
    \begin{split}
    \mathcal{Z}= \int \mathcal{D}[X] \, e^{i (S_{b,0 }[X]+S_{b,\mathrm{int}}[X])},
    \end{split}
    \label{eq:Z_int_electrons}
\end{equation}
where $S_{b,0 }[X]$ is given by \cref{eq:free_boson_action_t} and
\begin{equation}
    \begin{split}
    &S_{b,\mathrm{int}}[X] \\
    &= -\frac{1}{2} \int_{t,t'}  \begin{pmatrix}
    X^{c} \\
    X^{q}
    \end{pmatrix}^{T}_{t}
    \begin{pmatrix}
    0 & \Pi^A \\
    \Pi^R & \Pi^K
    \end{pmatrix}_{t-t'} \!
    \begin{pmatrix}
    X^{c} \\
    X^{q}
    \end{pmatrix}_{t'}.
    \end{split}
\label{eq:S_b_int}
\end{equation}
In the previous expression, the boson self-energy RAK components are given by
\begin{align}
    &\Pi^{R(A)}_{t-t'}= 4i g_c \sum_{\bm{r}} \left(G^{R(A)}_{+}G^{K}_{-}+G^{K}_{+}G^{A(R)}_{-}\right), \\
    &\Pi^K_{t-t'}= 4i g_c \sum_{\bm{r}} \left(G^{K}_{+}G^{K}_{-}+G^{A}_{+}G^{R}_{-}+G^{R}_{+}G^{A}_{-}\right), 
\end{align}
with the shorthand notation $G^\alpha_+ G^\beta_-=G^\alpha_{\bm{r},t-t'} G^\beta_{\bm{r},t'-t}$.
To have an estimate of the magnitude of this self-energy, we concentrate on the retarded component. In momentum-frequency space, it reads
\begin{equation}
    \Pi^R_{\omega}= 4ig \int_{k'}\left(G^{R}_{\bm{k}',\omega+\omega'}G^{K}_{\bm{k}',\omega'}+G^{K}_{\bm{k}',\omega+\omega'}G^{A}_{\bm{k}',\omega'}\right),
\end{equation}
where $G_{\bm{k},\omega}$ are the electronic propagators, and we used the shorthand continuum notation $\int_{k'}=V_d \int d\bm{k}'/(2\pi)^d  \int d\omega'/2\pi $, where $V_d$ is the volume of the system in $d$ dimensions.
Note that $\Pi^R_\omega$ does not have momentum dependence because in our model the bosons carry zero momentum. 
To further simplify the calculation, we perform a quasi-particle approximation on the electronic propagators as $G^R_\omega \approx 1/(\omega+ i \eta)$. 
Within this approximation, one has
\begin{equation}
    \Pi^R_\omega= 16g_c \eta  \int_{k'} \frac{\omega' F^e_{\omega'+\xi_{\bm{k}'}} 
    }{\left[\omega'^2+\left(\omega+i\eta\right)^2\right]\left(\omega'^2+\eta^2\right)}.
\end{equation}
Evaluating this self-energy on-shell, i.e.~at the frequency of the bosonic mode $\omega=\omega_0$, performing the change of variables $\tilde{\varepsilon}=\bm{k}^2/2m\omega_0$ and $\tilde{\omega}=\omega'/\omega_0$ and assuming $\eta \ll \omega_0$, one obtains
\begin{equation}
    \Pi^R_{\omega_0}= g_c \eta \frac{C_d}{\omega_0^2} \left( \frac{\omega_0}{\omega_{V}} \right)^{d/2} ,
\end{equation}
where we defined the energy scale $\omega_V= 1/\left(2m V_d^{2/d}\right)$ and the dimensionless constant
\begin{equation}
    \begin{split}
        C_d &=   \frac{8\Omega_d}{(2\pi)^{d+1}} \int_0^{\infty} \! \! d \tilde{\varepsilon} \, \tilde{\varepsilon}^{\frac{d}{2}-1} \\ &\times \mathrm{P.V.} \int_{-\infty}^{\infty} \! \! d \tilde{\omega}  \frac{
         F^e_{(\tilde{\omega}+\tilde{\varepsilon})\omega_0-\mu}
        }{\tilde{\omega} \left(\tilde{\omega}^2+1\right)},
    \end{split}
\end{equation}
with $\Omega_d$ the solid angle in $d$ dimensions and P.V.~denotes that the integral is taken in the Cauchy principal value sense.
We find that at lowest order for small $g_c$ and $\eta$, $\Pi^R_{\omega_0}$ scales as $g_c \eta/\omega_0^2$. This has to be compared to the real part of the on-shell inverse retarded propagator $\Re \left\{[D^{-1}]^R_{\omega_0}\right\}=2\kappa^2/\omega_0$. This means that, for finite $\kappa$ and $\omega_0$, there is always a small-$g_c$ or small-$\eta$ regime where the photon self-energy can be neglected. While the value of the constants $\omega_V$ and $C_d$ 
depends on the specific model parameters, we at least made sure that the condition $g_c \eta \ll \kappa^2 \omega_0$ is always satisfied in our numerical results.

\section{Boson propagator and environment-induced self-energy}
\label{app:QuadInt}
In this appendix, we first show how, starting from the action for a complex boson field, one can integrate away the $\mathcal{P}$-quadrature to obtain the boson propagator for the $\mathcal{X}$-quadrature. We then show how to derive the environment-induced self-energy for the system bosons in the case of a linear coupling between the $\mathcal{X}$-quadrature of the environment and the $X$-quadrature of the system boson. 
\subsection{$\mathcal{P}$-quadrature integration and boson propagator for the $\mathcal{X}$-quadrature}
Let us start by showing how the $\mathcal{P}$-quadrature of a boson field can be integrated away. This procedure can be applied to both the multi-mode environment bosons and the single-mode system bosons.
The required structure for this procedure is that the bosonic field only interacts with other fields through its $\mathcal{X}$-quadrature. 
The generic starting point is therefore a Hamiltonian of the form
\begin{equation}\label{eq:ham0_quad}
    \hat{H}=\hat{H}_{\mathrm{sys}}+\sum_s \left(\nu_s \hat{b}^\dagger_s \hat{b}_s+\frac{t_s}{\sqrt{2}} \hat{\mathcal{X}}_s \hat{\mathcal{O}}\right),
\end{equation}
where $\hat{b}^\dagger_s$ is the bosonic creation operator for the bare mode $s$ with energy $\nu_s$ and
\begin{equation}
    \hat{\mathcal{X}}_s=\frac{1}{\sqrt{2Y_b}}\left(\hat{b}_s+\hat{b}_s^\dagger\right)
    \label{eq:Xs_def}
\end{equation}
is the real quadrature operator, with $Y_b$ being a constant that is chosen based on convenience. 
The operators $\hat{H}_{\mathrm{sys}}$ and $\hat{\mathcal{O}}$ do not include any additional $\hat{b}_s$, $\hat{b}^\dag_s$ operators.
Lastly, $t_s$ is the coupling strength between the \mbox{$\mathcal{X}_s$-quadrature} and the $\hat{\mathcal{O}}$ operator.
The $\mathcal{P}_s$-quadrature of the boson is defined as
\begin{equation}
    \hat{\mathcal{P}}_s = \frac{1}{i\sqrt{2Y_b}}\left(\hat{b}_s-\hat{b}_s^\dagger\right).
    \label{eq:Ps_def}
\end{equation}
This quadrature only appears in the bare quadratic part of the Hamiltonian and can therefore be integrated away. 
The bare part of the environment action is given by \cite{KamenevBook}
\begin{equation}\label{eq:bAction}
\begin{aligned}
    S_{0}\left[b_s^*,b_s\right]=\int_\omega\sum_s &\begin{pmatrix}
    b_s^{c} \\
    b_s^{q}
    \end{pmatrix}^{\dag}_{\omega}\mathds{M}_{s,\omega}
    \begin{pmatrix}
   b_s^{c} \\
    b_s^{q}
    \end{pmatrix}_{\omega},
    \end{aligned}
\end{equation}
where $b^{c/q}_s$ are the classical and quantum components of the complex bosonic fields resulting from translating the term that is quadratic in $\hat{b}$-operators in the Hamiltonian \cref{eq:ham0_quad} to the $\pm$ time contour, and then performing the Keldysh rotation $b_s^{c/q}=(b^+_s \pm b^-_s)/\sqrt{2}$, analogously to \cref{eq:Keldysh_rot_fermions,eq:Keldysh_rot_bosons}.
The matrix $\mathds{M}_{s,\omega}$ is the inverse complex boson propagator, and is given by
\begin{equation}
    \mathds{M}_{s,\omega}=\begin{pmatrix}
    0 & \omega-\nu_s-i0^+ \\
    \omega-\nu_s+i0^+ & i2 F_\omega 0^+
    \end{pmatrix},
\end{equation}
with $0^+$ being a positive infinitesimal and $F_\omega$ being the distribution function of the bosons (assumed here to be $s$-independent, like in the thermal case). 
The coupling to the $\hat{\mathcal{O}}$ operator in \cref{eq:ham0_quad} gives rise to a term in the action of the form 
\begin{equation}
    S_{\text{c}}\left[\mathcal{X},\mathcal{O}\right]=-\int_\omega\sum_s t_s\begin{pmatrix}
    \bar{\mathcal{X}}_s^{c} \\
    \bar{\mathcal{X}}_s^{q}
    \end{pmatrix}^{T}_{\!-\omega}
    \begin{pmatrix}
   \mathcal{O}_{1} \\
    \mathcal{O}_{2}
    \end{pmatrix}_{\omega},
\end{equation}
where $\mathcal{O}_{1,2}$ are two objects that depend on the explicit form of the operator $\hat{\mathcal{O}}$. 
Inverting \cref{eq:Xs_def,eq:Ps_def}, the $b^{c/q}_\omega$ and $\bar{b}^{c/q}_\omega$ fields can be written in terms of the $\mathcal{X}^{c/q}$ and $\mathcal{P}^{c/q}$ quadratures as 
\begin{equation}\label{eq:b_quad_relation}
    \begin{aligned}
        b^{c/q}_\omega&=\sqrt{Y_b}\left(\mathcal{X}^{c/q}_{s,\omega}+i\mathcal{P}^{c/q}_{s,\omega}\right),\\
        \left(b^{c/q}_\omega\right)^*&=\sqrt{Y_b}\left(\mathcal{X}^{c/q}_{s,-\omega}-i\mathcal{P}^{c/q}_{s,-\omega}\right).
    \end{aligned}
\end{equation}
As this is nothing but a linear transformation, it is straightforward to rewrite $S_{0}$ in terms of the quadratures. 
Since both quadratures are real fields, one can rewrite the action into the following symmetric form
\begin{equation}
    \begin{aligned}
        &S_{0}\left[\mathcal{X},\mathcal{P}\right]=\int_\omega \sum_{s;\alpha,\beta=c,q}\bigg(\frac{1}{2}\mathcal{X}^\alpha_{s,-\omega}\left(\mathds{M}^{\text{S}}_{s,\omega}\right)_{\alpha,\beta}\mathcal{X}^\beta_{s,\omega}\\
        &\quad \quad+\frac{1}{2}\mathcal{P}^\alpha_{s,-\omega}\left(\mathds{M}^{\text{S}}_{s,\omega}\right)_{\alpha,\beta}\mathcal{P}^\beta_{s,\omega}-i\mathcal{P}^\alpha_{s,-\omega}\left(\mathds{M}^{\text{A}}_{s,\omega}\right)_{\alpha,\beta}\mathcal{X}^\beta_{s,\omega}\bigg),
    \end{aligned}
\end{equation}
where the symmetric and anti-symmetric inverse propagators are defined as
\begin{equation}
    \begin{aligned}
        \mathds{M}^\text{S}_{s,\omega}&=Y_b\left(\mathds{M}_{s,\omega}+\mathds{M}_{s,-\omega}^\text{T}\right),\\
        \mathds{M}_{s,\omega}^\text{A}&=Y_b\left(\mathds{M}_{s,\omega}-\mathds{M}_{s,-\omega}^\text{T}\right).
    \end{aligned}
\end{equation}
The functional integrals over the $\mathcal{P}$-fields can now be performed
\begin{equation}
\begin{aligned}
    \mathcal{Z}_b&=\int \mathcal{D}\left[\mathcal{P},\mathcal{X}\right]\exp\left[i\left(S_{0}[\mathcal{X},\mathcal{P}]+S_{\text{c}}[\mathcal{X},\mathcal{O}_A]\right)\right]\\
    &=\int \mathcal{D}\left[\mathcal{X}\right]\exp\left[i\left(S_{\mathcal{X},0}[\mathcal{X}]+S_{\text{c}}[\mathcal{X},\mathcal{O}_A]\right)\right],
    \end{aligned}
\end{equation}
where the $\mathcal{X}$-quadrature action is found to be
\begin{equation}\label{eq:XquadAction}
    S_{\mathcal{X},0}\left[\mathcal{X}\right]=\int_\omega\sum_{s;\alpha,\beta=c,q}\frac{1}{2}\mathcal{X}^\alpha_{s,-\omega}\left(\mathds{Q}_{s,\omega}\right)_{\alpha,\beta}\mathcal{X}^\beta_{s,\omega},
\end{equation}
and the bare inverse $\mathcal{X}$ propagator is given by
\begin{equation}
\begin{aligned}
\mathds{Q}_{s,\omega}&=\mathds{M}^\text{S}_{s,\omega}+\left[\mathds{M}^\text{A}_{s,-\omega}\right]^\text{T} \left[\mathds{M}^\text{S}_{s,\omega}\right]^{-1}\mathds{M}^\text{A}_{s,-\omega}\\
    =&\begin{pmatrix}
        0 & Q^A_{s,\omega}\\ Q^R_{s,\omega}& Q^K_{s,\omega}
    \end{pmatrix},
    \end{aligned}
\end{equation}
where the elements are given by
\begin{equation}\label{eq:inverseBareBosonProp}
    \begin{aligned}
        Q^R_{s,\omega}&=\left(Q^A_{s,\omega}\right)^*=2Y_b\frac{\left(\omega+i0^+\right)^2-\nu_s^2}{\nu_s},\\
        Q^K_{s,\omega}&=\frac{2Y_bi0^+F_{e,\omega}}{\nu_s^2}\left[\omega^2+\nu_s^2+(0^+)^2\right]+\frac{4Y_b i 0^+\omega F_{o,\omega}}{\nu_s}.
    \end{aligned}
\end{equation}
Here the even and odd parts of the boson distribution function have been separated out such that 
\begin{equation}
    \begin{aligned}
        F_{e,\omega}&=F_\omega+F_{-\omega},\\
        F_{o,\omega}&=F_\omega-F_{-\omega}.
    \end{aligned}
\end{equation}
 
\subsection{Environment-induced self-energy}
\label{app:bath_renormalization}
We consider now a linear coupling between the \mbox{$\mathcal{X}$-quadrature} of the environment bosons and the \mbox{$X$-quadrature} of the system bosons with bare frequency $\tilde{\omega}_0$, resulting in $\hat{\mathcal{O}}=\left(\hat{a}^\dagger+\hat{a}\right)/\sqrt{2 Y_a}=\hat{X}$.
The linear nature of the coupling means that the Gaussian functional integrals over the environment $\mathcal{X}_s$-fields can be directly performed
\begin{equation}
    \int\mathcal{D}\left[\mathcal{X}\right] \exp\left[i\left(S_{\mathcal{X},0}[\mathcal{X}]+S_\text{c}[\mathcal{X},X]\right)\right]=\exp\left(iS_{X,\mathrm{env}}[X]\right),
\end{equation}
with 
\begin{equation}
    S_{X,\mathrm{env}}\left[X\right]=-\frac{1}{2}\int_\omega\sum_{\alpha,\beta=c,q} X^\alpha_{-\omega} \left(\mathbbm{\Pi}^{\mathrm{env}}_\omega\right)_{\alpha,\beta} X^\beta_\omega,
    \label{eq:S_X_self-energy}
\end{equation}
being the resulting action describing the effect of the environment on the system bosons.
The environment-induced self-energy for the system bosons is then given by 
\begin{equation}
\begin{aligned}
    \mathbbm{\Pi}^\mathrm{env}_\omega&=\begin{pmatrix}
        0&\Pi^{\mathrm{env},A}_\omega\\
        \Pi^{\mathrm{env},R}_\omega & \Pi^{\mathrm{env},K}_\omega
    \end{pmatrix},\\&=\sum_s t_s^2 \sigma_x\left[\mathds{Q}_{s,\omega}\right]^{-1}\sigma_x,\\
    &=\sum_s t_s^2\begin{pmatrix}
        0&C^A_{s,\omega}\\
        C^R_{s,\omega} & C^K_{s,\omega}
    \end{pmatrix}.
    \end{aligned}
    \label{eq:Sigma_b_RAK}
\end{equation}
Inverting $\mathds{Q}_{s,\omega}$ leads to the following RAK components of the environment boson propagators
\begin{equation}
    \label{eq:C_RAK}
    \begin{aligned}
        C^R_{s,\omega}&=\left(C^A_{s,\omega}\right)^*=\frac{1}{2Y_b}\frac{\nu_s}{\left(\omega+i0^+\right)^2-\nu_s^2},\\
        C^K_{s,\omega}&= \frac{- i2\nu_s\omega 0^+ F^{\mathrm{env}}_{\mathcal{X},s,\omega}}{Y_b\abs{\left(\omega+i0^+\right)^2-\nu_s^2}^2},
    \end{aligned}
\end{equation}
where the environment $\mathcal{X}$-quadrature distribution function has been defined as 
\begin{equation}
    F^\mathrm{env}_{\mathcal{X},s,\omega}=\frac{F^\mathrm{env}_{o,\omega}}{2}+F^\mathrm{env}_{e,\omega}\frac{\omega^2+\nu_s^2+\left(0^+\right)^2}{4\omega \nu_s}.
\end{equation}
As this distribution function is odd in $\omega$, the Keldysh component $C^K_{s,\omega}$ is even by construction, which is a consequence of the formulation in terms of real fields \cite{KamenevBook}. 

To make further progress, the environment energies $\nu_s$ and couplings $t_s$ must be specified. 
This is conveniently done by introducing the spectral density
\begin{equation}
    J_\omega=\sum_s \frac{t_s^2A_{s,\omega}}{2\pi},
    \label{eq:J_omega_def}
\end{equation}
where $A_{s,\omega}=-2\Im C^R_{s,\omega}$ is the environment spectral function.
As all the $\omega$-dependence of $J_\omega$ originates from $A_{s,\omega}$ and given the expression for $C^R_{s,\omega}$ in \cref{eq:C_RAK}, $J_\omega$ is antisymmetric in $\omega$.
Using the spectral density as a weight, the discrete sum over the environment energies in $\Pi^{\mathrm{env},R}_\omega$ can be written as an integral  
\begin{equation}
    \Pi^{\mathrm{env},R}_\omega=\int_\epsilon  \frac{2\pi \epsilon J_\epsilon}{(\omega+i0^+)^2-\epsilon^2}.
\end{equation}
If $J_\omega$ is analytic on the real axis and
$\lim_{\abs{\omega}\rightarrow \infty}\abs{J_\omega}=\mathrm{const.}$, then the integral can be evaluated directly by infinitesimally shifting the integration contour into the complex plane. The result is 
\begin{equation}\label{eq:bosonEnvSigR_PV}
\begin{aligned}
    \Pi^{\mathrm{env},R}_\omega = -&\frac{i \pi}{2}\left(J_\omega-J_{-\omega}\right)\\-& \mathrm{P.V.}\int_{\epsilon} \frac{2\pi\epsilon J_\epsilon}{(\epsilon-\omega)(\epsilon+\omega)}.
\end{aligned}
\end{equation}
For $\Pi^{\mathrm{env},K}_\omega$ one notices that, as $J_\omega$ is an antisymmetric function, then $C^K_{s,\omega}=-iF^\mathrm{env}_{o,\omega} A_{s,\omega}/2$.
Using the definition \eqref{eq:J_omega_def} of $J_\omega$, the Keldysh component of the self-energy takes the form 
\begin{equation}\label{eq:bosonEnvSigK}
    \Pi^{\mathrm{env},K}_\omega=-i\pi J_\omega F^\mathrm{env}_{o,\omega}.
\end{equation}
The odd distribution function can be rewritten in terms of the environment occupation $n^{\mathrm{env}}_\omega=(F^\mathrm{env}_\omega-1)/2$ as
\begin{equation}
    F^\mathrm{env}_{o,\omega}=2\left(n^{\mathrm{env}}_\omega-n^{\mathrm{env}}_{-\omega}\right).
\end{equation}
For the spectral density we consider the simplest choice that satisfies the criteria described above
\begin{equation}
J_\omega=\mathcal{N}\frac{\kappa \, \omega}{\pi} \exp\left[-\left(\frac{\omega}{\omega_c}\right)^2\right],
\end{equation}
where $\mathcal{N}$ is a constant with the units of the inverse of an energy that we will choose later and $\omega_c$ is a cutoff frequency that ensures the convergence of \cref{eq:bosonEnvSigR_PV}. 
We consider the regime where the cutoff $(\omega_c)$ is by far the largest energy scale of the system, such that the system is unaffected by the cutoff.
In this regime the real part of the self-energy becomes a constant (on the energy scale of the system) and can therefore be absorbed into the bare frequency. 
Under this approximation the effective retarded self-energy simplifies to
\begin{equation}
    \Pi^{\mathrm{env},R}_\omega=-i \mathcal{N} \omega \kappa, 
\end{equation}
and the Keldysh self-energy takes the form
\begin{equation}\label{eq:keldyshBosonSE}
    \Pi^{\mathrm{env},K}_\omega=-i \mathcal{N}\omega \kappa  F_{o,\omega}.
\end{equation}
Due to the quadratic nature of the action~\eqref{eq:S_X_self-energy} in the $X$-fields, one can immediately combine it with the bare action for the single-mode system bosons, which has a form analogous to \cref{eq:XquadAction} with the replacements $\mathcal{X}_s\to X$ and $\nu_s\to \tilde{\omega}_0$, and without the sum over $s$. This results in the total quadratic action for the $X$-fields
\begin{equation}
    \begin{split}
    &S_{b,0}\left[X\right] \\
    &= \frac{1}{2} \int_{\omega} \begin{pmatrix}
    X^{c} \\
    X^{q}
    \end{pmatrix}^{T}_{-\omega}
    \begin{pmatrix}
    0 & [D^{-1}]^{A} \\
    [D^{-1}]^{R} & [D^{-1}]^{K}
    \end{pmatrix}_{\omega}
    \begin{pmatrix}
    X^{c} \\
    X^{q}
    \end{pmatrix}_{\omega}
    \end{split}
    \label{eq:Sb0_envselfenergy_app}
\end{equation}
with the inverse retarded propagator given by
\begin{equation}\begin{aligned}
    \left[D^{-1}\right]^{R}_\omega&=\left(D^R_\omega\right)^{-1}=2Y_a\frac{\omega^2-\tilde{\omega}_0^2}{\tilde{\omega}_0}-\Pi^{\mathrm{env},R}_\omega,\\&=\frac{2Y_a}{\tilde{\omega}_0}\left(\omega^2-\tilde{\omega}_0^2+i\frac{\mathcal{N} \tilde{\omega}_0}{2Y_a}\omega \kappa\right),
\end{aligned}
\label{eq:DinvR_app}
\end{equation}
the advanced inverse propagator given by the complex conjugate of \cref{eq:DinvR_app},  and the Keldysh inverse propagator given by $[D^{-1}]^{K}=-\Pi^{\mathrm{env},K}_\omega$, where the Keldysh self-energy is defined in \cref{eq:Sigma_b_RAK}. 
To have a form of the boson propagator that resembles that of thermal equilibrium, we scale the system quadrature $Y_a=\tilde{\omega}_0/\omega_0$ and chose the constant in the spectral density to be $\mathcal{N}=4/\omega_0$, where $\omega_0=\sqrt{\tilde{\omega}_0^2-\kappa^2}$ is the renormalized energy scale of the boson including the bath coupling.
With these choices, the retarded boson propagator obtained by inverting the matrix structure in \cref{eq:Sb0_envselfenergy_app} takes the form
\begin{equation}\label{eq:ret_boson}
    D^R_\omega=\frac{1}{2}\frac{\omega_0}{\left(\omega+i\kappa\right)^2-\omega_0^2},
\end{equation}
while the Keldysh propagator is given by
\begin{equation}
    D^K_\omega=\abs{D^R_\omega}^2 \Pi^{\mathrm{env},K}_\omega=F^b_\omega \left(D^R_\omega-D^A_\omega\right).
\end{equation}
Inserting \cref{eq:ret_boson,eq:keldyshBosonSE} one identifies the boson distribution function as
\begin{equation}\label{eq:BosonEnvDist}
    F^b_\omega = \frac{F^\mathrm{env}_{o,\omega}}{2}=n^{\mathrm{env}}_\omega-n^{\mathrm{env}}_{-\omega}.
\end{equation}
For a thermally occupied environment one has $n^{\mathrm{env}}_\omega=(e^{\omega/T}-1)^{-1}$, which leads to the expected result of $F^b_\omega=\coth\left(\omega/2T\right)$. More generally, we find that, through a linear quadrature coupling, the distribution of the system bosons $F^b_\omega$ can be directly tuned by engineering the occupation function of the environment $n^{\mathrm{env}}_\omega$.

\section{Cryostat-induced self-energy}\label{app:cryostat_model}
In this appendix, we show how to derive the cryostat-induced self-energy for the electrons shown in \cref{eq:Sigma^cryo}.
We model the cryostat as a thermal bath of electrons at temperature $T_0$ that is linearly coupled to the system electrons.
This is achieved by considering a Hamiltonian of the form 
\begin{equation}
\begin{split}
\hat{H}_{\text{cryo}}&=\sum_{s,\sigma,\bm{k}}\epsilon_{s,\sigma,\bm{k}}\hat{f}^\dagger_{s,\sigma,\bm{k}}\hat{f}^{}_{s,\sigma,\bm{k}}\\
&+\sum_{s,\sigma,\bm{k}}\left(t_{s,\sigma,\bm{k}}\,\hat{c}^\dagger_{\sigma,\bm{k}}\hat{f}^{}_{s,\sigma,\bm{k}}+\mathrm{h.c.}\right),
\end{split}
\label{eq:H_cryo}
\end{equation}
where $\hat{f}^\dagger_{s,\sigma,\bm{k}}$ creates a cryostat electron in the mode $s$ with spin $\sigma$ and momentum $\bm{k}$, while $\hat{c}^{\dagger}_{\sigma,\bm{k}}$ creates a system electron with spin $\sigma$ and momentum $\bm{k}$ as in \cref{H_0e}, $\epsilon_{s,\sigma,\bm{k}}$ is the dispersion of the bath electrons and $t_{s,\sigma,\bm{k}}$ is the coupling strength between cryostat electrons and the system electrons. We note that, by coupling the electrons at same momentum and spin, this interaction does not generate scattering in momentum but only an energetic redistribution of the electrons. 

From the Hamiltonian \eqref{eq:H_cryo} we can write down the part of the partition function that involves the cryostat
\begin{equation}
    \mathcal{Z}_{\text{cryo}}= \int \mathcal{D}\left[\bar{f}, f\right] \, e^{iS_{\text{cryo}}\left[\psi,\bar{\psi},f,\bar{f}\right]},
\end{equation}
where $f$ $\left(\psi\right)$ and $\bar{f}$ $\left(\bar{\psi}\right)$  are the Grassmann fields related to the cryostat (system) electron annihilation and creation operators. 
The action has the form
\begin{equation}
    \begin{aligned}
    & S_{\text{cryo}}\left[\psi,\bar{\psi},f,\bar{f}\right]=\sum_{\bm{k},s,\sigma}\int_\omega \\
    &\quad\Bigg[\begin{pmatrix}
    \bar{f}_{s,\sigma,\bm{k}}^{c} \\
    \bar{f}_{s,\sigma,\bm{k}}^{q}
\end{pmatrix}^{T}_{\omega}[(G^{f})^{-1}]_{s,\sigma,\bm{k},\omega}
    \cdot\begin{pmatrix}
   f^c_{s,\sigma,\bm{k}} \\
    f^q_{s,\sigma,\bm{k}}
    \end{pmatrix}_{\omega}\\
    &\quad+t_{s,\sigma,\bm{k}}\begin{pmatrix}
    \bar{f}_{s,\sigma,\bm{k}}^{c} \\
    \bar{f}_{s,\sigma,\bm{k}}^{q}
\end{pmatrix}^{T}_{\omega}\sigma_x
    \begin{pmatrix}
   \psi^c_{\sigma,\bm{k}} \\
    \psi^q_{\sigma,\bm{k}}
    \end{pmatrix}_{\omega}+\mathrm{h.c.}\Bigg].
    \end{aligned}
    \label{eq:S_cryo}
\end{equation} 
The cryostat electron propagator assumes the matrix form
\begin{equation}
    G^f_{s,\sigma,\bm{k},\omega}=\begin{pmatrix}
        G^{f,K} & G^{f,R} \\
        G^{f,A} & 0 \\
    \end{pmatrix}_{s,\sigma,\bm{k},\omega}.
\end{equation}
As the feedback of the system on the cryostat can be neglected, the retarded propagator is simply given by the non-interacting form
\begin{equation}
    G^{f,R}_{s,\sigma,\bm{k},\omega}=\frac{1}{\omega-\epsilon_{s,\sigma,\bm{k}}+i 0^+},
\end{equation}
the advanced propagator is given by its complex conjugate, and the Keldysh propagator takes the form 
\begin{equation}
\begin{aligned}
G^{f,K}_{s,\sigma,\bm{k},\omega}&=F^{\text{cryo}}_\omega\left(G^{f,R}_{s,\sigma,\bm{k},\omega}-G^{f,A}_{s,\sigma,\bm{k},\omega}\right),\\&=2i F^{\text{cryo}}_\omega \text{Im}G^\text{cryo,R}_{s,\sigma,\bm{k},\omega},
\end{aligned}
\end{equation}
where $F^{\text{cryo}}_\omega=\tanh \left(\omega/2 T_0\right)$.
As the action \eqref{eq:S_cryo} is quadratic, the Gaussian functional integral over the cryostat fields can be calculated exactly. The resulting contribution to the action of the system electrons is quadratic in the $\psi \,\left(\bar{\psi}\right)$ fields and has the matrix form
\begin{equation}
    S_{e-\mathrm{cryo}}\left[\bar{\psi},\psi\right]=-\sum_{\sigma,\bm{k}}\int_\omega \begin{pmatrix}
   \bar{\psi}^c_{\sigma,\bm{k}} \\
    \bar{\psi}^q_{\sigma,\bm{k}}
    \end{pmatrix}^T_{\omega}\mathbbm{\Sigma}^{\text{cryo}}_{\sigma,\bm{k},\omega}\begin{pmatrix}
   \psi^c_{\sigma,\bm{k}} \\
    \psi^q_{\sigma,\bm{k}}
    \end{pmatrix}_\omega,
\end{equation}
where the cryostat-induced self-energy $\mathbbm{\Sigma}^{\text{cryo}}$ has a similar causal structure as \cref{eq:Causal_self-energy}. 
The retarded component of the self-energy has the form
\begin{equation}
    \Sigma^{\text{cryo},R}_{\sigma,\bm{k},\omega}=\sum_s\abs{t_{s,\sigma,\bm{k}}}^2G^{f,R}_{s,\sigma,\bm{k},\omega}=\sum_s\frac{\abs{t_{s,\sigma,\bm{k}}}^2}{\omega-\epsilon_{s,\sigma,\bm{k}}+i0^+},
\end{equation}
while the Keldysh self-energy is given by
\begin{equation}
\begin{aligned}
\Sigma^{\text{cryo},K}_{\sigma,\bm{k},\omega}&=\sum_s\abs{t_{s,\sigma,\bm{k}}}^2G^{f,K}_{s,\sigma,\bm{k},\omega},\\&=-2\pi i \sum_s \abs{t_{s,\sigma,\bm{k}}}^2 F^\text{cryo}_\omega \delta(\omega-\epsilon_{s,\sigma,\bm{k}}),
\end{aligned}
\end{equation}
where the infinitesimal linewidth ($0^+$) of the cryostat spectrum has been used to write the cryostat spectral function as a $\delta$-function.
Following the same logic as in \cref{app:QuadInt}, one can introduce the cryostat spectral density as 
\begin{equation}
J^\text{cryo}_{\sigma,\bm{k},\omega}=\sum_s\abs{t_{s,\sigma,\bm{k}}}^2 \delta(\omega-\epsilon_{s,\sigma,\bm{k}}).
\end{equation}
Using this spectral density, the cryostat-induced self-energy takes a form similar to the one induced by the boson environment in \cref{eq:bosonEnvSigR_PV,eq:bosonEnvSigK}:
\begin{equation}
\begin{aligned}
\Sigma^{\text{cryo},R}_{\sigma,\bm{k},\omega}&=-i\pi J^\text{cryo}_{\sigma,\bm{k},\omega}-\mathrm{P.V.}\int_\epsilon \frac{J^\text{cryo}_{\sigma,\bm{k},\epsilon}}{\epsilon-\omega},\\
\Sigma^{\text{cryo},K}_{\sigma,\bm{k},\omega}&=-2\pi i J^\text{cryo}_{\sigma,\bm{k},\omega} F^\text{cryo}_\omega.
\end{aligned}
\end{equation}
Differently from the spectral function for the boson propagator, the cryostat spectral function is positive \cite{Fetter&Walecka}.
Considering a simple spectral density of the form 
\begin{equation}
J^\text{cryo}_{\sigma,\bm{k},\omega}=\frac{\eta_0}{\pi}\exp\left(-\frac{\omega^2}{\omega_c^2}\right),
\end{equation}
and considering the limit where $\omega_c$ is a very large energy scale, one finds that real part of the retarded self-energy just gives rise to a constant shift which we absorb into the chemical potential. 
Instead, the imaginary part gives rise to a spectral width of the system electrons
\begin{equation}\label{eq:cryoSigR}
\Sigma^{\text{cryo},R}_{\sigma,\bm{k},\omega}=-i\eta_0,
\end{equation}
whereas the Keldysh self-energy has form
\begin{equation}\label{eq:cryoSigK}
\Sigma^{\text{cryo},K}_{\sigma,\bm{k},\omega}=-2i\eta_0 F^\text{cryo}_\omega.
\end{equation}
These self-energies are added to the boson-induced normal self-energies, as shown in \cref{eq:cryo_sub}. 
Without the boson self-energies, the electrons are redistributed according to the cryostat distributions, such that electron distribution function in \cref{eq:electronFe} becomes
\begin{equation}
F^e_\omega=F^\text{cryo}_\omega,    
\end{equation}
and thus thermalizes with the cryostat.

\section{Numerical method to solve the NESS-Eliashberg equations at the phase transition}
\label{app:num_methods}
In this appendix, we show the numerical method employed to self-consistently solve the NESS-Eliashberg \cref{eq:RAK_Sigma_eqs,eq:RAK_Delta_eqs} at the phase transition. The critical coupling $g_c$ is defined as the smallest coupling at which these equations admit a non-zero solution for $\Delta^{R}_{\bm{k},\omega}$. As in \cref{sec:NESS-Eliasberg_eqs}, we set here the parametric momentum dependence to the Fermi momentum $|\bm{k}|=k_F$, as this is the first momentum channel to become unstable towards superconductivity, dropping the explicit momentum dependence and obtaining  \cref{eq:selfEnergy_EQs}. In the calculation, the Fermi momentum is self-consistently defined through \cref{eq:kF}, i.e.~it is the momentum at which the real part of the quasi-particle pole of the retarded normal Green's function $G^R_{\bm{k},\omega}$ is zero. 

To find the critical coupling $g_c$, the most naive approach would be solve the non-thermal Eliashberg equations iteratively and, coming from couplings $g>g_c$, find the coupling for which the order parameter $\Delta^R_\omega$ becomes zero. However, this procedure often has the problem that the iterative algorithm gets very slow close to the phase transition, requiring an extrapolation for $\Delta^R_\omega\to0$. A better alternative is to work with the linearized Eliashberg equations, obtained by assuming $\Delta^R_\omega$ to be infinitesimal and neglecting all the terms that are non-linear in $\Delta^R_\omega$ \cite{MarsiglioReview2020}. In \cref{eq:selfEnergy_EQs}, this linearization procedure corresponds to replacing $\mathcal{G}\to G$ everywhere. One can then formally express the linearized NESS-Eliashberg equations for the fields $\Delta_\omega=(\Delta^A_\omega,\Delta^R_\omega,\Delta^K_\omega)$ and $\Sigma_\omega=(\Sigma^A_\omega,\Sigma^R_\omega,\Sigma^K_\omega)$ as 
\begin{align}
    &\Sigma_{\omega} = g_c \int_{\omega'} \, \Xi_{\omega,\omega'} [\Sigma_{\omega'}],
    \label{eq:linearized_Sigma_kernel}\\
    &\Delta_{\omega} = g_c \int_{ \omega'} \, \mathcal{K}_{\omega,\omega'}  \Delta_{\omega'},
    \label{eq:linearized_Delta_kernel}
\end{align}
where the kernels $\mathcal{K}_{\omega,\omega'}$ and $\Xi_{\omega,\omega'}$  are respectively a $3\times 3$ matrix and a 3-dimensional vector in the RAK components with a continuous dependence on  frequency variables $(\omega,\omega')$ [see \cref{eq:selfEnergy_EQs} for their definitions]. The critical coupling $g_c$ is then the coupling for which a non-zero $\Delta_\omega$ solution of \cref{eq:linearized_Delta_kernel} appears, which corresponds to require
\begin{equation}
    \det \left[ \mathbbm{1}_3 \delta(\omega-\omega')-g_c \mathcal{K}_{\omega,\omega'}\right] = 0,
    \label{eq:Thouless_criterion}
\end{equation}
where $\mathbbm{1}_3$ is the $3\times3$ identity matrix, and the determinant is computed both in discrete RAK space and in continuum frequency space. This equation for $g_c$, often called ``Thouless criterion'' \cite{Thouless1960}, can be equivalently found from by looking for a divergence of the pairing susceptibility coming from the normal phase, as was attempted in Ref.~\cite{Chakraborty2021}. We remark however that in Ref.~\cite{Chakraborty2021} some important contributions in the RAK structure of the kernel were neglected, which are instead fully included here.

Direct solution of \cref{eq:Thouless_criterion} is however still quite troublesome, because it requires to discretize frequency space in a way to properly capture the relevant features of the kernel $\mathcal{K}_{\omega,\omega'}$ and calculate a determinant of a very large matrix. This procedure is typically much more straightforward in thermal equilibrium, where \cref{eq:Thouless_criterion} can be expressed on the naturally discretized imaginary Matsubara frequency axis, and a relatively small number of Matsubara frequencies is needed to fully describe the kernel (of the order of 1000 for our model). 
Here we employ an alternative method, originally proposed by McMillian in Ref.~\cite{McMillian1968} to solve real-frequency linearized Eliashberg equations at thermal equilibrium, which is based on an iterative algorithm applied directly to \cref{eq:linearized_Sigma_kernel,eq:linearized_Delta_kernel}, bypassing the calculation of the determinant of \cref{eq:Thouless_criterion}.
The iterative procedure is the following:
\begin{enumerate}
    \item Start from a guess for the coupling $g_c$ and the anomalous self-energy $\Delta_\omega$.
    \item Evaluate the normal self-energy $\Sigma_\omega$ through \cref{eq:linearized_Sigma_kernel} and the corresponding normal-state Green's function $G_\omega$ through \cref{eq:normalElectronProp}.
    \item  Evaluate the frequency integral on the r.h.s.~of \cref{eq:linearized_Delta_kernel} 
    and then update the coupling $g_c$ so that the updated $\Delta_{\omega}$ is rescaled to have $\Delta^{R}_{\omega=0}=1$.
    \item Repeat from step 2 until convergence on the coupling $g_c$ and the functions $\Delta_\omega$ and $\Sigma_\omega$ is reached.
\end{enumerate}
Imposing $\Delta^{R}_{\omega=0}=1$ in step 3 is equivalent to work in terms of the renormalized order parameter $\Delta_\omega/\Delta^{R}_{\omega=0}$, which remains finite also at the phase transition. This iterative method allows then to converge both on the coupling $g_c$ and on the real-frequency order parameter $\Delta_\omega$ at the same time \footnote{We remark that, at thermal equilibrium, McMillian's algorithm \cite{McMillian1968} is not the best method to obtain the real-frequency order parameter $\Delta^R_\omega/\Delta^R_{\omega=0}$ at the phase transition. A more advanced method, which first exploits the numerical advantage of the Matsubara frequency axis and then analytically continues the result to the real axis with a numerically exact procedure, has been developed by Marsiglio and Carbotte
\cite{Marsiglio1988,Marsiglio1991,MirabiMarsiglio2020}. The latter method, however, cannot be applied to our non-thermal case as it still relies on Matsubara frequencies.}. The most computationally expensive part of the algorithm is the calculation of the frequency integrals in the $\Delta_\omega$ and $\Sigma_\omega$ equations, which can be performed directly in frequency space or as a product in time space, by exploiting the convolution form of the integral.
Depending on the used algorithm, one can take advantage of the fact that the self-energies inherit symmetry properties from the real bosons and the particle-hole symmetry of the electrons 
\begin{equation}
    \begin{aligned}
    \Sigma^R_\omega&=-\left(\Sigma^R_{-\omega}\right)^*,\\
    \Sigma^K_\omega&=-\Sigma^K_{-\omega},\\
        \Delta^R_{\omega}&=\left(\Delta^R_{-\omega}\right)^*,\\
        \Delta^K_{\omega}&=\Delta^K_{-\omega}.
    \end{aligned}
\end{equation}
Sometimes it was necessary to partially slow down the iterations in order to converge, as it is often the case in the solution of iterative integral equations (see for example Refs.~\cite{Ferreira1980,Pini2019,Johansen2024}).

\section{Weak-coupling approximations in the temperature-bias example}\label{app:T-bias_wc}
In this appendix we derive weak-coupling analytical expressions for the temperature-bias example. We first find expressions for the electron spectral width $\eta$, the effective temperature $T_e$ and the non-thermal pairing parameter $\alpha_{\mathrm{nth}}$. Secondly, we use these results to derive a closed expression for the critical coupling $g_c$.  
In this setup, the bosons are strongly coupled to a bath with temperature $T_b$ and the electrons are weakly coupled to cryostat at the temperature $T_0$. These couplings can be modeled via the self-energies $\mathbbm{\Pi}^{\mathrm{env}}_\omega$ (see \cref{app:QuadInt}) and $\mathbbm{\Sigma}^{\mathrm{cryo}}_\omega$ (see \cref{app:cryostat_model}), respectively. 
As mentioned in \cref{eq:cryo_sub}, the latter self-energy can be introduced in our NESS-Eliashberg equations with the substitution $\Sigma_\omega\to\Sigma_\omega+\Sigma^{\mathrm{cryo}}_\omega$. Since we are interested in physical quantities at the phase transition, we work here with the linearized NESS-Eliashberg equations at $|\bm{k}|=k_F$.

\subsection{Electron spectral width $\eta$}
\label{app:eta_T-bias_wc}
The electron spectral width $\eta$ is given by the negative imaginary part of the retarded self-energy evaluated at the FS $\omega=0$. The coupling to the cryostat gives rise to a $g_c$-independent contribution $\eta_0=-\Im \Sigma^{\mathrm{cryo},R}_{0}$, while the interaction with the bosons contributes to the spectral width through $-\Im \Sigma^{R}_{0}$. The boson-induced broadening can be computed by taking the imaginary part of the linearized version of \cref{eq:selfEnergy_EQs} for $\Sigma^R_\omega$:
\begin{equation}
    \Im \Sigma^{R}_{\omega}=g_c\int_\epsilon \Re \left(D^R_{\omega-\epsilon}G^K_{\epsilon}+D^K_{\omega-\epsilon}G^R_{\epsilon}\right).
    \label{eq:ImSigmaR}
\end{equation}
To perform a weak-coupling approximation of \cref{eq:ImSigmaR}, one replaces the electron propagators with their quasi-particle approximation $G^R_{\epsilon}\approx 1/(\epsilon+i\eta)$ and $G^K_{\epsilon}= 2i F^e_\epsilon \Im (G^R_\epsilon)\approx-F^e_\epsilon (2 i \eta)/(\epsilon^2+\eta^2)$. Assuming that the electron quasiparticles are long-lived ($\eta \ll T_e,T_b,\omega_0$), the first term in parenthesis can be neglected since $G^K_{\epsilon}\approx -2 i \pi F^e_\epsilon \delta_\epsilon$ and $F^e_{\epsilon=0}=0$. Using the approximation $\Im G^R_{\epsilon}\approx -\pi \delta_\epsilon$, the remaining term $\Re \left(D^K_{\omega-\epsilon}G^R_{\epsilon}\right)=-\Im D^K_{\omega-\epsilon} \Im G^R_{\epsilon}$ gives the first-order contribution
\begin{equation}\label{eq:FirstOrderImSigmaR}
    \Im \Sigma^{R, (1)}_\omega
    =\frac{ g_c \Im D^K_\omega}{2}.
\end{equation}
Setting $\omega=0$ in \cref{eq:FirstOrderImSigmaR}, one obtains the first-order contribution for the electron spectral width
\begin{equation}\label{eq:eta1_2T}
    \begin{split}
    \eta^{(1)}&=-\Im \Sigma^{R, (1)}_{0}= -\frac{g_c\Im D^K_{0}}{2}\\
    &
    =2g_c\frac{\omega_0 \kappa T_b}{\left(\kappa^2+\omega_0^2\right)^2}=g_c \gamma \, T_b,
    \end{split}
\end{equation}
where we defined the constant
\begin{equation}
    \gamma = \frac{2 \omega_0 \kappa }{(\kappa^2+\omega_0^2)^2},
\end{equation}
which has the units of inverse energy squared, making $g_c \gamma$ a dimensionless quantity.
The spectral width of the FS at the leading order in $g_c$ is therefore given by
\begin{equation}
    \eta \approx \eta_0 + \eta^{(1)} = \eta_0 + g_c \gamma  \, T_b.
    \label{eq:eta_T-bias_wc_app}
\end{equation}

\subsection{Effective electron temperature $T_e$}
\label{app:Teff_T-bias_wc}
Let us now consider the effective electron temperature $T_e$, as defined in \cref{eq:F^e_approx_T_eff,eq:Te}.
First the substitution $\Sigma_\omega\to\Sigma_\omega+\Sigma^{\mathrm{cryo}}_\omega$ is used in \cref{eq:SigmaK_def_app}, obtaining
\begin{equation}
    F^e_\omega=F^{\mathrm{cryo}}_\omega-\frac{1}{2i \eta_0} \left(\Sigma^K_\omega - 2 i F^e_\omega \, \Im  \Sigma^R_\omega\right).
    \label{eq:SigmaK_def_app_cryo}
\end{equation}
Using the result in \cref{eq:SigmaK_dist_eq} and the imaginary part of \cref{eq:SigmaR_dist_eq} in \cref{eq:SigmaK_def_app_cryo}, one obtains the following kinetic equation
\begin{equation}
    \begin{split}
        F^e_\omega&=F^{\mathrm{cryo}}_\omega +\frac{2 g_c}{\eta_0} \int_\epsilon \Big[ F^b_{\omega-\epsilon} F^e_{\epsilon}+1 \\
        &\quad-F^e_{\omega} \left(F^e_{\epsilon}+F^b_{\omega-\epsilon}\right)\Big] \Im D^R_{\omega-\epsilon} \Im G^R_{\epsilon},
    \end{split}
    \label{eq:Fe_kin_eq_cryo}
\end{equation}
which corresponds to \cref{eq:Fe_kin_eq} specialized to the temperature-bias setup. 
We can now perform the weak-coupling approximation 
by replacing the electron retarded propagator with its quasiparticle approximation $G^R_{ \omega}\approx1/(\omega+i\eta)$ with $\eta$ given by \cref{eq:eta_T-bias_wc_app}, and consider an external frequency $\omega\to0$ in order to linearize the electron and cryostat distributions as $F^e_\omega \approx \omega/2T_e$ and $F^{\mathrm{cryo}}_\omega \approx \omega/2T_0$, respectively. We obtain then
\begin{equation}
    \begin{split}
        \frac{\omega}{2T_e}&= \frac{\omega}{2T_0}+\frac{2 g_c}{\eta_0} \int_\epsilon \Big[ F^b_{\omega-\epsilon} F^e_{\epsilon}+1\\
        &\quad- \frac{\omega}{2 T_e} \left(F^e_{\epsilon}+F^b_{\omega-\epsilon}\right)\Big]  \Im D^R_{\omega-\epsilon} \Im G^R_{\epsilon},
    \end{split}
\end{equation}
In the weak-coupling regime $\eta \ll T_0
,T_b,\omega_0$ such that $\Im G^R_{\epsilon} \approx -\pi\delta_\epsilon$, which makes it possible to evaluate the integral over $\epsilon$, obtaining
\begin{equation}
    \frac{1}{2T_e}= \frac{1}{2T_0}+  \frac{g_c \gamma}{2\eta_0} \Big( 1 - \frac{T_b}{T_e} \Big),
\end{equation}
where we have used the explicit form for $D^R_\omega$ from \cref{eq:D_omega_RAK}, simplified a common $\omega$ factor and then set $\omega\to0$ and expanded $F^b_{-\epsilon}\approx -2 T_b/\epsilon$ before the integration. Solving for $T_e$, one finally obtains
\begin{equation}
    T_{e}\approx \frac{\eta_0+g_c \gamma \, T_b}{\eta_0+g_c \gamma \, T_0}T_0.
    \label{eq:Teff_analytic_app}
\end{equation}

\subsection{Non-thermal pairing parameter $\anth$}
\label{subsec:alpha_nth_T-bias}
Let us now consider the linearized equation for $\Delta^K_\omega$ at \mbox{$|\bm{k}|=k_F$} and express the Keldysh component of the electron and boson propagators in terms of distributions as done in ~\cref{eq:F^Delta_ImDelta}, which with a slight rewriting takes the form
\begin{equation}
    \begin{split}
        \Delta^K_\omega &= 2 g_c \int_\epsilon \Im D^R_{\omega-\epsilon}  \\
        &\quad\times\Big\{ 2 i \left( F^b_{\omega-\epsilon}F^e_{\epsilon}+1 \right) \Im \left[\left(G^R_{\epsilon}\right)^2 \Delta^R_{\epsilon}\right] \\
        &\quad+F^b_{\omega-\epsilon} \left( \Delta^K_{\epsilon}-2i F^e_{\epsilon}  \Im \Delta^R_{\epsilon} \right) \abs{G^R_{\epsilon}}^2 \Big\}.
    \end{split}
\end{equation}
Taking the imaginary part of this equation, dividing it by $\Delta^R_\omega$ and considering the static limit $\omega\to 0$, one finds an equation for the non-thermal pairing parameter $\anth$ defined in \cref{eq:alpha_nth}:
\begin{equation}\label{eq:alpha_full}
    \begin{split}
        \anth &= 2g_c \int_\epsilon  \Im D^R_{\epsilon}\Bigg\{ 2 \left(F^b_{\epsilon}F^e_{\epsilon}-1\right) \Im  \left[\left(G^R_{\epsilon}\right)^2 \frac{\Delta^R_{\epsilon}}{\Delta^R_0} \right] \\
        &\quad+F^b_{\epsilon} \left[ \frac{\Im \Delta^K_{\epsilon}}{\Delta^R_0}-2F^e_{\epsilon}  \frac{\Im \Delta^R_{\epsilon}}{\Delta^R_0} \right] \abs{G^R_{\epsilon}}^2 \Bigg\},
    \end{split}
\end{equation}
where the odd nature of the Boson spectral function and distribution function has been used.  
To compute the integral in \cref{eq:alpha_full} we use the weak-coupling approximation $G^R_\epsilon \approx 1/(\epsilon+i\eta)$, with $\eta$ given by \cref{eq:eta_T-bias_wc_app}, together with a further assumption that $\eta \ll T_e, T_b, \omega_0$. Noticing that the functions $(G^R_{\epsilon})^2=1/(\epsilon+i\eta)^2$ and $|G^R_{\epsilon}|^2=1/({\epsilon}^2+\eta^2)$ are strongly peaked close to $\epsilon=0$, it is possible to approximate all the other multiplying function with their static limit, obtaining 
\begin{equation}\label{eq:alpha_T-bias_app_part_1}
    \begin{split}
        \anth &\approx  
        g_c \gamma
        \bigg\{  \left( \frac{T_b}{T_e}-1 \right) \int _\epsilon \frac{4 \eta \epsilon^2}{(\epsilon^2+\eta^2)^2}  \\
        &\quad + \frac{T_b}{\eta} \int _\epsilon \left[\frac{\epsilon}{ T_e}  \frac{\Im \Delta^R_{\epsilon}}{\Delta^R_0} -\anth\right] \frac{2\eta}{\epsilon^2+\eta^2} \bigg\}.
    \end{split}
\end{equation}
As $\epsilon \, \Im \Delta^R_\epsilon/\Delta^R_0 \to 0$ for $\epsilon \to 0$, the first term in the square brackets can be neglected. The remaining frequency integrals are $\int _\epsilon \frac{4\eta\epsilon^2}{(\epsilon^2+\eta^2)^2}=1$ and $\int_\epsilon \frac{2\eta}{\epsilon^2+\eta^2}=1$, leading to
\begin{equation}\label{eq:alpha_T-bias_app_part_2}
    \begin{split}
        \anth \approx g_c \gamma \left[ \left( \frac{T_b}{T_e}-1 \right) - \frac{T_b\anth}{\eta}  \right].
    \end{split}
\end{equation}
One can finally isolate $\anth$ and find the $g_c$-dependent expression
\begin{equation}\label{eq:alpha_T-bias_app}
    \begin{split}
        \anth &\approx\frac{g_c \gamma \left(\frac{T_b}{T_e}-1\right)}{1+ \frac{g_c \gamma \, T_b}{\eta}} =\frac{g_c \gamma \, \eta_0 (T_b-T_0)}{T_0(\eta_0+2 g_c \gamma \, T_b)},
    \end{split}
\end{equation}
where in the last equality \cref{eq:eta_T-bias_wc_app,eq:Teff_analytic_app} was used to simplify the expression. Moreover, by expanding the first equality in \cref{eq:alpha_T-bias_app} close to thermal equilibrium ($T_b\approx T_e\approx T_0$), one obtains the expression
\begin{equation}
    \anth \approx \mathcal{C} \, \frac{T_b-T_e}{T_e},
    \label{eq:alpha_expansion_app}
\end{equation}
where $\mathcal{C} = \frac{g_{c,\mathrm{th}}  \, \gamma \, (\eta_0+ g_{c,\mathrm{th}} \, \gamma \, T_b)}{\eta_0+2 g_{c,\mathrm{th}} \, \gamma \, T_b}$ is a positive constant, with $g_{c, \mathrm{th}}$ given by \cref{eq:gc_th}. This is the expression used in \cref{eq:alpha_expansion}.

\subsection{Closed-form expression for the critical coupling~$g_c$}
Having derived expressions for $\eta$ in \cref{eq:eta_T-bias_wc_app}, $T_e$ in \cref{eq:Teff_analytic_app} and $\anth$ in \cref{eq:alpha_T-bias_app}, these can now be substituted into the expression for the critical coupling of \cref{eq:gc}.
After some straightforward but tedious algebra, one finds a quadratic equation for $g_c$ with the following (positive) solution 
\begin{equation}
    \begin{split}
        g_c &= \frac{2 T_b}{|D^R_0|}+\frac{1}{2} \left( \frac{2 T_b}{|D^R_0|} + \frac{\eta_0}{2\gamma \, T_0} \right) \\
        &\quad\times \left( \! \sqrt{1+ \frac{4\eta_0(T_0-T_b)}{ \gamma \, T_0 |D^R_0|} \left( \frac{2 T_b}{|D^R_0|} + \frac{\eta_0}{2\gamma \, T_0 } \right)^{\!\!-2}} - 1 \! \right),
    \end{split}
    \label{eq:gc_closedform_T-bias_wc_app}
\end{equation}
which is closed-form expression for $g_c$ depending only on the external parameters $\eta_0$, $\kappa$, $\omega_0$, $T_0$ and $T_b$. 

\section{Weak-coupling approximations in the incoherent drive example}\label{app:incoherent_wc}
Similarly to \cref{app:T-bias_wc}, in this appendix we derive the weak-coupling analytical expressions for the electron spectral width $\eta$, the effective temperature $T_e$, the non-thermal pairing parameter $\alpha_{\mathrm{nth}}$ and the critical coupling $g_c$ for the incoherent drive example. In this setup, the bosons are strongly coupled to an environment with a thermal background at temperature $T_b$ and a incoherent drive at the frequency $\omega_d\gg T_b$, as described by the distribution in \cref{eq:incFb}. The electrons are instead decoupled from the cryostat ($\mathbbm{\Sigma}^{\mathrm{cryo}}=0$), so that their distribution $F^e_\omega$ is entirely determined by their coupling to the bosons.
\subsection{Zeroth-order electron distribution}
Since there is no cryostat, the electron distribution has to be self-consistently determined even for arbitrarily weak coupling. 
The distribution is determined by \cref{eq:Fe_kin_eq}. 
To zeroth order in $g_c$, the electrons have a vanishing linewidth such that $\Im G^{R,(0)}_{\epsilon}=-\pi \delta_\epsilon$.
The resulting equation for the distribution then takes the form
\begin{equation}\label{eq:Fe_kin}
        \big[ F^b_{\omega} F^{e,(0)}_{0}+1 -F^{e,(0)}_{\omega} (F^{e,(0)}_{0}+F^b_{\omega})\big]
        \Im D^R_{\omega}=0.
\end{equation}
At the FS $F^{e}_{0}=0$, which removes two terms in \cref{eq:Fe_kin}. Since $\Im D^R_{\omega}\neq 0$, which is guaranteed by the coupling to the external environment, the zeroth-order contribution to the electron distribution takes the form
\begin{equation}\label{eq:Fe0}
        F^{e,(0)}_{\omega}=\frac{1}{F^b_{\omega}}.
\end{equation}
This result resembles that of thermal equilibrium in \cref{eq:Fb_Fe_FDelta_theq}, but in this case it is true only at the zeroth order in $g_c$, rather than being self-consistently verified at all orders like at thermal equilibrium [cf.~\cref{eq:FDelta_kin_eq,eq:Fb_Fe_FDelta_theq,eq:magic_formula}]. 

\subsection{Electron spectral width $\eta$}\label{sec:app_inc_eta}
In the incoherent driving example, the spectral width of the electrons is solely due to the interaction with the bosons at the FS, i.e.~$\eta=-\Im \Sigma^R_0$. The first-order contribution to the normal self-energy can be calculated as the temperature-bias example, obtaining [cf.~\cref{eq:FirstOrderImSigmaR}]
\begin{equation}\label{eq:firstOrderSR}
    \Im \Sigma^{R,(1)}_{\omega}=\frac{g_c\Im D^K_{\omega}}{2}.
\end{equation}
Consequently, the electron spectral width at the FS, to first order in $g_c$, is  $\eta\approx\eta^{(1)}=g_c \gamma T_b$ found in \cref{eq:eta1_2T}. As $\omega_d\gg T_b$, there is a large energy separation between the incoherently driven and thermal occupations, and for this reason the incoherent drive only starts contributing to $\eta$ at second order in $g_c$.  However, since we are in the weak-coupling regime and at the phase transition, it follows that $g_c\sim T_b$, making $\eta^{(1)}$ comparable in magnitude to the second order term. It is therefore essential to also compute the second order term to capture the physics under incoherent driving at $\omega_d\gg T_b$. This can be done by considering again the full frequency integral defining $\Im \Sigma^R_\omega$ in \cref{eq:ImSigmaR}, setting $\omega=0$ and estimating the additional contribution from the frequency region around $\epsilon=\omega_d$.
To account for this contribution,
we insert $\Im \Sigma^{R,(1)}_{\epsilon}$ in the electron propagator 
\begin{equation}
    G^{R,(1)}_{\epsilon}=\left(\omega-i\Im \Sigma^{R,(1)}_{\epsilon}\right)^{-1}.
    \label{eq:G^{R,(1)}}
\end{equation}
Note that, in the weak-coupling regime, $\Re \Sigma^{R,(1)}_{\epsilon}$ can be neglected as the particle-hole symmetry forces it to vanish at the FS, while near $\omega_d$ its magnitude has to compared to $\omega_d$ itself and is therefore negligible.
The reason that it is sufficient to include \cref{eq:firstOrderSR} to capture effects of the drive, is the convolution nature of \cref{eq:ImSigmaR} for the retarded self-energy.
Using \cref{eq:G^{R,(1)}} inside \cref{eq:ImSigmaR} makes the self-energy sensitive to the behavior of the electron distribution near $\omega_d$, where $\Im \Sigma^{R,(1)}_{\omega}$ is peaked. 
For this reason, it is also necessary to include the zeroth-order result for the distribution found in \cref{eq:Fe0}. 
Within these approximations $\eta$ takes the form 
\begin{equation}\label{eq:eta_inc1}
\begin{aligned}
    \eta=&-\Im \Sigma^R_0 \\
    \approx&-g_c\int_\epsilon \Re \big(2i D^R_{-\epsilon} \, F^{e,(0)}_\epsilon\Im G^{R,(1)}_{\epsilon}\\&\hspace{19mm}+2i G^{R,(1)}_{\epsilon}F^b_{-\epsilon}\Im D^R_{-\epsilon}\big)\\
    =&-2g_c \int_\epsilon \left(\frac{1}{F^b_\epsilon}-F^b_\epsilon\right)\Im D^R_{\epsilon}\Im 
    G^{R,(1)}_{\epsilon}
    \\
    =&-2g_c \int_\epsilon \Bigg[\left(\frac{1}{F^{b,\text{th}}_\epsilon+F^{b,d}_\epsilon}-F^{b,\text{th}}_\epsilon-F^{b,d}_\epsilon\right)\\
    &\hspace{15mm}\times \Im D^R_{\epsilon} \Im 
    G^{R,(1)}_{\epsilon}\Bigg],
    \end{aligned}
\end{equation}
where the expression for the boson distribution in \cref{eq:incFb} and the symmetries of the real retarded boson propagator have been used. 
The large energy separation of the incoherent drive from the thermal part means that, for energies where $F^{b,d}_\epsilon$ is finite, $F^{b,\text{th}}_\epsilon$ looks constant, whereas near the FS $F^{b,d}_\epsilon$ vanishes.  This allows one to divide the integral in \cref{eq:eta_inc1} in two contributions: a contribution $\eta_<$ from frequencies around the FS ($\epsilon\approx 0$) and a contribution $\eta_d$ from frequencies around the drive frequency ($\epsilon \approx \omega_d$).
The first contribution can be identified as
\begin{equation}\label{eq:eta_lesser}
    \eta_< = -2 g_c  \int_\epsilon \left(\frac{1}{F^{b,\text{th}}_\epsilon}-F^{b,\text{th}}_\epsilon\right)\Im D^R_{\epsilon}\Im 
    G^{R,(1)}_{\epsilon}.
\end{equation}
As $F^{b,\text{th}}_\epsilon$ goes to unity for $\epsilon\gg T_b$, the integrand vanishes in this region and therefore only captures behavior near the FS. 
In the frequency region  around the FS, $\Im G^{R,(1)}_\epsilon$ can also be replaced with 
$\Im G^{R,(0)}_\epsilon=-\pi \delta_\epsilon$, such that one indeed recovers $\eta^<=\eta^{(1)}$.

Considering now the contribution from the frequency region around $\omega_d$, we find
\begin{equation}
\label{eq:eta_d_integral}
\begin{aligned}
    \eta_d = -2 g_c  \int_\epsilon \Bigg[&\left(\frac{1}{\mathrm{sign}(\epsilon)+F^{b,d}_\epsilon}-\mathrm{sign}(\epsilon)-F^{b,d}_\epsilon\right)\\\times&\Im D^R_{\epsilon}\Im 
    G^{R,(1)}_{\epsilon}\Bigg].
\end{aligned}
\end{equation}
Using \cref{eq:G^{R,(1)},eq:firstOrderSR}, $\Im G^{R,(1)}_\epsilon$ in this integral can be expressed as
\begin{equation}
    \Im G^{R,(1)}_\epsilon=\frac{2g_c \Im D^K_\epsilon}{4\epsilon^2+\left(g_c\Im D^K_\epsilon\right)^2}.
    \label{eq:ImG^{R,(1)}}
\end{equation}
Moreover, in weak-coupling regime and in the region where $\epsilon\sim\omega_d\gg g_c$, the $g_c$-dependent term in the denominator of \cref{eq:ImG^{R,(1)}} can be neglected.
With this approximation, \cref{eq:eta_d_integral} becomes
\begin{equation}
    \eta_d\approx g_c^2I_\eta,
\end{equation}
with the coupling-independent constant
\begin{equation}\label{eq:Ieta_inc}
\begin{aligned}
    I_\eta=&\int_0^\infty\frac{d\epsilon}{\pi} \left(1+F^{b,d}_\epsilon-\frac{1}{1+F^{b,d}_\epsilon}\right)\frac{\Im D^R_{\epsilon}\Im D^K_\epsilon}{\epsilon^2}\\
    =&2\int_0^\infty\frac{d\epsilon}{\pi} \left[2F^{b,d}_\epsilon+\left(F^{b,d}_\epsilon\right)^2\right]\frac{\left(\Im D^R_{\epsilon}\right)^2}{\epsilon^2},
\end{aligned}
\end{equation}
which, given the expression \cref{eq:Fd_inc} for $F^{b,d}_\epsilon$, vanishes for $A=0$ as expected. 
The total electron spectral width in the weak-coupling regime is then given by the sum of the two contributions
\begin{equation}\label{eq:eta_incFinal}
    \eta\approx\eta^{(1)}+\eta_d=g_c\left(\gamma T_b+g_cI_\eta\right).
\end{equation}

\begin{figure*}[tbp]
\centering
\includegraphics[width=0.8\textwidth]{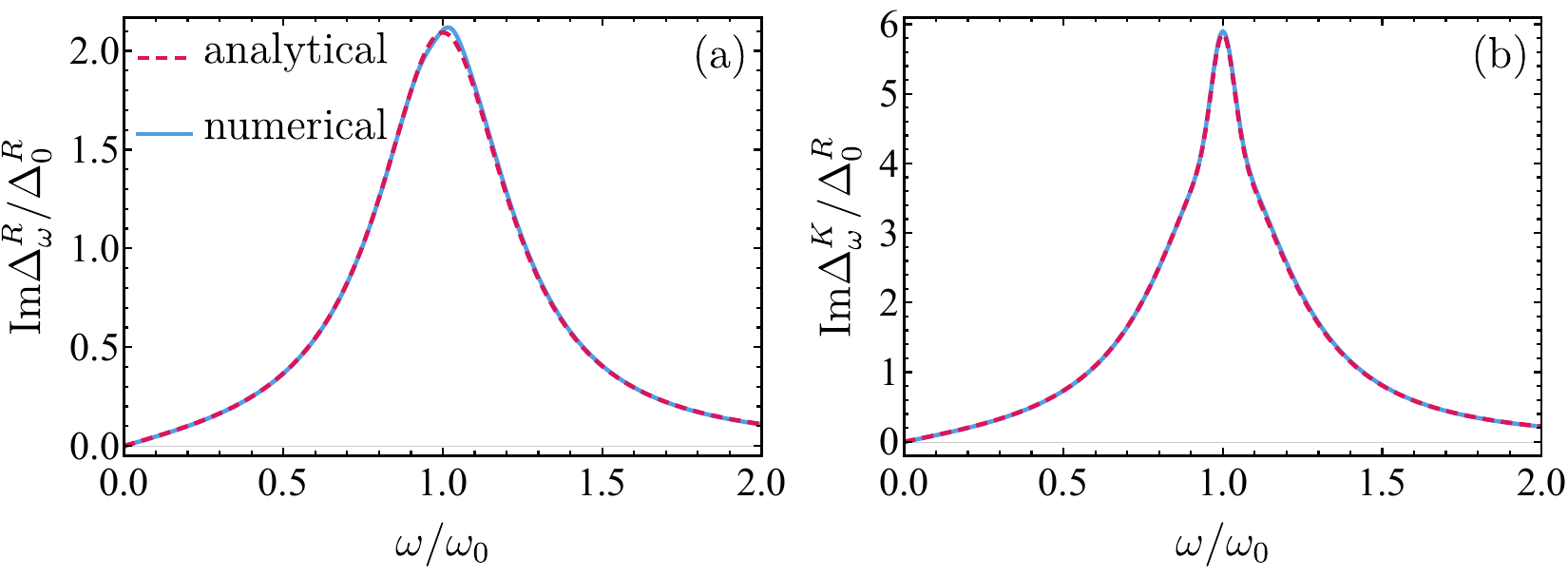}
\caption{Imaginary part of the (a) retarded and (b) Keldysh components of the anomalous self-energies for incoherent driving of the bosons with strength $A=0.4$. The converged self-consistent numerical solution of \cref{eq:selfEnergy_EQs} is labeled ``numerical'' while the approximation in \cref{eq:inc_anom} is denoted as ``analytical''.
The other parameters are set to the ones used in \pref{fig:KR}{b}.}
\label{fig:inc_anom}
\end{figure*}
\subsection{Effective temperature $T_e$}
From the zeroth-order result in \cref{eq:Fe0} it is clear that the electrons inherit the thermal background of the bosons. 
As discussed in \cref{eq:F^e_approx_T_eff,eq:Te}, the slope of the distribution near the FS determines the effective temperature $T_e$. Although the drive acts at high energies, interactions propagate its influence to lower energies, leading to a drive-induced modification of \(T_e\).  
To capture this effect we consider the derivative of the defining equation for the distribution function in \cref{eq:Fe_kin_eq} at the FS
\begin{equation}\label{eq:effTe_1}
\begin{aligned}
    \partial_\omega \int_\epsilon& \left[F^b_{\omega-\epsilon}F^e_\epsilon+1-F^e_{\omega}\left(F^e_\epsilon+F^b_{\omega-\epsilon}\right)\right]\\&\times\Im D^R_{\omega-\epsilon}\Im G^R_{\epsilon}\bigg\vert_{\omega=0}=0.
    \end{aligned}
\end{equation}
Mathematically, the interactions propagate the effects of the drive down to the FS due to the convolution structure of \cref{eq:effTe_1} and the contribution of the frequency integral around the drive frequency $\omega_d$. 
The finite value of $\kappa$ makes $\Im D^R_{\omega}$ a slowly changing function in the weak-coupling regime, such that its variation can be neglected compared to that of the thermal contribution near the FS.
Additionally, the change of electron temperature is going to be a small change on the integrated distributions  in \cref{eq:effTe_1} that depend only on $\epsilon$ (as $g_c$ is small). 
It is thus only in $F^e_\omega\approx\omega/2T_e$ where the effective electron temperature must be included.
One can now follow the same approach as for $\eta$, namely, using \cref{eq:G^{R,(1)},eq:Fe0} instead of the fully self-consistent electron propagators. 
With these approximations \cref{eq:effTe_1} takes the form
\begin{equation}
\begin{aligned}
    \int_\epsilon \Big[&\frac{\partial_\omega F^b_{\omega-\epsilon}\big\vert_{\omega=0}}{F^b_\epsilon}-F^e_{0}\partial_\omega F^b_{\omega-\epsilon}\big\vert_{\omega=0} \\& -\frac{1}{2T_e}\left(\frac{1}{F^b_\epsilon}-F^b_{\epsilon}\right)\Big]\Im D^R_{\epsilon}\Im G^{R,(1)}_{\epsilon}\approx0.
    \end{aligned}
\end{equation}
As $F^e_\epsilon$ is anti-symmetric around the FS, the second term vanishes and the last term can immediately be identified as being proportional to $\eta$ in \cref{eq:eta_inc1}, giving the result
\begin{equation}\label{eq:Te_2}
    I_0+\frac{\eta}{4 T_e g_c}=0,
\end{equation}
with 
\begin{equation}
    I_0=\int_\epsilon \frac{\partial_\omega F^b_{\omega-\epsilon}}{F^b_\epsilon}\Im D^R_{\epsilon}\,\Im G^{R,(1)}_{\epsilon}\bigg\vert_{\omega=0}.
\end{equation}
The region where $F^b_{\omega-\epsilon}$ has the largest gradient is for $\epsilon \lesssim T_b$, where it diverges as $2 T_b/\epsilon$. 
The large energy separation between the two terms in $F^b_\omega$ thus allows one to neglect the large energy contribution to $I_0$ and to approximate it as
\begin{equation}
\begin{aligned}
    I_0&\approx\int_{-\lambda}^{\lambda}\frac{d\epsilon}{2\pi} \frac{\partial_\omega \coth\left(\frac{\omega-\epsilon}{2T_b}\right)}{\coth\left(\frac{\epsilon}{2T_b}\right)}\Im D^R_{\epsilon}\Im G^{R,(1)}_{\epsilon}\bigg\vert_{\omega=0}\\
    &= -\frac{1}{T_b}\int_{-\lambda}^{\lambda}\frac{d\epsilon}{2\pi} \mathrm{csch}\left(\frac{\epsilon}{T_b}\right)\Im D^R_{\epsilon}\Im G^{R,(1)}_{\epsilon},
    \end{aligned}
\end{equation}
where $\lambda$ is a suitably chosen cutoff $T_b\ll\lambda\ll\omega_d$.
Being in the weak-coupling regime and only considering $\epsilon\leq \lambda$, the broadening $\eta^{(1)}$ is so small that it can again be approximated as infinitesimal, meaning that one can replace $\Im G^{R,(1)}_{\epsilon}\to \Im G^{R,(0)}_\epsilon=-\pi \delta_\epsilon$.
As the linear behavior of $\Im D^R_\epsilon$ cancels the divergence in $\mathrm{csch}(\epsilon/T_b)$, the integral is finite, giving the following approximate closed-form expression for $I_0$
\begin{equation}
    I_0\approx -\frac{\omega_0\kappa}{2\left(\kappa^2+\omega_0^2\right)^2}=-\frac{\gamma}{4}.
\end{equation}
Inserting this in \cref{eq:Te_2}, the approximation for the effective electron temperature can be written as 
\begin{equation}\label{eq:Te_inc_Final}
    T_e\approx \frac{\eta}{g_c \gamma}= \frac{T_b \eta}{\eta^{(1)}}=T_b\left(1+\frac{g_c I_\eta}{\gamma T_b}\right).
\end{equation} 
Combining this result with \cref{eq:Fe0}, the electron distribution can be approximated beyond the zeroth order as the inverse of the boson distribution, but with the thermal background at the temperature $T_e$
\begin{equation}\label{eq:incFe_app}
    F^e_\epsilon\approx\frac{1}{F^{b,\mathrm{th}}_{T_e,\omega}+F^{b,d}_\omega}.
\end{equation}
\subsection{Non-thermal pairing parameter $\alpha_{\text{nth}}$}
Having consistently included the effect of drive in both the spectral width and the effective temperature, one can use these to compute $\anth$. One starts by inserting the approximation \eqref{eq:incFe_app} for the electron distribution into the  \cref{eq:alpha_full} for $\anth$ and gets to the integral 
\begin{widetext}
\begin{equation}\label{eq:alpha_full_inc}
        \anth \!= \!2g_c\! \int_\epsilon \Im D^R_{\epsilon} \left\{2 \left(\frac{F^{b,\mathrm{th}}_{T_b,\epsilon}\!+\!F^{b,d}_\epsilon}{F^{b,\mathrm{th}}_{T_e,\omega}\!+\!F^{b,d}_\omega}\!-\!1\right)\Im  \!\left[(G^R_{\epsilon})^2 \frac{\Delta^R_{\epsilon}}{\Delta^R_0} \right]\! +\!\left(F^{b,\mathrm{th}}_{T_b,\epsilon}\!+\!F^{b,d}_\epsilon\right)\! \left[ \frac{\Im \Delta^K_{\epsilon}}{\Delta^R_0}\!-\! \frac{2\Im \Delta^R_{\epsilon}}{\left(F^{b,\mathrm{th}}_{T_e,\epsilon}+F^{b,d}_\epsilon\right)\Delta^R_0} \right] |G^R_{\epsilon}|^2 \right\}.
\end{equation}
Using the energy separation between $F^{b,\mathrm{th}}_{T_e,\omega}$ and $F^{b,d}_{\omega}$, together with the approximated behavior of the electron propagator near the FS, one can define a low-energy contribution to $\anth$ 
\begin{equation}
    \begin{split}
        \alpha_<\! =\! 2g_c\! \int_\epsilon&  \Im D^R_{\epsilon}\Bigg\{\! 2\! \left(\frac{F^{b,\mathrm{th}}_{T_b,\epsilon}}{F^{b,\mathrm{th}}_{T_e,\epsilon}}\!-\!1\right) \Im \! \left[\left(G^{R,(2)}_{\epsilon}\right)^2 \frac{\Delta^R_{\epsilon}}{\Delta^R_0} \right] 
        +F^{b,\mathrm{th}}_{T_b,\epsilon}\! \left(\! \frac{\Im \Delta^K_{\epsilon}}{\Delta^R_0}\!-\!\frac{2}{F^{b,\mathrm{th}}_{T_b,\epsilon}}  \frac{\Im \Delta^R_{\epsilon}}{\Delta^R_0} \right) \abs{G^{R,(2)}_{\epsilon}}^2 \Bigg\},
    \end{split}
\end{equation}
where $G^{R,(2)}_\omega=\left(\epsilon+i \eta\right)^{-1}$ ensures that the integrand vanishes for large energy, allowing us to extend the integral to the full energy axis.
Similarly to what was found for $\eta_<$ in \cref{eq:eta_lesser}, one finds that $\alpha_<$ has a functional form analogous to the temperature-bias one in \cref{eq:alpha_T-bias_app_part_2}:
\begin{equation}
    \label{eq:alpha_lesser_fin}
    \alpha_<\approx g_c \gamma \left[ \left( \frac{T_b}{T_e}-1 \right) - \frac{T_b\anth}{\eta}  \right].
\end{equation}
The contribution to the non-thermal pairing parameter due to the incoherent drive is instead captured by the region of the integral around $\epsilon\approx\omega_d$
\begin{equation}\label{eq:alpha_greater}
\begin{aligned}
        \alpha_d &= 2g_c \!\int_\epsilon\! \Im D^R_{\epsilon} \Bigg[ 2\! \left(\frac{\mathrm{sign}(\epsilon)\!+\!F^{b,d}_\epsilon}{\mathrm{sign}(\epsilon)\!+\!F^{b,d}_\omega}\!-\!1\right) \Im  \!\left[(G^R_{\epsilon})^2 \frac{\Delta^R_{\epsilon}}{\Delta^R_0} \right]\!+\!\left(\mathrm{sign}(\epsilon)\!+\!F^{b,d}_\epsilon\right) \!\left( \!\frac{\Im \Delta^K_{\epsilon}}{\Delta^R_0}\!-\! \frac{2\Im \Delta^R_{\epsilon}}{\left(\mathrm{sign}(\epsilon)+F^{b,d}_\epsilon\right)\Delta^R_0} \right) |G^R_{\epsilon}|^2 \Bigg]\\
        &=2g_c \int_\epsilon \Im D^R_{\epsilon} \left[ \left(\mathrm{sign}(\epsilon)+F^{b,d}_\epsilon\right)\frac{\Im \Delta^K_{\epsilon}}{\Delta^R_0}-2 \frac{\Im \Delta^R_{\epsilon}}{\Delta^R_0} \right] |G^R_{\epsilon}|^2,
\end{aligned}
\end{equation}
\end{widetext}
To compute this contribution, one needs approximations for the behavior of the retarded and Keldysh anomalous self-energies near $\omega_d$. These approximation must be obtained from their one-loop definition in Eqs.~\eqref{eq:selfEnergy_EQs}.
To this end, the crudest approximation one can perform is that the anomalous propagators $\mathcal{F}$ in Eqs.~\eqref{eq:selfEnergy_EQs} are strongly peaked at the FS, so that the bosonic propagators can be taken out of the convolution similarly to what was done in \cref{sec:nonThDeriv}. For the retarded anomalous self-energy, this gives
\begin{equation}\label{eq:DeltaRApp_incApp}
    \begin{aligned}
        \Delta^R_\omega&=ig\int_\epsilon \left(D^R_{\omega-\epsilon}\mathcal{F}^K_\epsilon+D^K_{\omega-\epsilon}\mathcal{F}^R_\epsilon\right)\\
        &\approx ig\left(D^R_{\omega}\int_\epsilon\mathcal{F}^K_\epsilon+D^K_{\omega}\int_\epsilon\mathcal{F}^R_\epsilon\right)\\
        &= ig D^R_{\omega}\int_\epsilon\mathcal{F}^K_\epsilon.\\
    \end{aligned}
\end{equation}
where the equal-time limit in \cref{eq:zeroRetAnomProp} was used in the last line. 
Equivalently, the behavior of the anomalous Keldysh self-energy around $\omega=\omega_d$ can be approximated as 
\begin{equation}\label{eq:DeltaKApp_incApp}
\begin{aligned}
\Delta^K_\omega&=ig\int_\epsilon \left(D^R_{\omega-\epsilon}\mathcal{F}^R_\epsilon+D^A_{\omega-\epsilon}\mathcal{F}^A_\epsilon+D^K_{\omega-\epsilon}\mathcal{F}^K_\epsilon\right)\\
&\approx ig\left(D^R_{\omega}\int_\epsilon\mathcal{F}^R_\epsilon+D^A_{\omega}\int_\epsilon\mathcal{F}^A_\epsilon+D^K_{\omega}\int_\epsilon\mathcal{F}^K_\epsilon\right)\\
&= igD^K_{\omega}\int_\epsilon\mathcal{F}^K_\epsilon.\\
\end{aligned}
\end{equation}
Using \cref{eq:DeltaRApp_incApp,eq:DeltaKApp_incApp} we find the approximations
\begin{equation}\label{eq:inc_anom}
    \begin{aligned}
        \frac{\Im \Delta^R_\omega}{\Delta^R_0}\Bigg\vert_{\omega\approx\omega_d}
        &\approx\frac{\Im D^R_\omega}{D^R_0},\\
        \frac{\Im \Delta^K_\omega}{\Delta^R_0}\Bigg\vert_{\omega\approx\omega_d}
        &\approx\frac{\Im D^K_\omega}{D^R_0}=\frac{2\left(\mathrm{sign}(\epsilon)+F^{b,d}_\epsilon\right)\Im D^R_\omega}{D^R_0}.
    \end{aligned}
\end{equation}
These approximations are compared with the numerical solutions in \cref{fig:inc_anom}, confirming that they work quite well in the weak-coupling regime \footnote{We note that a similar approximation for $\Delta^R_\omega$ is also found at zeroth order in the weak-coupling limit of standard Eliashberg theory at thermal equilibrium, see Ref.~\cite{MirabiMarsiglio2020}.}.
As the approximations~\eqref{eq:inc_anom} are used under an integral and multiplied by smooth functions, small deviations do not have a significant impact on the value of the integral.

Inserting the approximations~\eqref{eq:inc_anom} into \cref{eq:alpha_greater}, together with the approximation that the $\omega\approx\omega_d$ behavior of $\abs{G^R_\epsilon}^2\approx \epsilon^{-2}$, one obtains an approximation for $\alpha_d$ of the form
\begin{equation}
\label{eq:alpha_d_fin}
\begin{aligned}
    \alpha_d&\approx\frac{4g_c}{D^R_0} \int_\epsilon \frac{\left(\Im D^R_{\epsilon}\right)^2}{\epsilon^2} \left[ \left(\mathrm{sign}(\epsilon)+F^{b,d}_\epsilon\right)^2-1 \right]\\
    &=\frac{4g_c}{D^R_0} \int_\epsilon \frac{\left(\Im D^R_{\epsilon}\right)^2}{\epsilon^2} \left[2\,\mathrm{sign}(\epsilon)F^{b,d}_\epsilon+\left(F^{b,d}_\epsilon\right)^2\right]\\
    &=\frac{2g_c}{D^R_0}I_\eta,
\end{aligned}
\end{equation}
where in the last line \cref{eq:Ieta_inc} was used to rewrite the result. 
Collecting the results of \cref{eq:alpha_lesser_fin,eq:alpha_d_fin} one can write the full weak-coupling approximation for $\anth$ 
\begin{equation}\label{eq:alpha_inc_final}
\begin{aligned}
    \anth &\approx \alpha_<+\alpha_d,\\
    &\approx g_c \gamma \left[ \left( \frac{T_b}{T_e}-1 \right) - \frac{T_b\anth}{\eta}  \right]+\frac{2g_c}{D^R_0}I_\eta\\
    &=\frac{g_c}{1+\frac{g_c \gamma T_b}{\eta}}\left[\gamma \left(\frac{T_b}{T_e}-1\right)+\frac{2I_\eta}{D^R_0}\right].
\end{aligned}
\end{equation}
\subsection{Closed-form expression for the critical coupling~$g_c$}
Having derived expressions for $\eta$ in \cref{eq:eta_incFinal}, $T_e$ in \cref{eq:Te_inc_Final} and $\anth$ in \cref{eq:alpha_inc_final}, these can now be substituted into the expression for the critical coupling in \cref{eq:gc} to obtain
\begin{equation}\label{eq:gc_inc_t1}
    \begin{aligned}
        g_c&=\left[\frac{D^R_0}{2}\left(\frac{\anth}{\eta}-\frac{1}{T_e}\right)\right]^{-1}\\
        &\approx\frac{2 \left(g_c I_\eta+\gamma T_b\right)}{D^R_0\left[\frac{\frac{2 I_\eta}{D^R_0}+\gamma\left(\frac{T_b \gamma}{g_c I_\eta+ \gamma T_b}-1\right)}{1+\frac{T_b \gamma}{g_c I_\eta+ \gamma T_b}}-\gamma\right]}\\
        &=\frac{2 \left(g_c I_\eta+\gamma T_b\right)}{D^R_0\frac{2\left(I_\eta-D^R_0 \gamma\right)\left(g_c I_\eta +\gamma T_b\right)}{D^R_0\left(g_c I_\eta +2 \gamma T_b\right)}}\\
        &=\frac{g_c I_\eta +2 \gamma T_b}{I_\eta-D^R_0 \gamma},
    \end{aligned}
\end{equation}
Isolating $g_c$, one arrives at the self-consistent analytical approximation
\begin{equation}\label{eq:gc_inc_final}
        g_c\approx-\frac{2 T_b}{D^R_0}=g_{c,\text{th}}(T_b).
\end{equation}

\section{Quasi-classical approximation and momentum averaging}
\label{app:quasi-classical_approx}
In most common implementations of Eliashberg theory for phonon-mediated superconductivity, the phonon-electron interaction is not forward-scattering as in \cref{eq:H_bf}, but rather has a smooth momentum dependence. 
Considering a more generic momentum-dependent interaction, the linearized version of the NESS-Eliashberg equation for $\Delta^R$ in \cref{eq:Delta_matrix_RAK} would generalize to the form
\begin{equation}
    \begin{split}
        \Delta^R_{\bm{k},\omega}=& i \int_{\bm{k}'} \int_{\omega'}  g_{\bm{k},\bm{k}'} \Big[ D^R G^K G^R_{-} \, \Delta^A_{\bm{k}',\omega'}\\
        &+ \left(D^K G^R G^A_{-} + D^R G^R G^R_{-}\right) \Delta^R_{\bm{k}',\omega'} \\
        &+D^K G^R G^R_{-} \, \Delta^K_{\bm{k}',\omega'} \Big],
    \end{split}
    \label{eq:DeltaR_momentum_average_app}
\end{equation}
where $D^{\alpha} G^{\beta} G^{\gamma}_{-}=D^{\alpha}_{\bm{k}-\bm{k}',\omega-\omega'} G^{\beta}_{\bm{k}',\omega'} G^{\gamma}_{-\bm{k}',-\omega'}$ and appropriate dimensionality constants have been reabsorbed into the coupling $g_{\bm{k},\bm{k}'}$.

To simplify the calculation, the momentum dependence of the coupling $g_{\bm{k},\bm{k}'}$ and of the boson propagators $D^{\alpha}_{\bm{k}-\bm{k}',\omega-\omega'}$ is often dropped, and this in turn makes the anomalous self-energy momentum independent, i.e.~$\Delta_{\bm{k},\omega}\equiv\Delta_{\omega}$ \cite{MarsiglioReview2020}. This corresponds to the so-called quasi-classical approximation \cite{KopninBook}, which is usually applicable in the regime $ T_c \ll \omega_0 \ll \mu$, where $T_c$ is the critical temperature, $\omega_0$ is a typical phonon energy scale and $\mu$ is the chemical potential
\cite{MarsiglioReview2020,AllenMitrovicBook}. Within this approximation, the only momentum dependent quantities in \cref{eq:DeltaR_momentum_average_app} are the electron propagators $G$. One can then perform the momentum integral by rewriting it as an integral over the energy 
\begin{equation}
    \int_{\bm{k}'} G^{\alpha}_{\bm{k}',\omega'} G^{\beta}_{-\bm{k}',-\omega'}=\int^{\varepsilon_{\mathrm{max}}}_{-\varepsilon_{\mathrm{min}}} \! d\varepsilon \, \nu(\varepsilon) \, G^{\alpha}_{\varepsilon,\omega'}
    G^{\beta}_{\varepsilon,-\omega'},
\end{equation}
where $\nu(\varepsilon)$ is the density of states at the energy $\varepsilon$, and $\varepsilon_{\mathrm{min}}$ and $\varepsilon_{\mathrm{max}}$ are some energy cutoffs of the electronic bands. The integral is typically estimated by approximating the density of states $\nu(\varepsilon)$ with its value at the FS $\nu_F$, by assuming particle-hole symmetry ($\varepsilon_{\rm{min}}=-\varepsilon_{\rm{max}}$) and by taking the limit for $\varepsilon_{\rm{max}}\to \infty$ \cite{MarsiglioReview2020}. The final integral to be computed is of the form
\begin{equation}
    \int^{\infty}_{-\infty} \! d\varepsilon \, G^{\alpha}_{\varepsilon,\omega'}
    G^{\beta}_{\varepsilon,-\omega'}.
\end{equation}
Focusing on the combination $G^R G^R_{-}$, which appears in \cref{eq:Delta_matrix_RAK} multiplying $\Delta^K$, and replacing the electron  propagator with its quasi-particle approximation $G^R_{\varepsilon,\omega}=1/(\omega-\varepsilon+i\eta)$, the integral becomes
\begin{equation}
    \begin{split}
        \int^{\infty}_{-\infty} \! d\varepsilon \, G^{R}_{\varepsilon,\omega'}
        G^{R}_{\varepsilon,-\omega'}
    =\int^{\infty}_{-\infty} \! d\varepsilon \frac{1}{(\epsilon-i\eta)^2-\omega'^2}=0.
    \end{split}
\end{equation}
This implies that, within the quasi-classical approximation, $\Delta^R$ and $\Delta^K$ are essentially decoupled at the phase transition and any non-thermal contribution to the critical coupling coming from $\Delta^K$, like the $\anth$ term in \cref{eq:gc}, is going to be suppressed. 
On the other hand, for any model where the momentum structure of the coupling function $g_{\bf{k},\bf{k}'}D^R_{\bf{k}-\bf{k}',\omega-\omega'}$ (and subsequently of the order parameter) is important, like in our forward-scattering interaction model of \cref{eq:H_bf}, $\Delta^R$ and $\Delta^K$ remain coupled and a non-thermal contribution from $\Delta^K$ always appears.

\bibliography{Eliashberg_bibliography.bib}

\clearpage
\end{document}